\documentclass[fleqn,usenatbib]{mnras}
\usepackage{fix-cm}
\usepackage{multirow}
\usepackage{booktabs}
\usepackage{lmodern}
\usepackage{newtxtext,newtxmath}

\usepackage[T1]{fontenc}

\DeclareRobustCommand{\VAN}[3]{#2}
\let\VANthebibliography\thebibliography
\def\thebibliography{\DeclareRobustCommand{\VAN}[3]{##3}\VANthebibliography}

\usepackage{graphicx}	
\usepackage{amsmath}

\usepackage{orcidlink}

\title[Manifold learning and JWST filters]{COSMOS-Web: Estimating Physical Parameters of Galaxies Using Self-Organizing Maps}

\author[F. Abedini et al.]{
Fatemeh Abedini\,\orcidlink{0009-0000-5827-5435}$^{1}$ \thanks{E-mail: fatemeh.abedini@iasbs.ac.ir},
Ghassem Gozaliasl\,\orcidlink{0000-0002-0236-919X}$^{2,3}$\thanks{E-mail: ghassem.gozaliasl@aalto.fi},
Akram Hasani Zonoozi\,\orcidlink{0000-0002-0322-9957}$^{1,4}$,
Atousa Kalantari\,\orcidlink{0009-0006-2285-6792}$^{1}$,\newauthor
Maarit Korpi-Lagg\,\orcidlink{0000-0002-9614-2200}$^{2}$,
Olivier Ilbert\,\orcidlink{0000-0002-7303-4397}$^{5}$,
Hollis B. Akins\,\orcidlink{0000-0003-3596-8794}$^{6}$,
Natalie Allen\,\orcidlink{0000-0001-9610-7950}$^{7,8}$,
Rafael C. Arango-Toro\,\orcidlink{0000-0002-0569-5222}$^{5}$,\newauthor
Caitlin M. Casey\,\orcidlink{0000-0002-0930-6466}$^{9,6,7}$,
Nicole E. Drakos\,\orcidlink{0000-0003-4761-2197}$^{10}$,
Andreas L. Faisst\,\orcidlink{0000-0002-9382-9832}$^{11}$,
Carter Flayhart\,\orcidlink{0009-0000-0272-5468}$^{12}$,\newauthor
Maximilien Franco\,\orcidlink{0000-0002-3560-8599}$^{13}$,
Hosein Haghi\,\orcidlink{0000-0002-9058-9677}$^{1,4}$,
Aryana Haghjoo$^{14}$,
Santosh Harish\,\orcidlink{0000-0003-0129-2079}$^{12}$,
Hossein Hatamnia\,\orcidlink{0009-0007-3673-4523}$^{14}$,\newauthor
Jeyhan S. Kartaltepe\,\orcidlink{0000-0001-9187-3605}$^{12}$,
Ali Ahmad Khostovan\,\orcidlink{0000-0002-0101-336X}$^{15,12}$,
Anton M. Koekemoer\,\orcidlink{0000-0002-6610-2048}$^{16}$,
Vasily Kokorev\,\orcidlink{0000-0002-5588-9156}$^{6}$,\newauthor
Rebecca L. Larson\,\orcidlink{0000-0003-2366-8858}$^{16}$,
Gavin Leroy\,\orcidlink{0009-0004-2523-4425}$^{17}$,
Daizhong Liu\,\orcidlink{0000-0001-9773-7479}$^{18}$,
Henry Joy McCracken\,\orcidlink{0000-0002-9489-7765}$^{19}$,
Jed McKinney\,\orcidlink{0000-0002-6149-8178}$^{6}$,\newauthor
Nicolas McMahon\,\orcidlink{0009-0003-9297-6587}$^{12}$,
Wilfried Mercier\,\orcidlink{0000-0001-6865-499X}$^{5}$,
Bahram Mobasher$^{14}$,
Sophie Newman\,\orcidlink{0009-0001-3422-3048}$^{20}$,\newauthor
Louise Paquereau\,\orcidlink{0000-0003-2397-0360}$^{19}$,
Jason Rhodes\,\orcidlink{0000-0002-4485-8549}$^{21}$, 
Brant E. Robertson\,\orcidlink{0000-0002-4271-0364}$^{22}$,
Sogol Sanjaripour\,\orcidlink{0009-0009-3048-9090}$^{14}$,\newauthor
Marko Shuntov\,\orcidlink{0000-0002-7087-0701}$^{7,8,23}$,
Sina Taamoli\,\orcidlink{0000-0003-0749-4667}$^{14}$,
Sune Toft\,\orcidlink{0000-0003-3631-7176}$^{7,8}$,
Francesco Valentino\,\orcidlink{0000-0001-6477-4011}$^{7,24}$,
Eleni Vardoulaki\,\orcidlink{0000-0002-4437-1773}$^{25}$,\newauthor
John R. Weaver\,\orcidlink{0000-0003-1614-196X}$^{26}$ \\
(Affiliations can be found after the references)
}

\date{Accepted XXX. Received YYY; in original form ZZZ}
\pubyear{\the\year{}}
\begin{document}
\label{firstpage}
\pagerange{\pageref{firstpage}--\pageref{lastpage}}
\maketitle

\begin{abstract}
The \textit{COSMOS-Web} survey, with its unparalleled combination of multiband data, notably, near-infrared imaging from \textit{JWST}'s NIRCam (F115W, F150W, F277W, and F444W), provides a transformative dataset down to $\sim28$ mag (F444W) for studying galaxy evolution. In this work, we employ Self-Organizing Maps (SOMs), an unsupervised machine learning method, to estimate key physical parameters of galaxies---redshift, stellar mass, star formation rate (SFR), specific SFR (sSFR), and age---directly from photometric data out to $z=3.5$. SOMs efficiently project high-dimensional galaxy color information onto 2D maps, showing how physical properties vary among galaxies with similar spectral energy distributions.
We first validate our approach using mock galaxy catalogs from the \textit{HORIZON-AGN} simulation, where the SOM accurately recovers the true parameters, demonstrating its robustness. Applying the method to \textit{COSMOS-Web} observations, we find that the SOM delivers robust estimates despite the increased complexity of real galaxy populations. Performance metrics ($\sigma_{\mathrm{NMAD}}$ typically between $0.1$--$0.3$, and Pearson correlation between $0.7$ and $0.9$) confirm the precision of the method, with $\sim$ $70\%$ of predictions within 1$\sigma$ dex of reference values. Although redshift estimation in COSMOS-Web remains challenging (median $\sigma_{\mathrm{NMAD}} = 0.04$), the overall success of the highlights its potential as a powerful and interpretable tool for galaxy parameter estimation. A key advance of this work is the use of JWST/NIRCam photometry, particularly the F444W band, which enhances SOM training and allows more accurate estimation of stellar mass, SFR, and age compared to previous studies using IRAC/Spitzer filters.

\end{abstract}

\begin{keywords}
galaxies: evolution – galaxies: fundamental parameters – method: data analysis – method: statistical
\end{keywords}



\section{Introduction}

The physical parameters of galaxies, such as redshift, stellar mass, star formation activity, and age, are crucial for studying the formation and evolution of galaxies through cosmic time. Multi-wavelength datasets combined with spectroscopic samples have advanced our understanding of galaxy formation, evolutionary stages, environments, and cosmic epochs. These surveys utilize various filters and depths to observe large numbers of galaxies and provide valuable observational data. In recent decades, ground- and space-based galaxy surveys such as COSMOS \citep{Scoville2007, laigle2016, weaver2022}, CANDELS \citep{grogin2011, Koekemoer2011}, and UltraVISTA \citep{McCracken2012, muzzin2013} have been conducted. Spectroscopic surveys have also targeted narrower wavelength ranges with higher resolution, including SDSS \citep{York2000}, GAMA \citep{driver2011}, AGES \citep{kochanek2012}, and VANDELS \citep{mclure2018}.


Recent and upcoming surveys include COSMOS-Web \citep{shuntov2025}, the largest James Webb Space Telescope (JWST) program in both area and total prime-time allocation, covering 0.54 ${\mathrm{deg}}^2$ with a 5$\sigma$ point-source depth of approximately 27.5 to 28.2 magnitudes in the NIRCam filters; Euclid \citep{laureijs2011, amendola2018}, covering 14,000 ${\mathrm{deg}}^2$ with a 5$\sigma$ depth of 26.2 in the I-band; LSST \citep{lsstsciencecollaboration, Ivezi2019}, which will cover 18,000 ${\mathrm{deg}}^2$ and reach a 5$\sigma$ depth of 27.5 in the r-band; and the Nancy Grace Roman Space Telescope \citep{spergel2015, akeson2019}, which will cover 2,400 ${\text{deg}}^2$ with a 5$\sigma$ depth of 27.9 in the F062 (red) band.

Furthermore, future spectrographs will be capable of capturing thousands of spectra simultaneously across wide fields of view. The volume of spectroscopic data will grow significantly, depending on the methods used for object selection and the biases and limitations inherent in constructing spectroscopic galaxy samples \citep{wang2022, jin2024}. These surveys will observe billions of galaxies across multiple filters, which are essential for developing efficient and accurate methods to estimate the physical parameters of galaxies.

One of the standard methods for estimating the physical properties of galaxies, such as redshift, stellar mass, star formation rate (SFR), and specific SFR (sSFR), is the spectral energy distribution (SED) fitting \citep{gallazzi}. This method involves comparing a galaxy's observed light across multiple wavelengths, or its SED, to a set of theoretical or empirical templates. These templates are built by modeling the effects of various physical properties on the SED and are constructed using stellar population synthesis models. These models simulate the light emitted by galaxies with different ages, metallicities, star formation histories, and dust attenuation \citep{sawicki1998,walcher2011,conroy2013,leja2017,pacific2023}.

By determining the probability distribution functions, the best-fit templates are identified, and the properties of the galaxy are derived by analyzing these fits to the observational data. SED fitting is particularly useful when spectroscopic data are limited, especially in large-scale galaxy surveys. Although SED fitting is widely used, there are also challenges associated with this physically motivated method. Uncertainties in stellar population models can introduce biases \citep{mitchell2013,dahlen2013,mobasher2015,laigle2019}, and the similarity of templates with different combinations of physical parameters can lead to ambiguities in the fitting process. Additionally, SED fitting depends on models \citep{lower2020} and template libraries may be incomplete, failing to account for all types of galaxies in the universe \citep{marchesini2010,muzzin2013}. Finally, for large-scale observational data and complex template libraries, the process can be computationally expensive \citep{conroy2013,speagle2016}.

These limitations motivate the use of alternative approaches, such as machine learning methods. In recent years, the use of machine learning techniques has been increasingly explored for various purposes \citep{ball2010, baron2019, FOTOPOULOU2024}, including classification and morphological analysis \citep{dieleman2015, huertas2015, aussel2024, signor2024}, modeling of strong lensing systems \citep{Hezaveh2017, Leuzzi2024}, and especially to estimate physical parameters from observed photometry \citep{collister2004, salvato2019, mucesh2021, alsing2024, enia2024, aufort2024, kovavic2025}.

\citet{dobbels2020} used shallow neural networks to predict far-infrared fluxes, dust luminosity, dust mass, and effective temperature across the UV to MIR bands, based on a low-redshift sample from the Herschel-ATLAS (H-ATLAS) DR1 data \citep{valiante2016} and DustPedia \citep{davies2017}. They also noted that random forest and linear models are effective for flux predictions. \citet{bisigello2023} implemented deep learning neural networks and convolutional neural networks to estimate redshifts, stellar masses, and SFRs using mock data from the Euclid and Rubin/LSST surveys. \citet{gilda2021} presented MIRKWOOD, a supervised machine learning model based on the NGBOOST algorithm, to derive stellar masses, dust masses, stellar metallicities, and instantaneous SFR using multiband data from GALEX, HST, Spitzer, Herschel, and JWST, applied to mock galaxy SEDs. \citet{surana2020} employed deep neural networks to predict stellar mass, SFR, and dust luminosity from multiband data in the GAMA survey.

Due to their data-driven nature, machine learning methods are less dependent on models and templates to estimate physical parameters. Machine learning is also computationally efficient and capable of processing large datasets, an important advantage in the era of large-scale astronomical surveys. Furthermore, it can effectively handle complex, non-linear relationships in galaxy data. However, the performance of machine learning predictions is highly dependent on the quality and size of the input data. Observational features may contain errors, for example from the photometric process, which can cause machine learning methods to incorrectly estimate physical parameters related to galaxy evolution. It is therefore important to objectively evaluate machine learning predictions and compare them with standard methods, such as SED fitting, to assess their reliability and potential biases \citep{jafariyazani2024, enia2024}.

One of the machine learning algorithms is Self-Organizing Map (SOM), an unsupervised learning method in the class of manifold learning algorithms that reduces the dimensionality of complex data while preserving its topological structure \citep{kohonen1981hierarchical, kohonen2001self}. In the context of galaxy surveys, SOM is typically trained on high-dimensional data, such as galaxy colors. Once the SOM is trained, similar input data in the high-dimensional space are mapped to nearby locations on a lower-dimensional (typically 2D) grid. SOM have been used for tasks such as the classification of galaxy spectra, galaxy morphologies \citep{galvin2019, faisst2019, sanjaripour2024}, and the estimation of physical parameters of galaxies, some of which are reviewed below.

In \citet{master2015}, SOM was used to visualize the high-dimensional color space of galaxies and identify areas requiring spectroscopic observations, with a focus on the Euclid survey. In \citet{master2017} and \citet{master2019}, the work continued by presenting the C3R2 survey, which conducted spectroscopic observations on the regions identified by SOM as needing further study. This approach was used to calibrate the color-redshift relation for the upcoming Euclid, LSST, and WFIRST surveys. \citet{hemmati2019a} employed SOM based on CANDELS observations to create a 2D map for efficient spectroscopic targeting and photometric calibration for the WFIRST survey. In another paper by \citet{hemmati2019b}, the same method was applied by training SOM with templates to compare the stellar mass of COSMOS2015 galaxies at redshift $\sim1$ with those derived from SED fitting. \citet{davidzon2019} used the Horizon AGN simulation \citet{laigle2019} as SOM input data to estimate redshift and SFR using galaxy colors. In \citet{davidzon2022}, SOM was trained on COSMOS 2020 data to estimate stellar mass and SFR for SXDF data. Other papers, such as \citet{jafariyazani2024} and \citet{latorre2024}, also applied SOM to estimate the physical parameters of galaxies.

The goal of this paper is to use SOM to estimate physical parameters of galaxies, such as redshift, stellar mass, SFR, sSFR, and age, by training the SOM separately on simulated and observational data and comparing the results. We also apply the SOM trained on simulated data to observational data to evaluate the reliability and limitations of the simulation and to understand how well the simulated data generalize to real observations. SOMs are trained across low-, medium-, and high-redshift bins to assess their performance at different cosmic epochs.

We use two specific datasets: the HORIZON-AGN (HZ-AGN) simulation, a mock dataset matched to the COSMOS field, and the COSMOS-Web (CW) survey data as the observational counterpart. Two JWST filters, F277W and F444W, are used to investigate whether these NIRCam filters can improve the precision of estimating physical parameters compared to previous studies. Our findings provide insights into both the potential and the challenges of using SOMs for next-generation galaxy surveys.

This paper is organized as follows. Section~\ref{sec:data} provides a brief description of the HZ-AGN and CW data. Section~\ref{sec:method} explains the SOM, the input data, data preparation, SOM training, test data, and the prediction phase. Section~\ref{sec:results} presents the results in three subsections: (\ref{subsec:som_sim}) training the SOM with HZ-AGN data and predicting on a separate HZ-AGN dataset; (\ref{subsec:som_obs}) training with CW data and predicting on a separate CW dataset; and (\ref{subsec:som_sim_obs}) applying the SOM trained on HZ-AGN data to CW data. In Section~\ref{sec:discussion}, we compare our results with previous SOM-based studies of physical parameter estimation, discuss the limitations of our predictions and possible ways to address them, evaluate the strengths and weaknesses of SOMs in this context, and provide practical guidelines for observers based on the key results of this work.
Finally, Section~\ref{sec:conclusions} summarizes the study and presents the main conclusions.

Throughout this paper, all the magnitudes are expressed in the AB system \citep{oke1974}, and we adopt a $\Lambda$CDM cosmology with $H_0 = 70 \, \text{km} \, \text{s}^{-1} \, \text{Mpc}^{-1}$, $\Omega_m = 0.3$, and $\Omega_\Lambda = 0.7$. We use the initial mass function (IMF) from \citep{chabrier2003}.

\section{Data}
\label{sec:data}
\subsection{The HORIZON-AGN Virtual Observatory}

The HZ-AGN simulation is run using the {\fontfamily{cmtt}\selectfont RAMSES} \citep{teyssier2002} code, which uses adaptive mesh refinement and includes a wide range of physical processes. These include gas cooling down to a temperature of ${10}^{4}$ K due to H and He collisions, as well as the contribution from metals \citep{sutherland1993}. Gas heating is included from a uniform UV background \citep{haardt1996}, assuming reionization occurred at redshift $z = 10$. Star formation is modeled using the Kennicutt-Schmidt law \citep{schmidt1959,kennicutt1998} with a constant star formation efficiency of 0.02. Stellar feedback is included through mechanical energy \citep{dubois2008} from stellar winds and Type II supernovae, using the {\fontfamily{cmtt}\selectfont STARBURST99} model. Type I supernovae are also considered, following the explosion rate from \citet{greggio1983}. AGN feedback is included in the simulation using the model by \citet{dubois2012}, which accounts for both radio and quasar AGN modes. This is based on the Eddington-limited Bondi-Hoyle-Lyttleton gas accretion rate onto massive black holes. This step is important for properly modeling AGN-driven early-type galaxies and capturing the variety of galaxy shapes in cosmological simulations \citep{dubois2013a,dubois2016}.

The HZ-AGN simulation provides reasonable basic statistics on galaxy populations, such as the mass and HI mass functions, cosmic SFR density, and morphology \citep{kaviraj2017,kokorev2021, Margalef2020}. These data represent a wealth of information on galaxy properties including redshift, stellar mass, SFR, and formation ages, covering the redshift range $0 < z < 4$. The BC03 \citet{bc2003} model is used for simple stellar populations (SSPs), with the same IMF and metallicity settings as those in the COSMOS2015 catalog \citep{laigle2016}. The mass of each stellar particle in the simulation is $2\times10^6,M_\odot$. The dust column density of each galaxy is computed for every stellar particle, using the gas metallicity to model dust distribution, assuming a constant dust-to-metal mass ratio, and a Milky Way extinction curve. Dust attenuation depends on the geometry and metal content of the galaxy. Nebular emission lines are not considered in the flux contamination of the mock photometry (see \citet{laigle2019, toro2024} for more information). The true values of the SFR and stellar mass at every age are available for each galaxy in the HZ-AGN data, with a time resolution of 1 Myr.  

In this study, we use the mock catalog by \citet{laigle2019}. The integrated galaxy spectra are derived by summing all dust-attenuated SSPs of each galaxy's stellar particles. The flux is calculated by integrating the redshifted galaxy spectra over the filter transmission curves. This catalog includes the COSMOS filter set, and the mock data are extended with JWST NIRCam and MIRI filters. Noise is added to the predicted fluxes by perturbing them according to flux uncertainties as a function of flux in the observational data.

\subsection{The COSMOS-Web Data}

\subsubsection{Photometric Catalog}

The Cosmic Evolution Survey (COSMOS) \citep{scoville2013,capak2007b} covers a two-square-degree area of the sky and includes multi-wavelength observations \citep{capak2007,ilbert2009,ilbert2013,laigle2016,weaver2022}, ranging from X-ray to radio \citep{civano2016,hasinger2007,smolcic2017}. Optical photometry data includes observations from CFHT's MegaCam in the u-band \citep{sawicki2019}, the Hubble Space Telescope (ACS-HST) in the F814W band \citep{koekemoer2007}, the g, r, i, z, and y bands from Hyper-Suprime-Cam (HSC), 13 intermediate and narrow bands from Subaru Suprime-Cam (SC) \citep{taniguchi2015}, and NIR observations from the UltraVISTA survey in the Y, J, H, and Ks bands \citep{McCracken2012}. These deep observations enable studies of galaxy formation and evolution. The COSMOS-Web (CW) survey is a new photometric catalog that includes the first 255 hours of observations of the COSMOS field using the JWST \citep{casey2022, shuntov2025}, and is designed to bridge the gap between deep-field surveys from the Hubble Space Telescope and shallower wide-area surveys, such as the upcoming Roman Space Telescope and Euclid missions.

The CW survey covers a contiguous area of $0.54$ deg$^2$. The NIRCam filters \citep{rieke2023} used are F115W, F150W, F277W, and F444W, achieving a $5\sigma$ point-source depth of approximately $27.5-28.2$ AB magnitudes. To create NIRCam mosaics, the survey includes 152 visits per filter. The high resolution of NIRCam helps to separate sources that were blended in ground-based observations. Additionally, the MIRI F770W filter \citep{wright2022} observes a non-contiguous $0.19$ deg$^2$ area with a depth of approximately $25.33-26.0$ AB magnitudes \citep{casey2022,shuntov2025}.

The {\fontfamily{cmtt}\selectfont SourceXtractor++} software \citep{bertin2022} is used to model Sérsic surface brightness profiles \citep{seric1963} for sources in native-resolution NIRCam images \citep{toro2024, shuntov2025}. In total, the CW survey has detected nearly $800,000$ galaxies. Multi-wavelength data from this survey aid in the detection of more galaxies within large-scale structures and improve the estimation of their physical parameters.

\subsubsection{Spectroscopic Redshift for COSMOS-Web Data}

Although photometric redshifts derived from SED fitting provide essential information about galaxies, spectroscopic redshifts are crucial for calibrating and validating them. Incorporating spectroscopic redshifts along with photometric redshifts during SED fitting leads to more reliable measurements of physical parameters. This is particularly beneficial for investigating various aspects of galaxy evolution and how physical properties change over cosmic time. Additionally, it helps in studying the 3D large-scale structure, allowing researchers to examine how the environment affects galaxy evolution.

However, measuring spectroscopic redshifts for all galaxies in a field is challenging because of the time and cost involved in obtaining measurements for every galaxy. As a result, photometric redshifts are often used in conjunction with spectroscopic redshifts. For the COSMOS field, multiple surveys have been conducted to measure spectroscopic redshifts \citet{khostovan2025}. 

These surveys include zCOSMOS-bright \citep{lilly2007}, FORS2/VLT\citep{comparat2015}, DEIMOS/Keck II \citep{kartaptepe2010}, FMOS/Subaru \citep{roseboom2012}, MOIRC/Subaru \citep{onodera2012}, WFC3-grism/HST \citep{krogager2014}, zCOSMOS-deep, MOSDEF \citep{kriek2015}, XSHOOTER/VLT, VUDS \citep{lefvre2015}, and DEIMOS 10K \citep{hasinger2018}. These spectroscopic surveys cover a wide range of redshifts. For more information on these surveys and their redshift coverage, see Table 2 of \citet{gozaliasl2024}.

\section{Method}
\label{sec:method}
The goal of this work is to predict the physical parameters of galaxies from broadband photometry using a conditional density estimator based on a SOM. In this section, we first focus on the SOM method and objective of the prediction task, which is the fundamental approach we used, and then describe the steps taken to achieve this goal.

\subsection{Self-Organizing Map (SOM)}

The SOM \citep{kohonen1981hierarchical,kohonen2001self,Kohonen2014} is an unsupervised learning algorithm that projects high-dimensional data onto a lower-dimensional, typically 2D map, while preserving the data topology. The training procedure can be formalized as follows:

Each cell $i$ of the SOM grid is associated with a weight vector $w_i \in \mathbb{R}^d$, where $d$ is the dimensionality of the input space ($d = |\mathcal{D}|$ in our case). The training begins with randomly initialized weight vectors. For each input feature vector $x(t)$, the best matching unit (BMU) is identified by finding the closest weight vector in Euclidean distance:
\begin{equation}
c = {\arg\min}_i { \lVert x(t) - w_i(t) \rVert },
\end{equation}
that is, the cell whose weight vector is closest to $x(t)$.

At each training step $t$, the weights are updated according to
\begin{equation}
w_i(t+1) = w_i(t) + h_{ci}(t)[x(t) - w_i(t)],
\end{equation}
where $h_{ci}(t)$ is the neighborhood function that decreases with the distance between cell $i$ and the $\rm BMU_c$ on the SOM grid:
\begin{equation}
h_{ci}(t) = \alpha(t)\exp\left(-\frac{\lVert r_c - r_i \rVert^2}{2\sigma^2(t)}\right).
\end{equation}
Here, $\alpha(t)$ is the learning rate and $\sigma(t)$ is the neighborhood radius, both of which are monotonically decreasing functions of time. In the final convergence phase, the neighborhood radius $\sigma(t)$ is typically reduced to zero (or a very small value), so that ultimately only the BMU itself is updated. This fine-tunes the map while preserving the topographic order established during the initial coarse ordering phase. This iterative procedure adjusts the weights so that the SOM progressively learns to represent the distribution and topology of the input data.

We use the {\fontfamily{cmtt}\selectfont SOMPY}\footnote{\url{https://github.com/sevamoo/SOMPY}} package \citep{moosavi2014sompy} to construct and train the SOM. In SOMPY, the training data $x$ are first normalized to zero mean and unit variance:
\begin{equation}
    \tilde{x} = \frac{x - \mu}{\sigma},
\end{equation}
where $\mu$ and $\sigma$ are the mean and standard deviation of the training set, respectively.

The weight vectors of each cell are initialized using principal component analysis (PCA) \citep{chatfield1980multivariate}. Given the covariance matrix of the training data
\begin{equation}
    C = \frac{1}{N}\sum_{n=1}^N \big(x_n - \mu\big)\big(x_n - \mu\big)^\top ,
\end{equation}
the eigenvalue decomposition
\begin{equation}
    Cu_k = \lambda_k u_k
\end{equation}
provides eigenvectors $u_k$ (principal components). The initial weights $w_i(0)$ are then aligned along the first two principal components,
\begin{equation}
    w_i(0) \in \mathrm{span}\{u_1, u_2\},
\end{equation}
ensuring that the initial weights are close to the distribution of the input data. During training, input data with similar characteristics are grouped into neighboring cells. This allows for meaningful analysis of high-dimensional data, through a 2D projection, making SOM a powerful tool for visualizing and organizing complex datasets.

We denote the target variables as:
\begin{equation}
    \theta = \{ z, M_*, \mathrm{SFR}, \mathrm{sSFR}, \mathrm{age}_\mathrm{mw} \}
\end{equation}
where $z$ is the redshift, $M_*$ is the stellar mass, SFR and sSFR denote the star formation rate and the specific star formation rate, respectively. The input data consist of the colors:
\begin{equation}
\begin{aligned}
 \mathcal{D} = \{ & \mathrm{C}_{\mathrm{ug}}, \, \mathrm{C}_{\mathrm{gr}}, \, \mathrm{C}_{\mathrm{ri}}, \, \mathrm{C}_{\mathrm{iz}}, \, \mathrm{C}_{\mathrm{zY}}, \, \mathrm{C}_{\mathrm{YJ}}, \, \mathrm{C}_{\mathrm{JH}}, 
                      \mathrm{C}_{\mathrm{HK}}, \, \mathrm{C}_{\mathrm{KF277}}, \, \\ & \mathrm{C}_{\mathrm{F277F444}} \}
\end{aligned}
\end{equation}

The prediction task can be formally expressed using the conditional density
\begin{equation}
p(\theta \mid \mathcal{D}),
\end{equation}
which is described in Subsection \ref{subsec:Prediction}.

\begin{figure*}
    \includegraphics[width=1\textwidth]{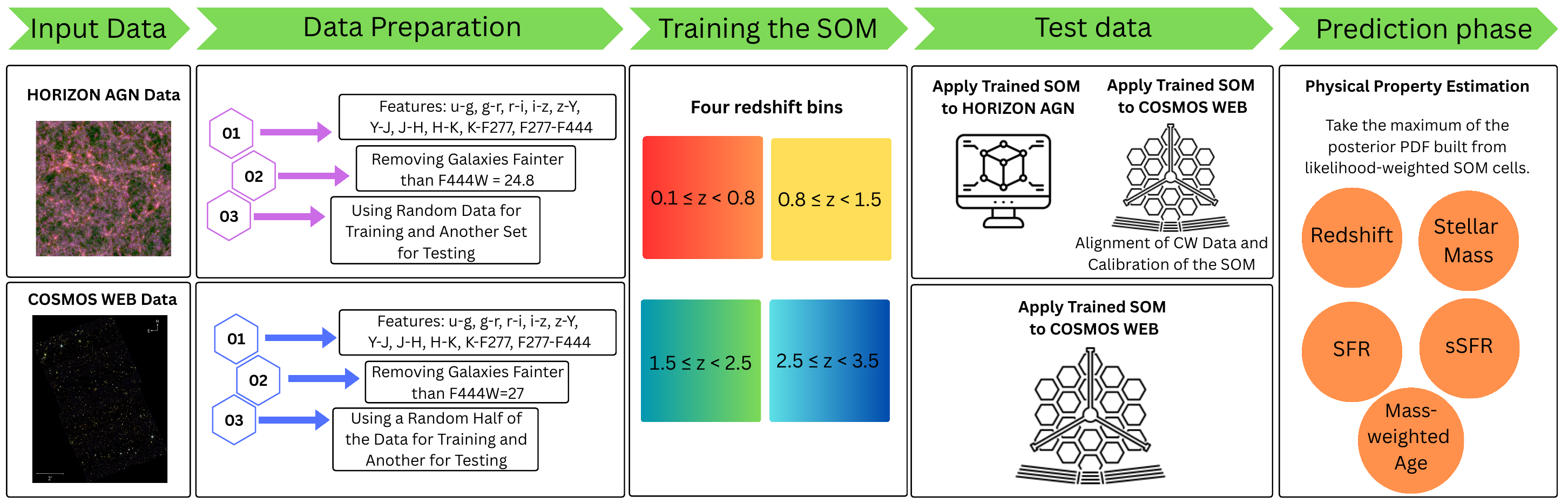}
    \caption{Workflow for estimating galaxy physical properties using SOMs. The process involves two datasets: HZ-AGN simulations and CW observations. After initial preprocessing, including feature selection, magnitude cuts, and splitting the data into training and testing sets, SOMs are trained separately in four redshift bins. The SOM trained on the HZ-AGN data is applied to both the HZ-AGN and CW data, while the SOM trained on the CW data is applied to the CW data. Physical parameters such as redshift, stellar mass, SFR, sSFR, and mass-weighted age are estimated from the peak of the posterior PDF constructed using likelihood-weighted SOM cells.
}
    \label{fig:flowchart}
\end{figure*}

Figure \ref{fig:flowchart} shows a schematic overview of the proposed method. The steps proceed from (i) input data, (ii) data preparation, (iii) training the SOM, (iv) test data, and (v) prediction phase. In the following subsections, we formalize each step mathematically.

\subsection{Input Data}
\label{subsec:input_data}

We utilize two datasets as the input data: the HZ-AGN simulation and the CW observational data. The filters used in this study are: MegaCam/CFHT u, HSC/Subaru g, r, i, z, VIRCAM/VISTA Y, J, H, $K_s$, and two NIRCam filters from JWST, F277W, and F444W. We specifically chose JWST F277W and F444W instead of the Spitzer/IRAC ch1 and ch2 bands used in earlier studies, allowing for a more direct comparison. Additionally, using only these two filters improves data completeness. Although F115W and F150W are included in HZ-AGN and CW, we chose not to use them to maintain consistency and avoid introducing inhomogeneities due to partial coverage with other filters such as Y, J, and H. The impact of including F115W and F150W on improving estimates of physical parameters will be explored by Kalantari et al. (in preparation).

Table \ref{tab:filters} presents the bands used for both simulation and observational data, covering wavelengths from the optical to the near-infrared. The table also includes the $5\sigma$ empty-aperture depth of the bands \citep{shuntov2025}. For CW data, we use the model magnitudes derived from fitting Sérsic surface brightness profiles \citep{seric1963} with the {\fontfamily{cmtt}\selectfont SourceXtractor++} software \citep{bertin2022,shuntov2025}. To maintain consistency with the CW data, we use magnitudes with added errors in the simulated data, making them more realistic under observational conditions \citep{laigle2019}.

\begin{table}
\caption{Optical and NIR data in observed-frame colors used for training the SOM \citep{shuntov2025}}
\label{tab:filters}
\begin{tabular}{|l|l|l|l|l|}
\hline
Instrument & Band  & Central\textsuperscript{(a)} & Width\textsuperscript{(b)} & CW Depth\textsuperscript{(c)} \\
           &      & $\lambda$ [\AA]             & $\lambda$ [\AA]     & [mag]       \\
\hline
\multirow{1}{*}{MegaCam/CFHT} & $u^*$ & 3858 & 598 & 27.3 \\ 
\hline
\multirow{5}{*}{Subaru/HSC} 
& g & 4847 & 1383 & 27.6 \\
& r & 6219 & 1547 & 27.2 \\
& i & 7699 & 1471 & 27.0 \\
& z & 8894 & 766 & 26.6 \\
\hline
\multirow{4}{*}{VISTA/VIRCAM} 
& Y & 10216 & 923 & 25.8 \\
& J & 12525 & 1718 & 25.8 \\
& H & 16466 & 2905 & 25.5 \\
& Ks & 21557 & 3074 & 25.3 \\
\hline
\multirow{2}{*}{JWST/NIRCam} 
& F277 & 28001 & 6999 & 28.1 \\
& F444 & 44366 & 11109 & 28.0 \\
\hline
\end{tabular}

\textsuperscript{a} Median of the transmission curve

\textsuperscript{b} Full width at half maximum (FWHM) of the transmission curve 

\textsuperscript{c} 5$\sigma$ depth computed in empty apertures with diameters  of $1.0^{\prime\prime}$ for the ground-based, $0.15^{\prime\prime}$ for the space-based JWST/NIRCam and HST/ACS.
\end{table}

\begin{figure}
	\includegraphics[width=1\columnwidth]{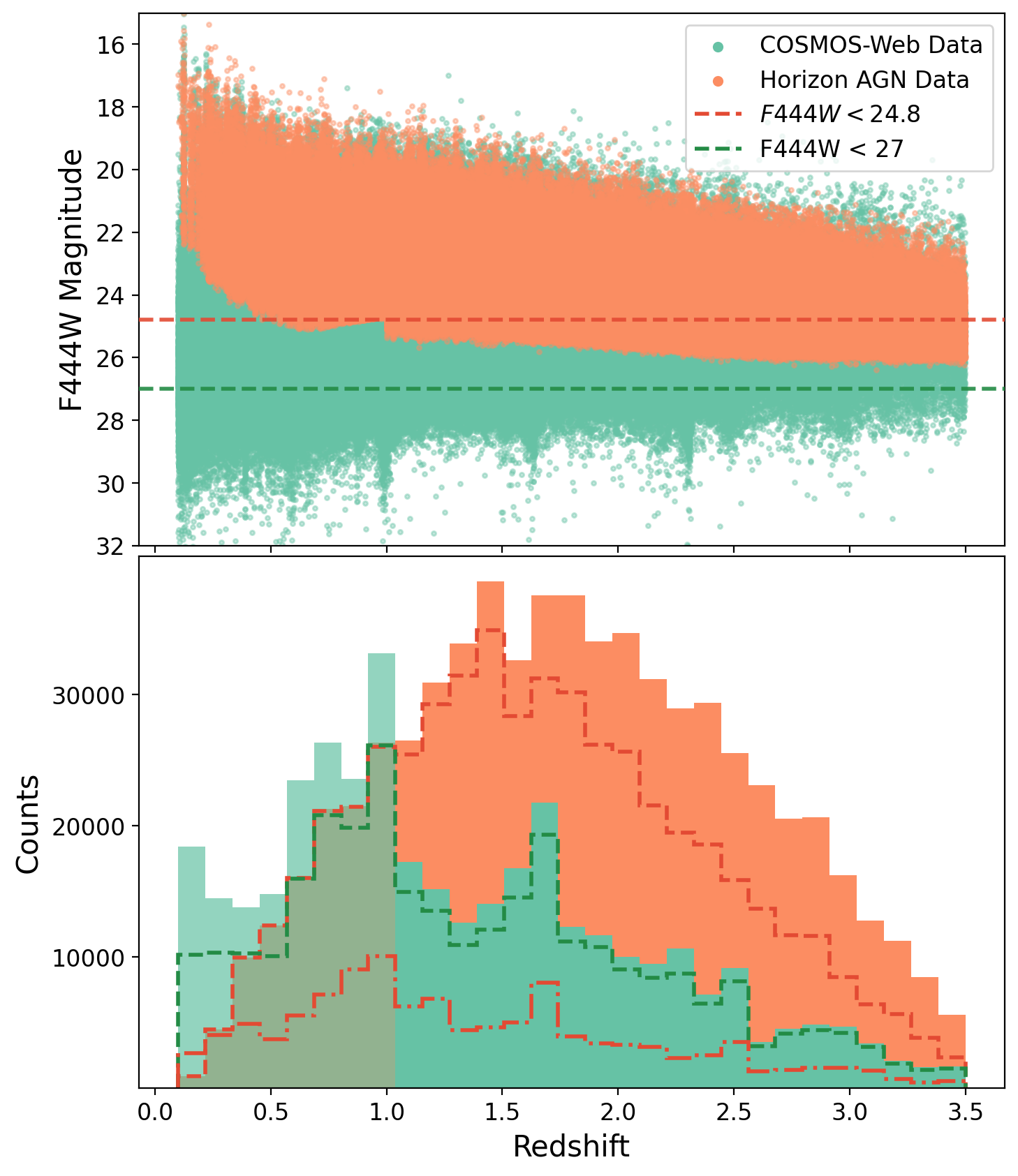}
    \caption{Comparison of the HZ-AGN catalog and the CW catalog in the redshift range $0.1 \leq z < 3.5$. Upper panel: Magnitude-redshift distribution in F444W magnitudes. The dashed lines show the magnitude completeness for the simulation and observational data. Lower panel: Redshift distribution of the simulation and observational data. The dashed lines represent the simulation and observational data with $m_\mathrm{F444W} < 24.8$ and $m_\mathrm{F444W} < 27$, respectively. The dot-dashed line represents the $m_\mathrm{F444W} < 24.8$ magnitude cut applied to the observational data prior to applying it to the SOM trained on simulation data.
    }
    \label{fig:distribution}
\end{figure}

The redshift range considered in this study is $0.1 \leq z \leq 3.5$ for both datasets. 
Due to the lack of observational data beyond $z=3.5$, galaxies above this threshold were excluded. The total number of simulated data points is $652,962$, and the number of observational data points is  $362,298$. The upper panel of Figure \ref{fig:distribution} shows the redshift distribution of all simulated and observational data from $0.1$ to $3.5$, demonstrating that the COSMOS-Web data have deeper magnitudes with greater dispersion. The lower panel shows the number of objects across the redshift range, indicating that below a redshift of 1, the number of observational data points is higher, while above a redshift of 1, the number of simulated data points is higher than that of the observational data.

\subsection{Data Preparation}
\label{subsec:data_prepration}
The preparation of the HZ-AGN and CW datasets for SOM training proceeds in the following steps:

\begin{enumerate}
    \item \textbf{Feature assignment.}  
    The features provided for both HZ-AGN and CW are colors, $\mathcal{D}(\mathrm{C})$, defined as: $u-g$, $g-r$, $r-i$, $i-z$, $z-Y$,$Y-J$, $J-H$, $H-K_\mathrm{S}$, $K_\mathrm{S}-\mathrm{F277}$, and $\mathrm{F277W}-\mathrm{F444W}$.
    
    \item \textbf{Physical parameters.}  
    For HZ-AGN, the physical parameters are the intrinsic simulation values. For CW, we used photometric redshifts by default and substituted spectroscopic redshifts when available. We cross-matched the spectroscopic dataset ($166{,}506$ objects) with the COSMOS-Web catalog ($783{,}846$ objects) using a $1^{\prime\prime}$ radius and excluded NaN entries, yielding $33{,}704$ galaxies with spectroscopic redshifts. Throughout this paper, “redshift” for CW refers to either photometric or spectroscopic values depending on availability. Stellar mass, SFR, and sSFR for CW are taken from the median of the {\fontfamily{cmtt}\selectfont LePHARE} SED fitting outputs \citep{arnouts2002,ilbert2006}. Since the HZ-AGN data provide the mass-weighted age, we compute the mass-weighted age ($\mathrm{age}_\mathrm{mw}$) following the method described in \citet{Wuyts2011,gozaliasl2024}, using the equation:
    \begin{equation}
    \mathrm{age}_{\mathrm{mw}} =
    \frac{\displaystyle \int_{0}^{t_{\mathrm{obs}}}
    \mathrm{SFR}(t)(t_{\mathrm{obs}} - t)\mathrm{d}t}
    {\displaystyle \int_{0}^{t_{\mathrm{obs}}}
    \mathrm{SFR}(t) \mathrm{d}t}.
    \label{eq:massweighted_age}
    \end{equation}
    Here, the star formation rate (SFR) can follow either an exponential model ($e^{-t/\tau}$) or a delayed-$\tau$ model ($\tau^{-2} t e^{-t/\tau}$), and $t_{\mathrm{obs}}$ denotes the age of the galaxy.

    \item \textbf{Magnitude completeness cuts.}  
    A completeness cut in the F444W filter is applied to both datasets: $m_\mathrm{F444W}<24.8$ for HZ-AGN and $m_\mathrm{F444W}<27$ for CW \citep{toro2024}, with the limits shown as dashed lines in Fig.~\ref{fig:distribution}. The total number of objects in the HZ-AGN dataset is approximately $790,000$ within $1~\mathrm{deg}^{2}$, while the CW dataset contains about $780,000$ objects over $0.54~\mathrm{deg}^{2}$. Consequently, the surface number density of the two datasets (counts/$\mathrm{degree}^2$) differ by approximately a factor of two \citep{laigle2016, shuntov2025}. After applying the magnitude completeness cut, the total number of objects in the redshift range $0.1 \leq z < 3.5$ is $514{,}223$ for the HZ-AGN dataset and $295{,}769$ for the CW dataset, with roughly similar surface number densities. This completeness cut is applied to both HZ-AGN and CW whenever a SOM is trained and then applied to the same dataset (HZ-AGN$\rightarrow$HZ-AGN, CW$\rightarrow$CW), where the arrow indicates the dataset used for training and the dataset used for application, respectively. Since we also apply the SOM trained on simulation data to observational data (HZ-AGN$\rightarrow$CW), we impose the same cut of $m_\mathrm{F444W} < 24.8$ on the observational test data to match the simulation’s completeness. This is illustrated by the dot-dashed line in Figure~\ref{fig:distribution}, which shows a reduction in the number of observational data points within this limit.
    
    \item \textbf{Redshift binning.}  
    We train the SOM for each assigned redshift bin: $0.1$–$0.8$, $0.8$–$1.5$, $1.5$–$2.5$, and $2.5$–$3.5$. Training separately in each bin allows us to better assess the performance of predicting physical parameters at different redshifts and reduces degeneracies across cosmic time.
    
    \item \textbf{Train/test splits.} In SOM training, it is important that the number and distribution of training and testing data are similar, to ensure reliable mapping and preservation of the data topology. Because the SOM trained on HZ-AGN is applied to the CW data, and the CW dataset contains $113{,}651$ galaxies within the adopted magnitude completeness limit ($m_\mathrm{F444W} < 24.8$), we downsample the HZ-AGN simulation (which initially contains $514{,}233$ galaxies after the completeness cut) to the same size. Specifically, we randomly select an equal number of observational data points for training and another dataset of the same size for testing in each redshift bin. This downsampling ensures that the simulation and observational datasets used in the SOM have comparable sizes and distributions, minimizing biases and maintaining consistency between training and testing. As a result, the total number of HZ-AGN data is $N_{\mathrm{HZ,train}}^{\mathrm{HZ}} = N_{\mathrm{HZ,test}}^{\mathrm{HZ}} = 113{,}650$.

    For CW data, we also randomly split half the data into training and test sets in each redshift bin. For the training data, we used all objects in the CW training set, including galaxies, stars, and X-ray sources, to avoid biases that could arise from training only on galaxies. As a result, the total numbers are $N_{\mathrm{CW,train}}^{\mathrm{CW}} = N_{\mathrm{CW,test}}^{\mathrm{CW}} = 147{,}884$. The number of CW test data points when applying the SOM trained with HZ-AGN is also $N_{\mathrm{CW,test}}^{\mathrm{HZ}} = 113{,}651$.

\end{enumerate}

Table \ref{tab:summay_data} presents the median redshift in the training data, as well as the number of training and testing data points in each redshift bin. The sample sizes are approximately balanced overall. For the test data, the numbers refer to all sources considered for projection onto the SOM. Some sources may be excluded from prediction if they are identified as other types of objects (e.g., stars or X-ray sources) during the prediction stage.

\begin{table}
    \caption{Number of training data and test data in each redshift bin}
    \label{tab:summay_data}
\begin{tabular}{|c|c|c|c|c|c|c}
\hline
z range & $<z>_{\mathrm{train}}$ & Train Set & $N_\mathrm{train}$ & Test Set & $N_\mathrm{test}$ \\

\hline
\multirow{4}{*}{0.1 $\leq$ z < 0.8} & 0.6 & HZ-AGN & 27925 & HZ-AGN & 27925
\\
                       & 0.5 & CW & 38568 & CW & 38568  \\
                       & 0.6 & HZ-AGN & 27925 & CW & 27925 \\
\hline
\multirow{4}{*}{0.8 $\leq$ z < 1.5}  & 1.2 & HZ-AGN & 41115 & HZ-AGN & 41115
\\
                       & 1.1 & CW & 48422 & CW & 48422  \\
                       & 1.2 & HZ-AGN & 41115 & CW & 41115 \\
\hline
\multirow{4}{*}{1.5 $\leq$ z < 2.5}  & 1.9 & HZ-AGN & 34296 & HZ-AGN & 34296
\\
                       & 1.9 & CW & 47392 & CW & 47392  \\
                       & 1.9 & HZ-AGN & 34296 & CW & 34296 \\
\hline
\multirow{4}{*}{2.5 $\leq$ z < 3.5} & 2.8 & HZ-AGN & 10314 & HZ-AGN & 10314
\\
                       & 2.9 & CW & 13502 & CW & 13502  \\
                       & 2.8 & HZ-AGN & 10314 & CW & 10314 \\
\hline
\end{tabular}

\end{table}

\subsection{Training the SOM}
\label{subsec:training_som}

The first step in running the SOM is to define the grid size. We followed the procedure of \citet{davidzon2019}. We consider three metrics: quantization error ($\mathrm{qe}$), defined as the mean Euclidean distance between the training data and the BMU weights; training time ($\mathrm{t}$); and cell fraction ($\mathrm{cf}$), defined as the fraction of cells that have more than $5$ data points assigned to them. The threshold of $5$ was chosen because the training dataset is relatively small and this value produced stable results.

We define a quality score (S) to evaluate how well the SOM is trained based on three metrics. Equation~\ref{eq:quality_score} shows how it is calculated. These metrics are normalized to scale between $0$ and $1$ across all possible grid sizes. We assign the greatest emphasis to the quantization error (coefficient of 0.6), followed by the cell fraction (coefficient of 0.3), and then the training time (coefficient of 0.1). Using these weights, we compute an overall quality score.

It should be noted that this quality score depends on the specific dataset and may not be applicable globally. When testing different values for the coefficients and the maximum grid size, these coefficients help preserve acceptable levels of quantization error, cell fraction, and computation time, ensuring that all these parameters are considered to achieve the maximum quality score. Although other parameters, such as topographic error (defined as the fraction of data points for which the first and second BMUs are not adjacent), can be considered, the selected metrics are more important in the context of estimating physical parameters, as shown in previous studies \citet{davidzon2019, khostovan2025}. Therefore, we chose to focus only on these parameters to better evaluate the reliability of the predicted values.

\begin{equation}
\begin{split}
S =\ & 0.6 \times \left(1 - \frac{\mathrm{qe} - \mathrm{qe}_{\min}}{\mathrm{qe}_{\max} - \mathrm{qe}_{\min}}\right) + 0.3 \times \left(\frac{\mathrm{cf} - \mathrm{cf}_{\min}}{\mathrm{cf}_{\max} - \mathrm{cf}_{\min}}\right) \\
&+ 0.1 \times \left(1 - \frac{t - t_{\min}}{t_{\max} - t_{\min}}\right)
\end{split}
\label{eq:quality_score}
\end{equation}

\begin{figure}
	\includegraphics[width=1\columnwidth]{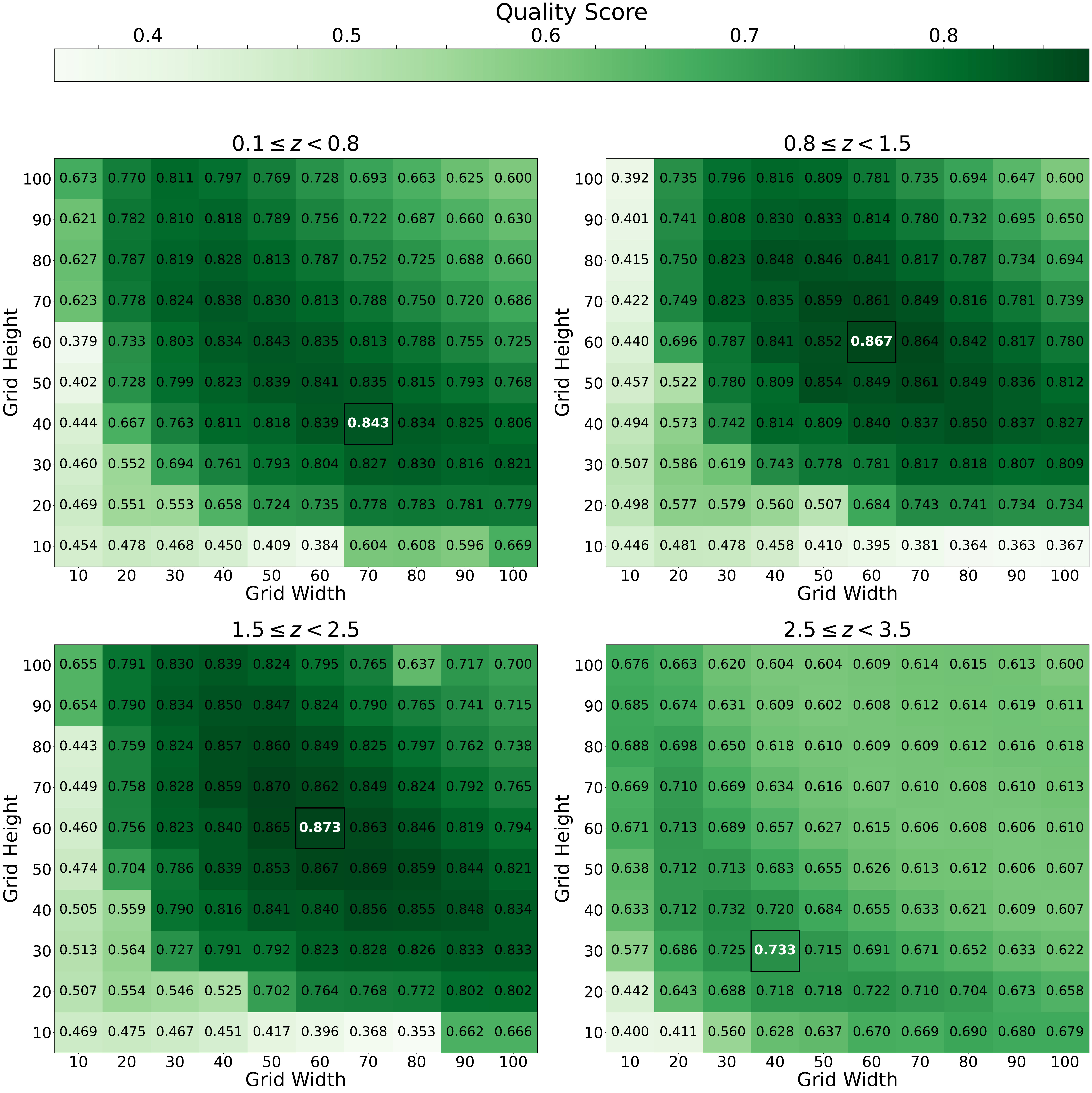}
    \caption{Quality score for different grid sizes, ranging from $10\times10$ to $100\times100$, for the SOM trained on simulation data. The optimal grid sizes are as follows: $70\times40$ for the redshift bin $0.1$–$0.8$, $60\times60$ for $0.8$–$1.5$ and $1.5$–$2.5$, and $40\times30$ for $2.5$–$3.5$. After calculating the quality scores for the SOM trained on observational data, the optimal grid sizes are as follows: $50\times60$ for the redshift bin $0.1$–$0.8$, $50\times80$ for $0.8$–$1.5$,  $60\times60$ for $1.5$–$2.5$, and $40\times40$ for $2.5$–$3.5$.
    }
    \label{fig:quality_sim}
\end{figure}

We started with grid sizes ranging from $10\times10$ to $100\times100$, increasing by 10 units in both dimensions. We considered all possible grid sizes, including both square and rectangular shapes, to find the optimal grid size. Figure~\ref{fig:quality_sim} shows the quality score for all grid sizes in each redshift bin when the SOM is trained on simulation data. The optimal grid sizes are as follows: $70\times40$ for the redshift bin $0.1$–$0.8$, $60\times60$ for $0.8$–$1.5$ and $1.5$–$2.5$, and $40\times30$ for $2.5$–$3.5$. The final quantization errors at the optimal grid sizes are: $0.88$ for $0.1$–$0.8$, $1.04$ for $0.8$–$1.5$, $1.17$ for $1.5$–$2.5$, and $1.31$ for $2.5$–$3.5$. The topographic errors for the grid sizes are as follows: $0.1-0.8$, $0.13$, $0.8-1.5$, $0.17$, $1.5-2.5$, $0.23$, and $2.5-3.5$, $0.21$.

For the SOM trained on the observational data, the same analysis yields optimal grid sizes of $50\times60$ for $0.1$–$0.8$, $50\times80$ for $0.8$–$1.5$, $60\times60$ for $1.5$–$2.5$, and $40\times40$ for $2.5$–$3.5$. The final quantization errors at the optimal grid sizes are: $1.11$ for $0.1$–$0.8$, $1.06$ for $0.8$–$1.5$, $1.13$ for $1.5$–$2.5$, and $1.22$ for $2.5$–$3.5$. The topographic errors for the grid sizes are as follows: $0.1-0.8$, $0.19$, $0.8-1.5$, $0.18$, $1.5-2.5$, $0.19$, and $2.5-3.5$, $0.20$. By testing other grid sizes on both simulation and observational data, we observe that these values fall within the lower range of topographic error; significantly smaller or larger grids generally lead to increased error.

Appendix~\ref{ap:SOM} shows the number of data points and the average Euclidean distance between the input data and the weights in each cell for the optimal grid size, demonstrating the acceptable quantization error and occupation in the SOM trained on simulation and observational data. In addition, the distribution of features and weights, as noted in \citet{hemmati2019a}, is similar for SOMs trained on both simulation and observational data.

For the number of iterations in SOM training, we use the default values for both the rough training and fine-tuning phases, as provided by the {\fontfamily{cmtt}\selectfont SOMPY} package. These defaults are automatically determined based on the SOM grid size and the number of training samples. Manually setting these values does not significantly improve SOM performance, while the default settings typically result in faster training.

\subsubsection{Visualization of the SOM}
\label{subsec:visalization_of_som}
\begin{figure*}
    \includegraphics[width=1\textwidth]{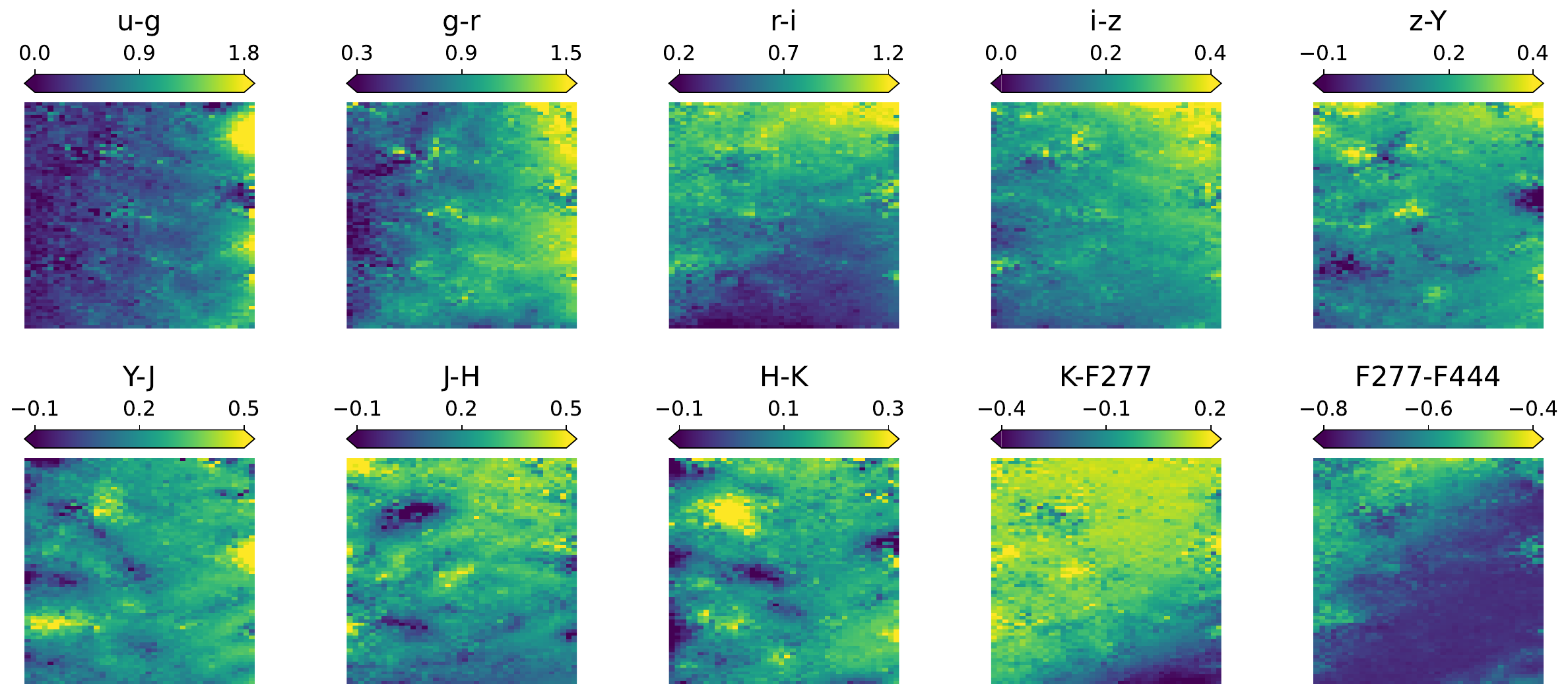}
    \caption{Interpolated distribution of 10-color features of the training data in the redshift bin $0.1$ to $0.8$, projected onto the SOM trained on HZ-AGN galaxies. Color ranges are normalized independently for each color, with minimum and maximum values set to the $2$nd and $98$th percentiles of each color's distribution respectively. The clustering of the projected features on the SOM trained on CW is also effective.
}
    \label{fig:features_sim}
\end{figure*}

For better visualization of the clustering on the SOM, we projected the input data onto the trained SOM and examined how the trend of each feature changes. Figure~\ref{fig:features_sim} shows the interpolated distribution of 10 color features from the training data, projected onto the SOM trained with HZ-AGN galaxies in the redshift bin $0.1$ to $0.8$. The figure shows that galaxies with similar colors are clustered together on the SOM. However, galaxies with higher and lower color values are dispersed across several regions. This pattern is evident in other redshift bins. For the SOM trained with CW galaxies, the feature trends look different due to the different training data, but the clustering appears acceptable. 

After training, the SOM can be visualized by assigning labels to each cell based on known properties of the input data, such as redshift, stellar mass, SFR, and mass-weighted age. After projecting the input data onto the trained SOM and finding the BMU, the physical parameters of the galaxies are assigned to each cell. Since each cell may contain more than one galaxy, we consider the median of the physical parameters, following the approach of \citet{davidzon2019}.

The physical parameters of each SOM cell are computed as
\begin{equation}
    \hat{\theta}_\mathrm{cell} = \mathrm{median}\{\theta_i \in \text{cell}\}.
\end{equation}

To assign physical parameters to the SOM, some parameters; such as stellar mass and SFR; require information about the total SED of galaxies. These parameters must be normalized to a reference magnitude. However, redshift, sSFR, and mass-weighted age do not require normalization. For the normalization of the stellar mass and SFR, we apply Equation \ref{eq:normalization} where $m_\text{ref}$ is the reference magnitude used for normalization, set to $m_\mathrm{F444}=22$ for simulation data and $m_\mathrm{F444}=23$ for observational data which about more than $90$ percent of the objects have magnitudes fainter than this threshold. We also note that this normalization assumes a constant M/L for all galaxies in the dataset, which may introduce a bias in cases of extreme SFH \citep{zibetti2009,ciesla2017}.

\begin{equation}
    \mathrm{X}_{\mathrm{norm}} = \mathrm{X} \times 10^{-0.4 (m_{\mathrm{ref}} - m_{\mathrm{F444}})}
    \label{eq:normalization}
\end{equation}

\begin{figure*}
    \includegraphics[width=1\textwidth]{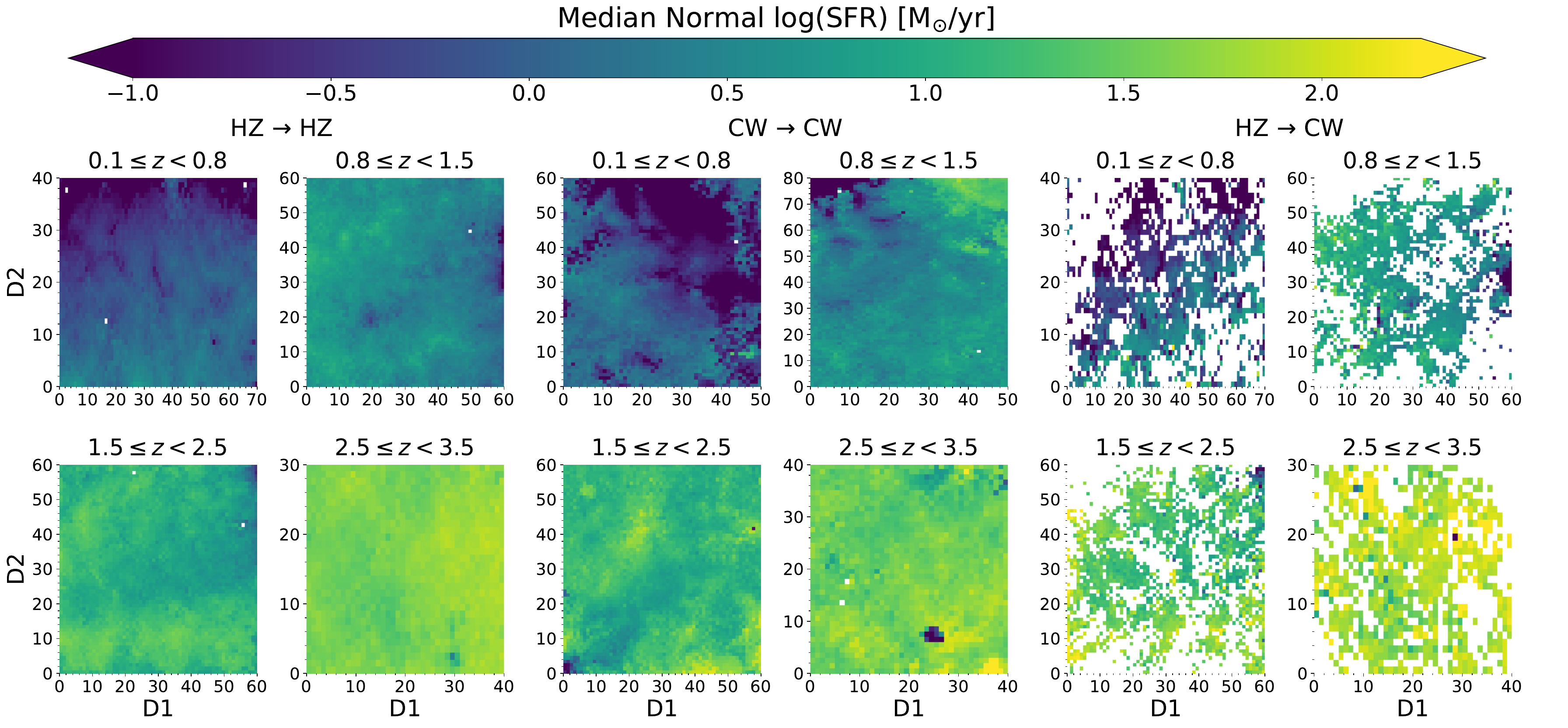}
    \caption{Labeling the SOM trained on HZ-AGN data (left panel), trained on CW data (center panel), and projecting the calibration of CW data onto the SOM trained on HZ-AGN data (right panel) for the median of the normalized SFR across different redshift bins. The same trend can be seen for other parameters (redshift, normalized stellar mass, mass-weighted age). At low redshift, galaxies exhibit lower stellar masses, lower SFRs, and older ages, whereas at high redshift, they have higher stellar masses, higher SFRs, and younger ages.  
}
    \label{fig:sfr_labels}
\end{figure*}

Figure~\ref{fig:sfr_labels} shows the trends in the median values of the normalized SFR across the SOM for three cases: the SOM trained on HZ-AGN data, the SOM trained on CW data, and CW data projected onto the SOM trained on HZ-AGN. Blank regions indicate cells to which no galaxies are assigned. For other physical parameters, namely redshift, normalized stellar mass, and mass-weighted age, the same trends are observed. At low redshifts, galaxies with lower stellar mass, lower SFR, and older ages are present, whereas at higher redshifts, galaxies with higher stellar mass, higher SFR, and younger ages appear. In the redshift bin $2.5$–$3.5$, in the SOM trained on the CW data (center panel of Figure~\ref{fig:sfr_labels}), there is a small area characterized by low stellar mass, low SFR, and older ages; these galaxies warrant further investigation.  




\begin{figure}
	\includegraphics[width=1\columnwidth]{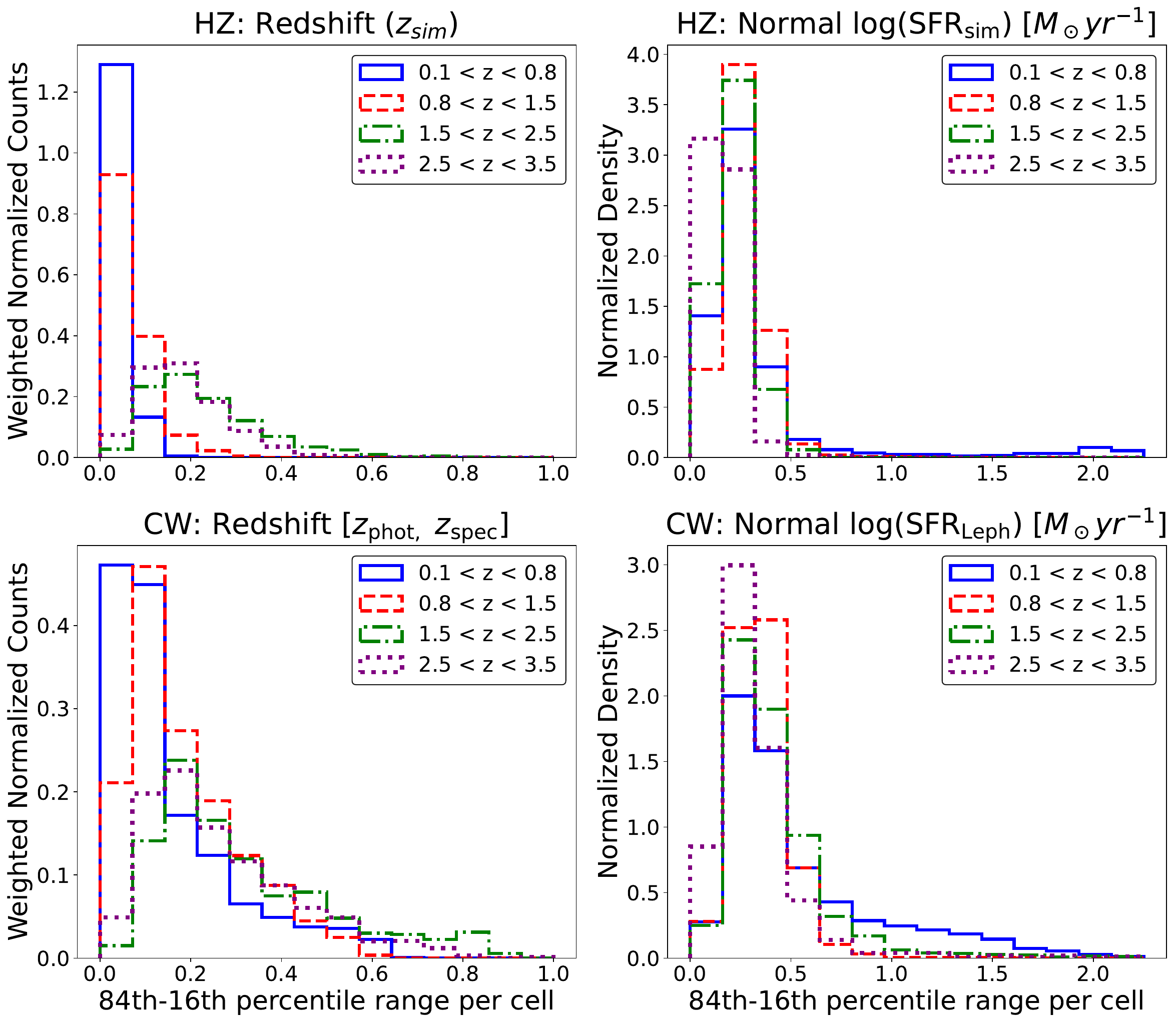}
    \caption{
    Histogram of the differences between the 16th and 84th percentiles for redshift and normalized SFR across different redshift bins. For redshift, the histogram is weighted by the inverse of the width of each redshift bin to obtain a normalized count per unit redshift. The top panel shows the SOM trained on HZ data, and the bottom panel shows the SOM trained on CW data.
}
    \label{fig:difference_two_cases}
\end{figure}

Since SOM labeling is based on the median of the physical parameters, understanding the variation within each cell is important. Figure~\ref{fig:difference_two_cases} shows the normalized distribution difference between the $84$th and $16$th percentiles for two parameters, redshift and normalized SFR, for SOMs trained on HZ-AGN and CW data. This plot highlights how consistently the SOM assigns galaxies with similar colors and physical parameters to the same cell. For redshift, because the widths of the bins vary in the redshift bins, we weight the histogram by the inverse of the width of the bin to obtain a normalized count per unit of redshift. The difference in redshift is somewhat larger in the higher redshift bins ($1.5-2.5$ and $2.5-3.5$). For the normalized $\log(\mathrm{SFR})$, the differences are slightly larger in the lower redshift bins, suggesting greater physical diversity; this trend is also visible for the normalized stellar mass and age. For the SOM trained on CW data (bottom panel in Figure~\ref{fig:difference_two_cases}), the spreads are noticeably larger for redshift and SFR (with a similar trend seen for normalized stellar mass and mass-weighted age) than those observed in the SOM trained on HZ-AGN data (top panel), suggesting that the CW dataset exhibits greater degeneracy. 


\subsection{Test Data}

In this subsection, we explain the process of projecting the test data onto the SOM and describe the corresponding steps.

The first item is projecting the HZ-AGN test data to the SOM trained on the HZ-AGN data. For the redshift bins $0.1$–$0.8$, $0.8$–$1.5$, $1.5$–$2.5$, and $2.5$–$3.5$, the numbers of galaxies used to estimate the physical parameters are $N^\mathrm{HZ}_\mathrm{HZ,pred} =\{27925,\ 41115,\ 34296,\ 10314 \}$, respectively, which are the same as those of the test data.

The second item is projecting the CW test data to the SOM trained on the CW data. It should be noted that, for predicting the physical parameters, we only consider galaxies ($\mathrm{flag}=0$). About $9,500$ of test data points are not galaxies. For the redshift bins $0.1-0.8$, $0.8-1.5$, $1.5-2.5$, and $2.5-3.5$, the numbers of galaxies for estimating the physical parameters are $N^\mathrm{CW}_\mathrm{CW,pred}=\{37510,\ 44460,\ 43531,\ 12836\}$, respectively.

The third item is projecting the CW data onto the SOM trained with galaxies from the HZ-AGN simulation. Since the color distribution of the CW data differs from that of the SOM weights or the trained colors \citet{hemmati2019a, hemmati2019b}), it is necessary to adjust the observational data to better match the SOM weights or the simulation training distribution.

\subsubsection{Covariate Shift Correction}
\label{subsec:covaraite_shift_correction}
To adjust the observational data to match the simulation data, we apply a covariate shift correction with the simulation serving as the reference. A feature-wise affine transformation is applied to the observational data, consisting of independent scaling and shifting for each feature:
\begin{equation}
    C^\mathrm{corr}_{n,f} = m_f C^\mathrm{obs}_{n,f} + c_f,
\end{equation}
where $C^\mathrm{obs}_{n,f}$ is the color feature $f \in \mathcal{D}$ for galaxy $n$ in the observational dataset, and $C^\mathrm{corr}_{n,f}$ is the corresponding transformed value after correction. The parameters $m_f$ and $c_f$ represent the scaling and shifting factors for feature $f$, respectively.

As an initial guess, we estimate $(m_f, c_f)$ using the 25th and 75th percentiles of each feature in the two datasets:
\begin{equation}
    m^{(0)}_f = \frac{Q_{0.75}(C^\mathrm{sim}_f)-Q_{0.25}(C^\mathrm{sim}_f)}{Q_{0.75}(C^\mathrm{obs}_f)-Q_{0.25}(C^\mathrm{obs}_f)},
\end{equation}
\begin{equation}
    c^{(0)}_f = \mathrm{median}(C^\mathrm{sim}_f) - m^{(0)}_f \, \mathrm{median}(C^\mathrm{obs}_f).
\end{equation}

The parameters $\{m_f,c_f\}_{f=1}^d$ are then optimized by minimizing a combined discrepancy between the transformed observational features and the corresponding simulation features. The loss function is defined as
\begin{equation}
    \mathcal{L}(m,c) = \sum_{f=1}^{10} \Big[ W_1(F^\mathrm{corr}_f,F^\mathrm{sim}_f) + \lambda D_\mathrm{KDE}(p^\mathrm{corr}_f,p^\mathrm{sim}_f) \Big].
\end{equation}
where $F^\mathrm{corr}_f$ and $F^\mathrm{sim}_f$ are the empirical cumulative distribution functions (CDFs) of the corrected observational features and the simulation features, respectively. The first term is the Wasserstein-1 distance:
\begin{equation}
    W_1(F^\mathrm{corr}_f,F^\mathrm{sim}_f) = \int_{-\infty}^{\infty} \big|F^\mathrm{corr}_f(u)-F^\mathrm{sim}_f(u)\big| \, du,
\end{equation}
and the second term compares kernel density estimates (KDEs) of the two distributions:
\begin{equation}
    D_\mathrm{KDE}(p,q) = \int_{-\infty}^{\infty} \big(p(u)-q(u)\big)^2 \, du,
\end{equation}
where $p$ and $q$ are Gaussian KDEs fitted to the normalized features. The hyperparameter $\lambda$ balances the contribution of the two terms; in this work we set $\lambda=1$. Optimization is performed using the L-BFGS-B algorithm implemented in the {\fontfamily{cmtt}\selectfont SciPy} library \citep{virtanen2020}, with multiple restarts to avoid local minima.

\begin{figure}
	\includegraphics[width=1\columnwidth]{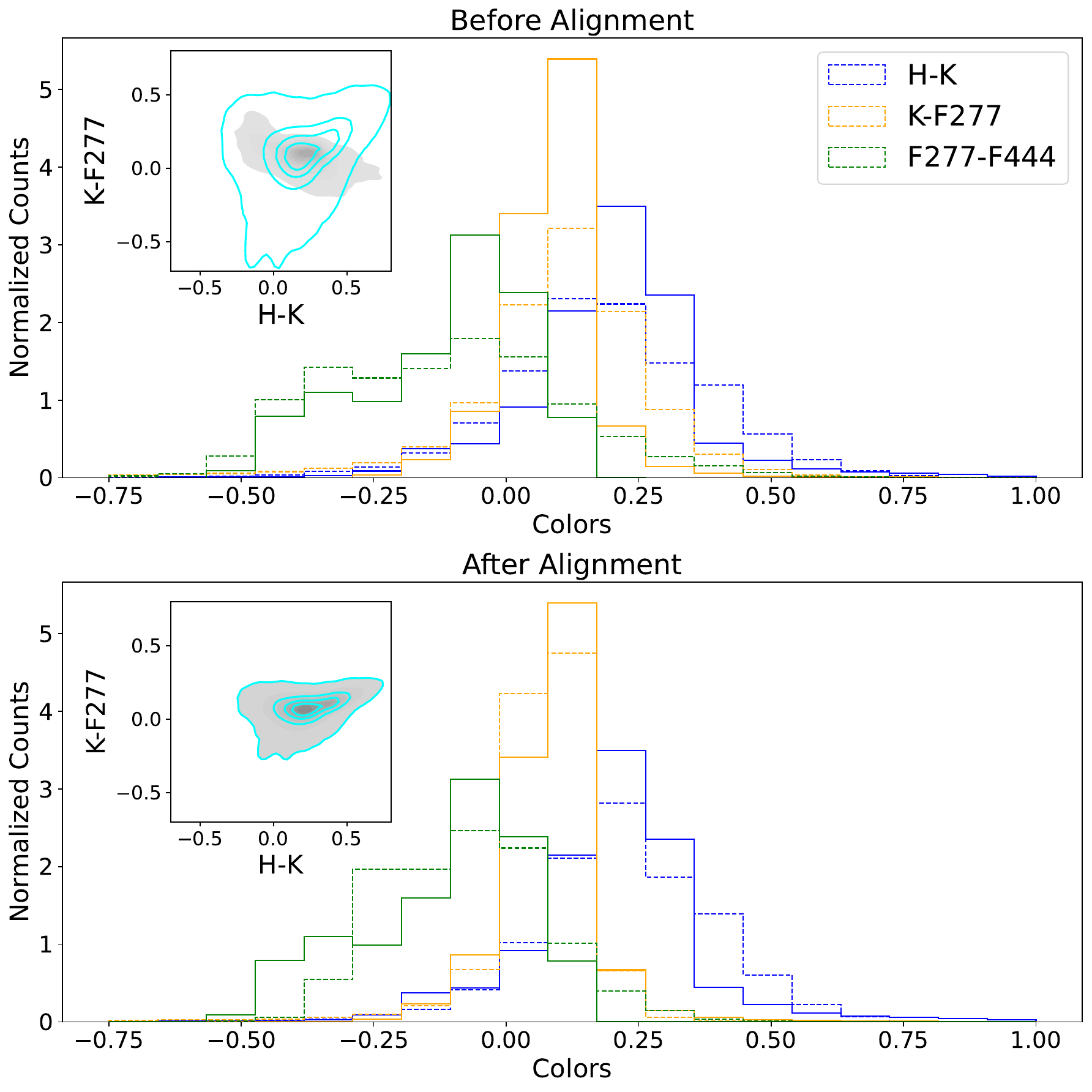}
    \caption{Normalized distribution of three colors (H–K, K–F277, F277–F444) for the redshift bin $0.8$ to $1.5$. The solid lines represent the SOM weights (trained with HZ-AGN galaxies and denormalized afterward), and the dashed lines represent the CW colors. The upper panel shows the distribution before alignment, and the lower panel shows the distribution after alignment. The inset figure displays the color-color distribution (H-K, K-F277) with the gray density representing the SOM weights and the aqua contours representing the CW colors before and after alignment, shown with 5 contour levels.}
    \label{fig:cosmos_alignmnet}
\end{figure}

This procedure ensures that the observational color distributions are statistically consistent with the distribution of the simulation training set, thereby addressing covariate shift and improving the reliability of subsequent SOM-based analyses. Figure~\ref{fig:cosmos_alignmnet} illustrates the alignment process for three colors (H–K, K–F277, and F277–F444) in the redshift bin $0.8 < z < 1.5$. Solid lines represent the SOM weights, while dashed lines represent the corrected observational colors. The inset plots show the color density distributions of the CW data and the SOM weights. The figure demonstrates that, after alignment, the color distributions of the SOM weights and the CW data are approximately matched.

In the simulation data, the stellar masses of galaxies are limited to $\log(M/M_\odot) \geq 9$, whereas in the observational data, we also have galaxies with masses below this threshold. Therefore, we calibrate the SOM trained on the simulation data to better match the physical parameters of the CW observations. To avoid biases, we randomly select $20\%$ of the observational data for calibration and exclude these sources from the dataset used to estimate physical parameters.

We define metrics to evaluate whether the covariate shift works. One of the metrics is the number of samples that fall into underpopulated regions of the training data, which shows whether the covariate shift moves the observational data correctly (see Section \ref{subsubsec:Covariate} for details). To rely more confidently on the covariate shift, we also remove the galaxies that lie in the underpopulated regions after applying the covariate shift and before starting the calibration and prediction phases. For the redshift bins $0.1-0.8$, $0.8-1.5$, $1.5-2.5$, and $2.5-3.5$, the numbers of galaxies to calibrate are $N^\mathrm{HZ}_\mathrm{CW,cal}=\{5381,\ 7757, 6453,\ 1852\}$, respectively.

The right panel of Figure~\ref{fig:sfr_labels} shows the median values of the calibrated data projected onto the SOM for the normalized SFR. We use the same reference magnitude for the simulation data to normalize the stellar mass and SFR ($m_\mathrm{F444}=22$). Compared to left panel of figure \ref{fig:sfr_labels}, which shows the SOM labeled using the training (simulation) data, we observe similar general trends.  In the $2.5$ to $3.5$ redshift bin, some cells also show lower stellar mass and SFR values that are not present in the simulation. Without aligning the distributions of the observational data and SOM weights, the projection of the observational data onto the SOM results in a sparser representation. In contrast, when the distributions are matched, the SOM is better filled, with smoother transitions and more continuous patterns.


For the redshift bins $0.1$–$0.8$, $0.8$–$1.5$, $1.5$–$2.5$, and $2.5$–$3.5$, the numbers of galaxies used to estimate the physical parameters, after excluding calibration objects and non-galaxy sources, are $N^\mathrm{HZ}_\mathrm{CW,pred}=\{21413,\ 30936,\ 25280,\ 7312\}$, respectively.

\subsection{Prediction Phase}
\label{subsec:Prediction}

After the SOM is labeled, test data can be projected onto it by identifying the likelihood function for each test object \citep{latorre2024}, which is defined as
\begin{equation}
    L_{i,j} = \exp\!\left(-\frac{1}{2} \chi^2_{i,j}\right),
    \label{eq:likelihood}
\end{equation}
where $\chi^2_{i,j}$ is given by
\begin{equation}
    \chi^2_{i,j} = \sum_{f=1}^{10}
    \left( \frac{C^{\mathrm{test}}_f - C^{\mathrm{median}}_{i,j,f}}{\sigma^{\mathrm{test}}_f} \right)^2.
\end{equation}
where, $C^{\mathrm{test}}_f$ is the color corresponding to the test object, $C^{\mathrm{median}}_{i,j,f}$ is the median color of training objects mapped to cell $(i,j)$, and $\sigma^{\mathrm{test}}_f$ is the photometric uncertainty which is defined as the square root of the sum of the squares of the magnitude errors corresponding to that specific color ($
\sigma^{\mathrm{test}}_f = \sqrt{\sigma_{m_a}^2 + \sigma_{m_b}^2}$
for magnitudes $m_a$ and $m_b$.) The likelihood is normalized such that $\sum_{i,j} L_{i,j} = 1$. In this way, for each test data, we obtain the normalized likelihood values for each cell of the SOM.

After this step, for each test object, we derive the posterior probability density function (PDF) of the target parameter by constructing a weighted histogram of the cell values $\hat{\theta}_{i,j}$, using the normalized likelihoods $L_{i,j}$ as weights:
\begin{equation}
p(\theta|\mathcal{D}) \approx p(p_k) =
\frac{1}{\Delta p} \cdot
\displaystyle\sum_{i,j} L_{i,j} \;
\Theta\left(\hat{\theta}_{i,j} \in [p_k, p_{k+1})\right),
\end{equation}
where $\hat{\theta}_{i,j}$ is the median parameter value in cell $(i,j)$, $\Delta p$ is the bin width, and $\Theta$ denotes the Iverson bracket ($\Theta(\text{true})=1$, $\Theta(\text{false})=0$).  
The factor $\frac{1}{\Delta p}$ ensures that $\int p(\theta|\mathcal{D})\,d\theta = 1$.
\begin{equation}
\hat{\theta}_{\rm SOM} = \arg\max_\theta \; p(\theta|\mathcal{D})
\end{equation}
The peak of the resulting PDF is adopted as the SOM prediction.

To evaluate the quality of these predictions, we compare them with the intrinsic physical parameters in the simulation data or with the results of SED fitting (e.g., LePhare) in the observational data, as well as with the ${\rm age}_{\rm mw}$ derived independently. To avoid biases in the sSFR prediction, we divided the predicted SFR by the predicted stellar mass, so the sSFR is not predicted directly. These reference values are denoted as $\theta_\mathrm{ref}$ for the physical parameters and $z_\mathrm{ref}$ for redshift. It should be noted that although we predict the redshift, for binning purposes, we use the true redshift in the simulation data, and the photometric or spectroscopic redshift for galaxies where available.

We used seven metrics to quantify this comparison for the physical parameter $\theta$. First, the normalized median absolute deviation (NMAD) is defined as $\sigma_{\text{NMAD}} = 1.48 \times \text{median}(|\log(\theta_{\text{SOM}}) - \log(\theta_{\text{ref}})|)$. For redshift, it is defined as $1.48 \times \text{median}(|z_{\text{SOM}} - z_{\text{ref}}|/(1+z_{\text{ref}}))$ \citet{ilbert2009}. Second, the outlier fraction is the percentage of objects where $|\log(\theta_{\text{SOM}}) - \log(\theta_{\text{ref}})| > 1\sigma$, with $\sigma$ being the standard deviation of the difference $\log(\theta_{\text{SOM}}) - \log(\theta_{\text{ref}})$. For the redshift, the outlier condition is defined as $|(z_{\text{SOM}} - z_{\text{ref}})|/(1+z_{\text{ref}}) > 1 \sigma_{\text{NMAD}}$. Fourth, RMSE is the root mean square error of the difference between the predicted and reference parameters ($|\Delta z|/(1+z_\mathrm{ref})$ for the redshift). Fifth, $\sigma_O$ is the root mean square error of the parameter differences after removing galaxies with outliers greater than 1$\sigma$ (or $1\sigma_{\text{NMAD}}$ for the redshift). Sixth, the bias is the mean difference between the parameter predicted by the SOM and the reference physical parameter ($|\Delta z|/(1+z_\mathrm{ref})$ for the redshift). Seventh, the Pearson correlation coefficient (r) between the parameter predicted by the SOM and the reference parameter shows that the closer it is to $1$, the stronger the linear correlation.

\section{Results}
\label{sec:results}
\subsection{Applying the SOM Trained on HORIZON-AGN to the HORIZON-AGN Data}
\label{subsec:som_sim}

\begin{figure}
	\includegraphics[width=1\columnwidth]{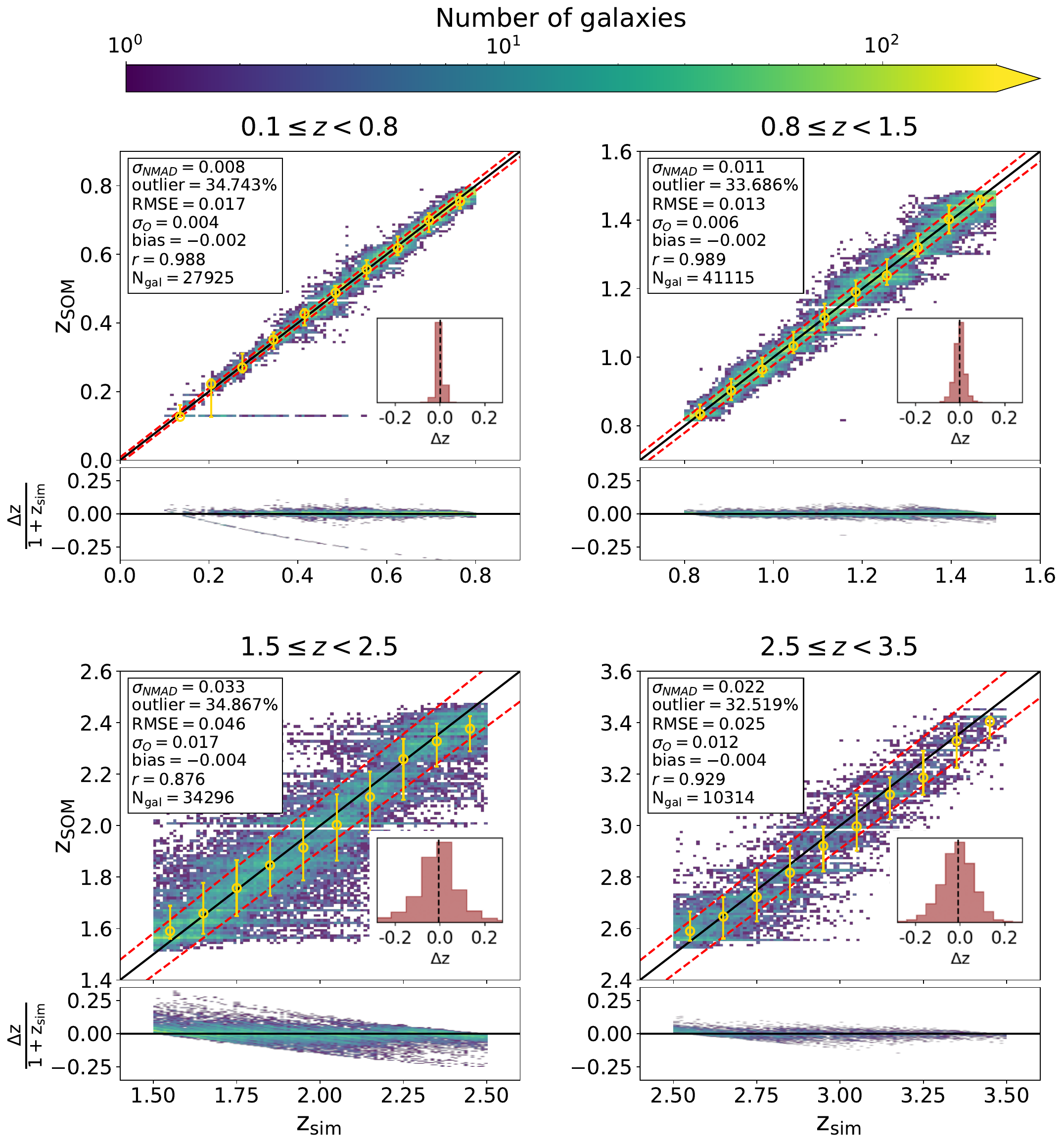}
    \caption{Comparison of the redshift predicted by the SOM trained on HZ-AGN data with the true values. The solid line represents the 1:1 relationship, and the red dashed lines indicate the $\pm 1\sigma_{\text{NMAD}}(1+z)$ boundaries. The open circles show the median of predicted redshift in each bin of the simulation data, while the error bars represent the difference between the 84th and 16th percentiles. The dark dashed line in the inset histogram shows the median of $\Delta z$. $N_{\text{gal}}$ denotes the number of galaxies in the test data within each redshift bin (see Subsection \ref{subsec:Prediction} for more information about the metrics).}
    \label{fig:redshift_sim}
\end{figure}

In this section, we present the predictions of the SOM trained on HZ-AGN data and applied to the HZ-AGN data. First, we predict the redshift of the simulation data and compare it with the intrinsic redshift values. Figure~\ref{fig:redshift_sim} shows the comparison between the redshift predicted by the SOM and the true redshift from the simulation. The open circles represent the SOM-predicted redshift medians, with error bars indicating the range between the 84th and 16th percentiles. The dark dashed line in the inset histogram indicates the median value of $\Delta z$.

The redshift bins $0.1-0.8$ and $0.8-1.5$ are generally accurate, with low $\sigma_{\mathrm{NMAD}}$, negligible bias, and a high Pearson correlation coefficient. However, in the $1.5-2.5$ and $2.5-3.5$ ranges, some points are more dispersed, as shown by the higher values of $\sigma_\mathrm{NMAD}$ and RMSE. Slight overestimation and underestimation are also seen at low and high redshifts, respectively.

\begin{figure}
     \includegraphics[width=1\columnwidth]{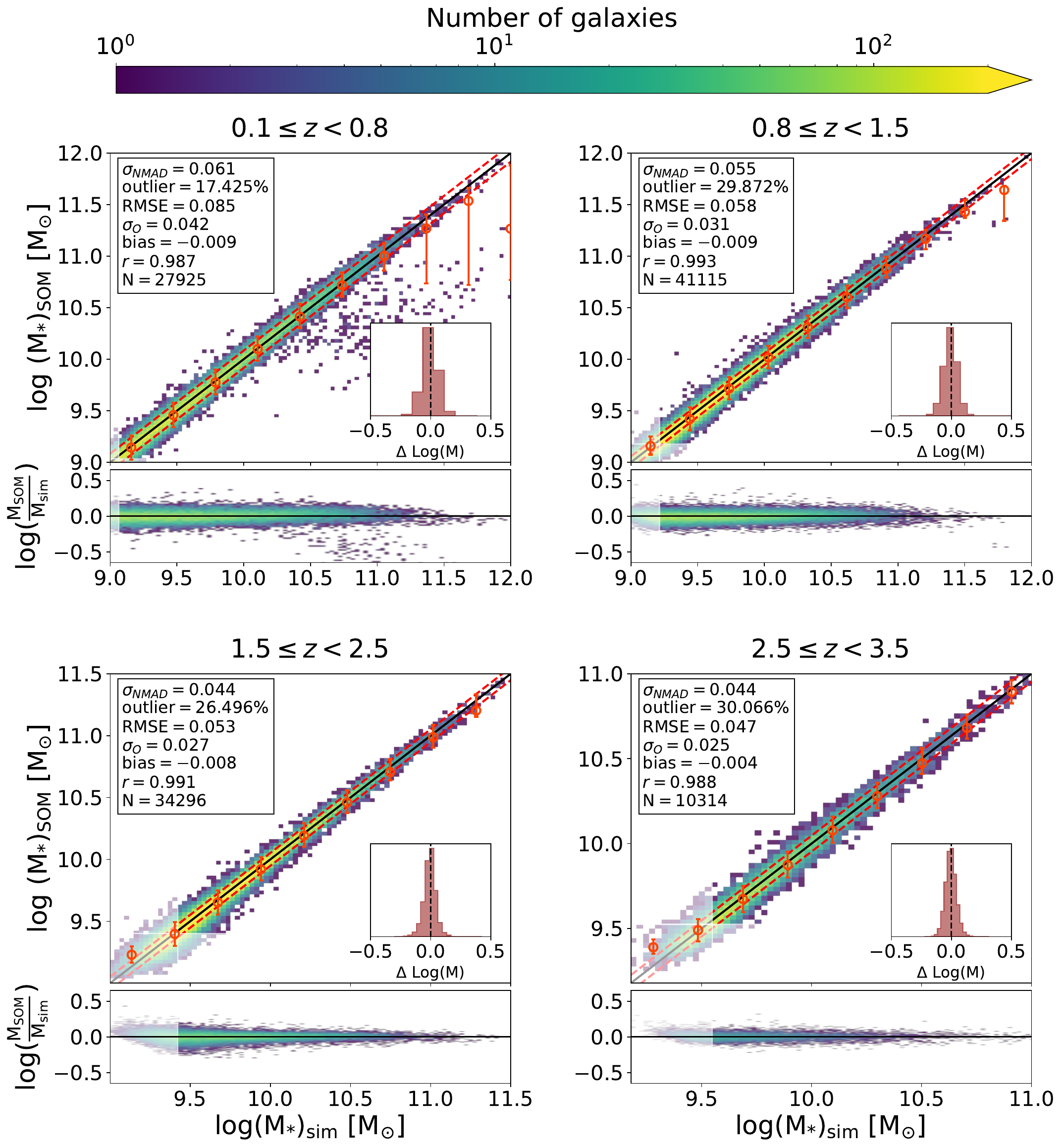}
    \caption{Comparison of the total stellar mass predicted by the SOM trained on HZ-AGN data with the true values. The solid line represents the 1:1 relation, and the dashed lines indicate the $\pm 1\sigma$ dex boundaries, where $\sigma$ is the standard deviation. The shaded regions show the mass completeness for both the true mass of the HZ-AGN data and the mass estimated by the SOM. The open circles show the median of the predicted parameters in each bin of the data, and the error bars represent the difference between the 84th and 16th percentiles. The dark dashed line in the inset histogram shows the median of the differences. $N_{\text{gal}}$ represents the number of galaxies in the test data within each redshift bin. (see Subsection \ref{subsec:Prediction} for more information about the metrics).}
    \label{fig:mass_sim}
\end{figure}

Second, for the stellar mass, after prediction, we first renormalized it to the reference magnitude $m_\mathrm{F444} = 22$. Figure~\ref{fig:mass_sim} shows the comparison between the predicted stellar masses from the SOM and the true simulated values. We also computed the mass completeness for each predicted and true stellar mass bin using the method described in \citet{pozzetti2010}, based on the reference magnitude $m_\mathrm{F444}^\mathrm{lim}= 24.8$. Masses below the completeness limit are shaded in the figure.

Across all redshift bins, the predictions closely match the simulated masses, with low $\sigma_{\text{NMAD}}$ and biases close to zero. A very strong linear correlation is also evident in the Pearson correlation coefficient, which is $0.99$. In the redshift bins $0.1$–$0.8$ and $0.8$–$1.5$, the $\sigma_\mathrm{NMAD}$ and RMSE are slightly higher than in the other redshift bins, indicating somewhat greater dispersion. In the higher redshift bins ($1.5$-$2.5$ and $2.5$-$3.5$), there is a slight overestimation for stellar masses below $\log(M/M_\odot) = 9.5$, which falls below the mass completeness threshold.

\begin{figure}
	\includegraphics[width=1\columnwidth]{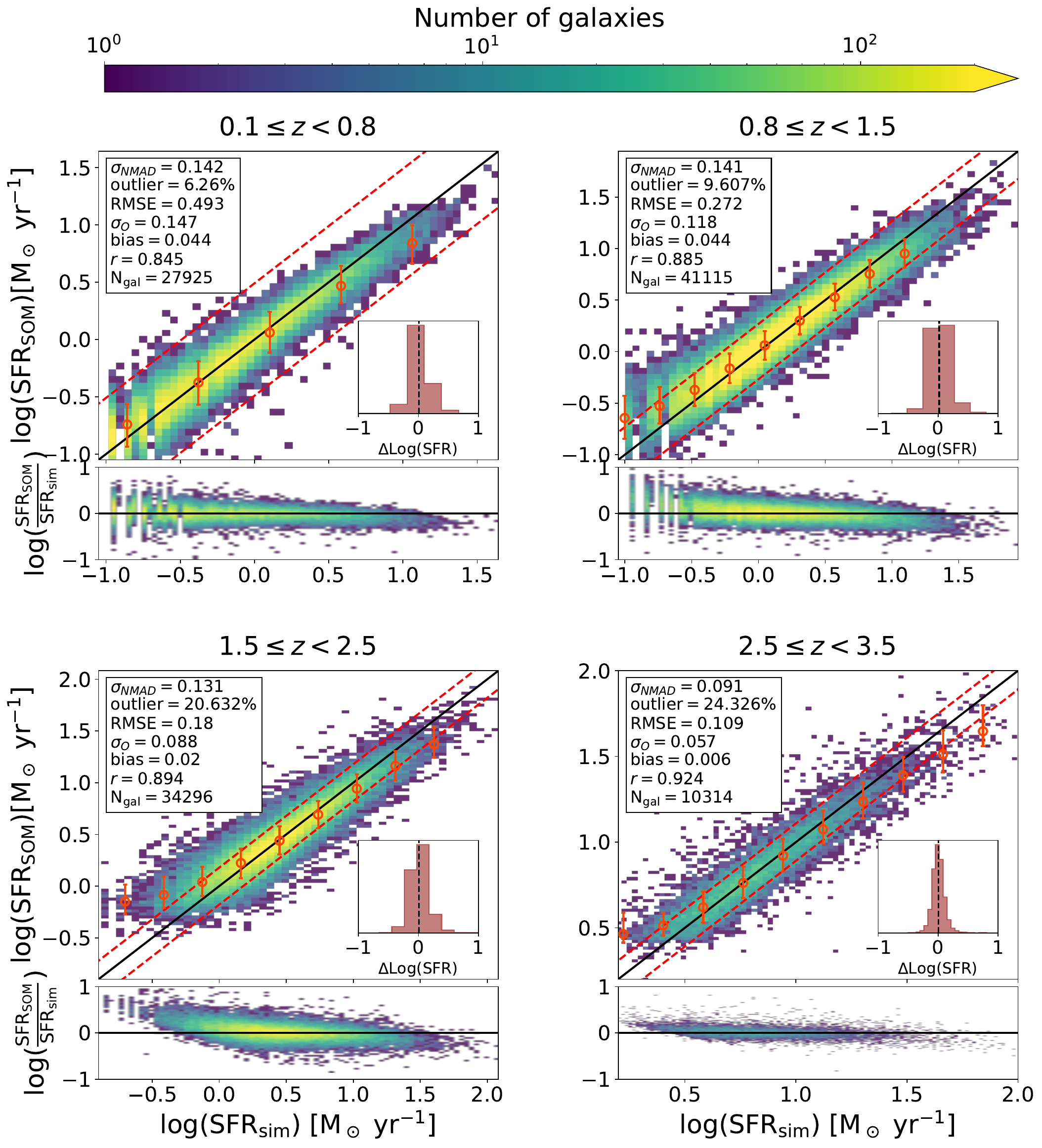}
    \caption{Comparison of the SFR predicted by the SOM trained on HZ-AGN data with the true values. The predictions are generally accurate, although overestimation and underestimation occur at low and high SFR values, respectively. See the caption of Figure~\ref{fig:mass_sim} for details about the lines. }
    \label{fig:sfr_sim}
\end{figure}

Third, after renormalizing the SFR estimated by the SOM, we compare the predictions with the intrinsic values of the simulation. Figure~\ref{fig:sfr_sim} shows the comparison between the predicted SFR by the SOM and the simulated SFR. Overall, the SFR predictions are good, as indicated by the low bias and high Pearson correlation coefficient.

In the redshift bin $0.1$–$0.8$, the RMSE is highest, with an underestimation in $\log(\text{SFR}[M_\odot,\text{yr}^{-1}]) > 0.5$. In the $0.8$-$1.5$ and $1.5$-$2.5$ redshift bins, there is a slight overestimation for galaxies with $\log(\text{SFR}[M_\odot,\text{yr}^{-1}]) < 0$ and a slight underestimation for those with $\log(\text{SFR}[M_\odot,\text{yr}^{-1}]) > 1$. In the $2.5$-$3.5$ bin, overestimation occurs for $\log(\text{SFR}[M_\odot,\text{yr}^{-1}]) < 0.5$, and an underestimation occurs for $\log(\text{SFR}[M_\odot,\text{yr}^{-1}]) > 1$

Fourth, we compare the estimated sSFR, obtained by dividing the predicted SFR by the predicted stellar mass, with the simulated sSFR, defined as ${\text{SFR}_{\text{sim}}/{M_*}_{\text{sim}}}$. Figure~\ref{fig:ssfr_sim} (see Appendix~\ref{ap:ssfr_estimation}) shows the comparison between the predicted $\log(\mathrm{sSFR})[{\mathrm{yr}}^{-1}]$ from the SOM and the simulated values, indicating that the predictions are reasonably accurate. 

The high RMSE occurs in the redshift bin $0.1$–$0.8$, where some points are more dispersed. However, the high Pearson correlation coefficient $(0.791)$ indicates that the predicted values are still strongly correlated with the true values. In other redshift bins, a similar pattern of overestimation at low sSFR and underestimation at high sSFR is observed, as seen in the SFR results.

\begin{figure}
	\includegraphics[width=1\columnwidth]{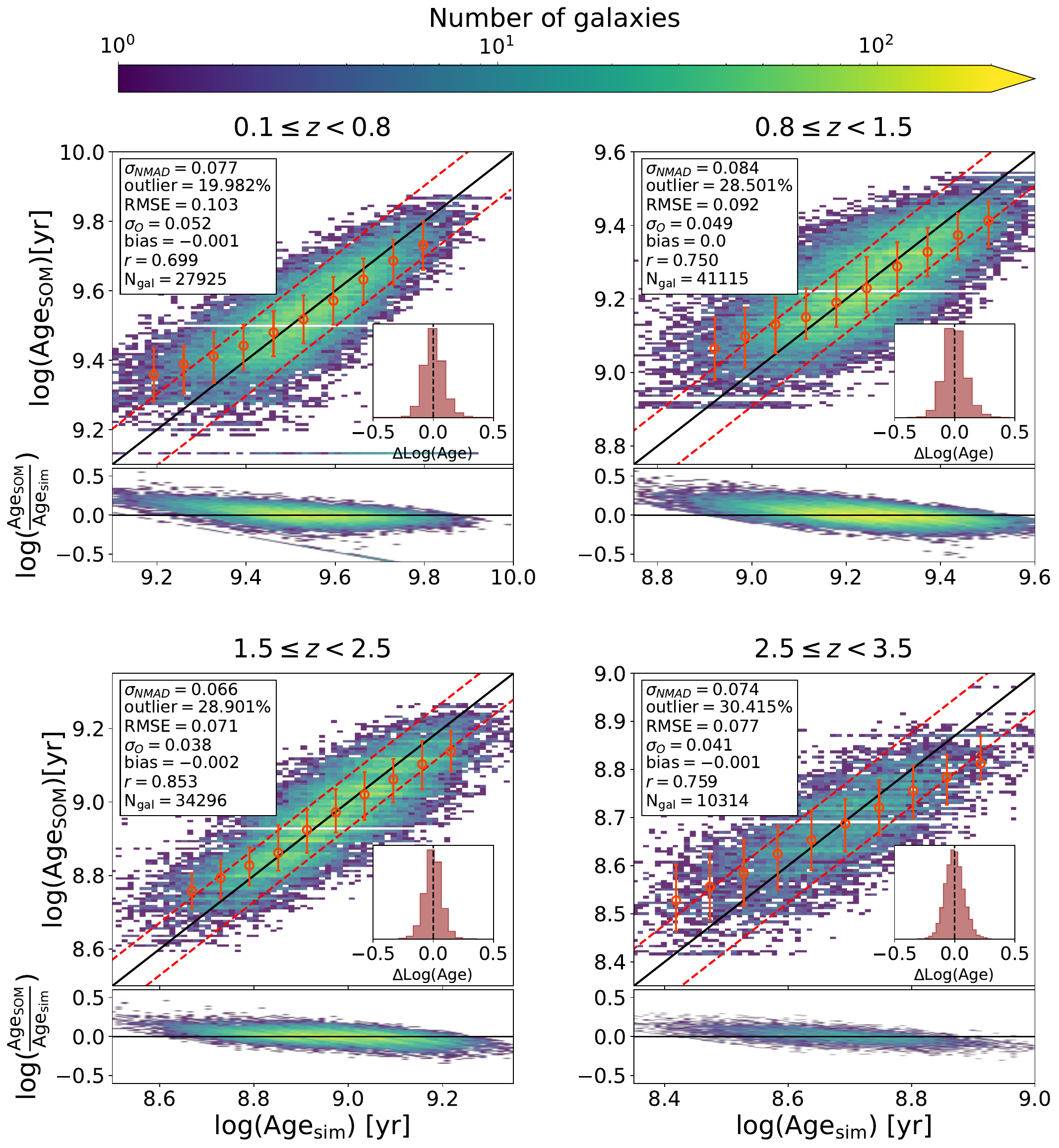}
    \caption{Comparison of the mass-weighted age predicted by the SOM trained on HZ-AGN data with the true values. Overestimation and underestimation occur at low and high age, respectively. See the caption of Figure~\ref{fig:mass_sim} for details about the lines.}
    \label{fig:age_sim}
\end{figure}

Fifth, the mass-weighted age estimated by the SOM is compared to the simulated age in the test data. Figure~\ref{fig:age_sim} shows the SOM predictions versus the true mass-weighted ages. Across all redshift bins, the predictions are acceptable, with a Pearson correlation coefficient of $0.7$. A consistent pattern is observed in all redshift bins: overestimation at lower ages and underestimation at higher ages. Approximately $30\%$ of the data fall outside the 1$\sigma$ range, as measured by the standard deviation.

In summary, predictions of the physical parameters in the HZ-AGN data, using the SOM trained on the simulated data, are generally accurate. For redshift, the predictions show some dispersion in higher-redshift galaxies, while for other parameters, the dispersion is more prominent in low-redshift galaxies. 

\subsection{Applying the SOM Trained on COSMOS-Web to the COSMOS-Web Data}
\label{subsec:som_obs}

\begin{figure}
	\includegraphics[width=1\columnwidth]{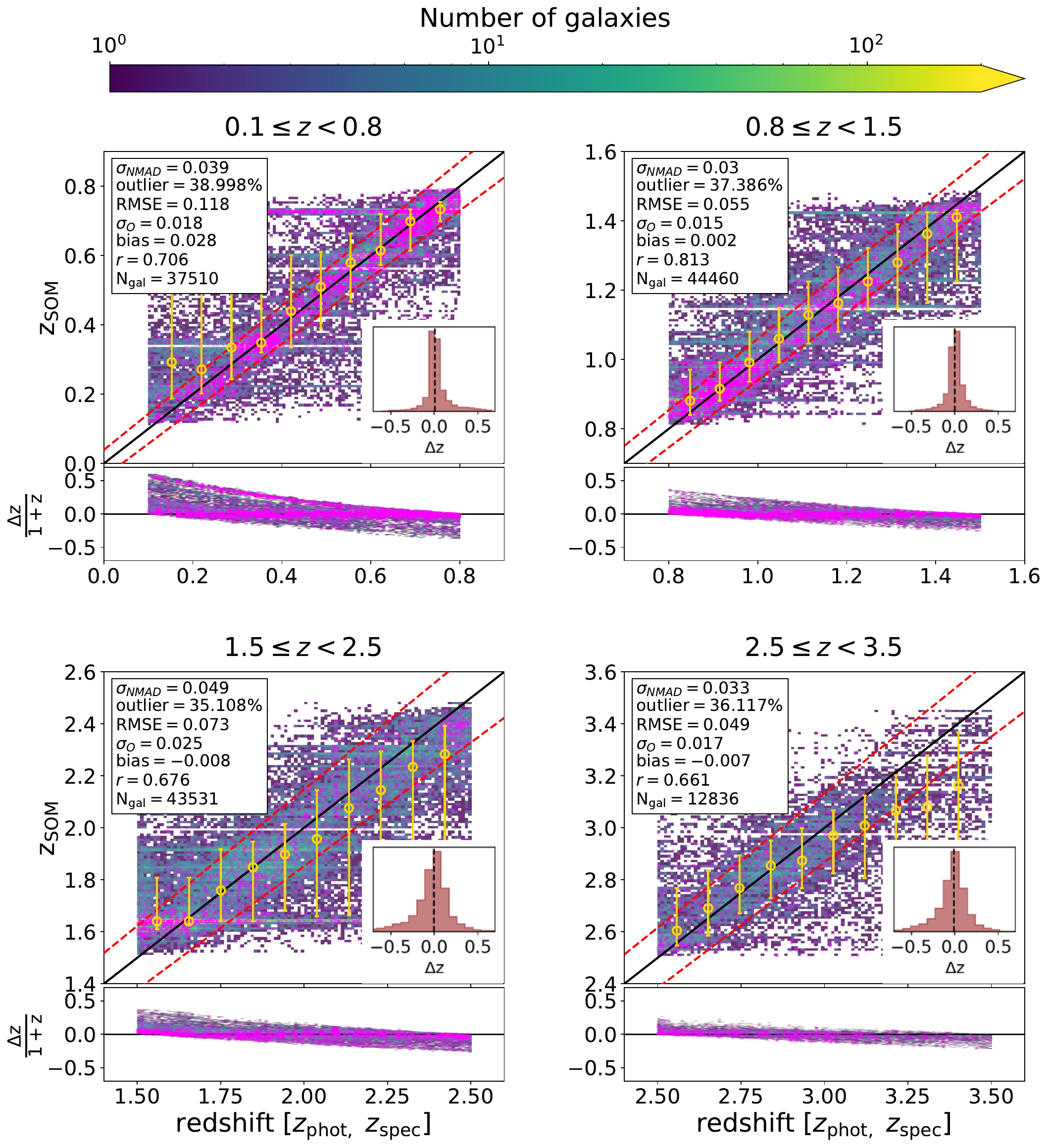}
    \caption{Comparison of the redshift predicted by the SOM trained on CW data with the redshift estimated using SED fitting (LePhare) and the spectroscopic redshifts, where available. Magenta points represent observational data with spectroscopic redshifts, which lie along the 1:1 relation line. See the caption of Figure~\ref{fig:redshift_sim} for details about the lines.}
    \label{fig:redshift_obs}
\end{figure}

In this section, we present the predictions of the SOM trained on CW data and applied to the CW data. The first parameter we estimated with the SOM is the redshift. Figure~\ref{fig:redshift_obs} shows the redshift predictions in each redshift bin. The open circles represent the median predicted redshift values, whereas the magenta points correspond to observational data with spectroscopic redshifts. The overall bias is close to zero, except in the redshift bin from $0.1$ to $0.8$, where it is slightly larger $(0.028)$. The Pearson correlation coefficient also indicates that the linear correlation is not very strong $(0.7<r<0.8)$.

Notably, galaxies with spectroscopic redshifts are predicted more accurately, especially in the $0.1$ to $0.8$ redshift range where more spectroscopic data are available (about $7,444$ galaxies).

Focusing more closely on each redshift bin in the range $0.1$ to $0.8$, the overestimation occurs primarily from $0.1$ to $0.3$. For redshifts above $0.3$, the medians align well with the 1:1 relation line, although a slight underestimation appears between $0.7$ and $0.8$. This bin also shows a higher RMSE, indicating greater dispersion in the redshift predictions, which is also evident in the residual plot. In the redshift range $0.8$ to $1.5$, the median of most data points is well predicted, with some underestimation evident between $1.2$ and $1.5$. In the $1.5$ to $2.5$ range, there is a slight overestimation between $1.5$ and $1.7$, followed by an increasing underestimation at redshifts higher than $1.9$. Finally, in the $2.5$ to $3.5$ bin, although the overall bias remains close to zero, the redshifts below $2.8$ tend to be slight overestimated, while those above $3$ are underestimated.

\begin{figure}
	\includegraphics[width=1\columnwidth]{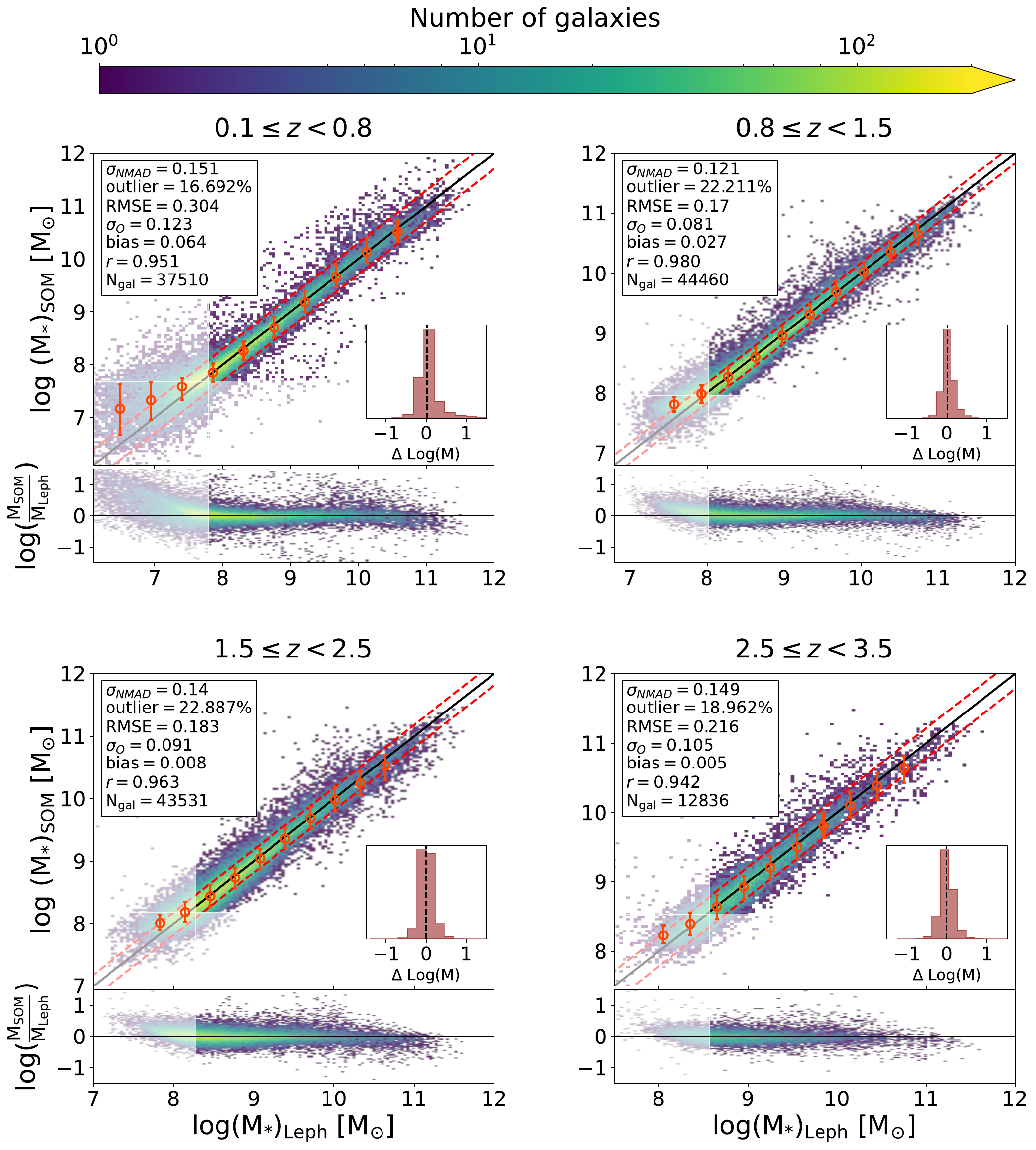}
    \caption{Comparison of the total stellar mass predicted by the SOM trained on CW data with the total stellar mass estimated using SED fitting (LePhare). The shaded regions indicate the mass completeness limits for the estimates from LePhare and the SOM. The SOM provides acceptable predictions with low scatter. See the caption Figure~\ref{fig:mass_sim} for details about the lines.
}
    \label{fig:mass_obs}
\end{figure}

For stellar mass, after deriving the SOM predictions, we renormalized them to the reference magnitude of $m_{\mathrm{F444W}} = 23$ and compared them to the output of the SED fitting. Figure~\ref{fig:mass_obs} shows the predicted stellar masses. The shaded regions indicate the limits of mass completeness in each redshift bin, based on the magnitude completeness of $m_{\mathrm{F444}}^{\mathrm{lim}} = 27$. In all redshift bins, the predicted stellar masses lie close to the 1:1 relation line, which is also evident in the strong correlation between the parameters, as indicated by the Pearson correlation coefficient ($0.9$). The largest dispersion is observed for stellar masses below the completeness threshold.

In the redshift bin $0.1 \leq z < 0.8$, a significant overestimation appears for masses below $\log(M/M_\odot) < 8$, and the bias is relatively larger compared to the other redshift bins $(0.064)$. For the other redshift bins, a slight overestimation is also observed at the low-mass end. In the redshift bin $0.8 \leq z < 1.5$, the RMSE is lower, indicating fewer points with large deviations from the LePhare predictions. In the bin $1.5 \leq z < 2.5$, about $23\%$ of galaxies fall outside the $1\sigma$ range, with some scattered points. In the highest redshift bin, $2.5 \leq z < 3.5$, the $\sigma_{\mathrm{NMAD}}$ and RMSE are slightly high, indicating that some points are scattered relative to the LePhare stellar mass estimates.

\begin{figure}
	\includegraphics[width=1\columnwidth]{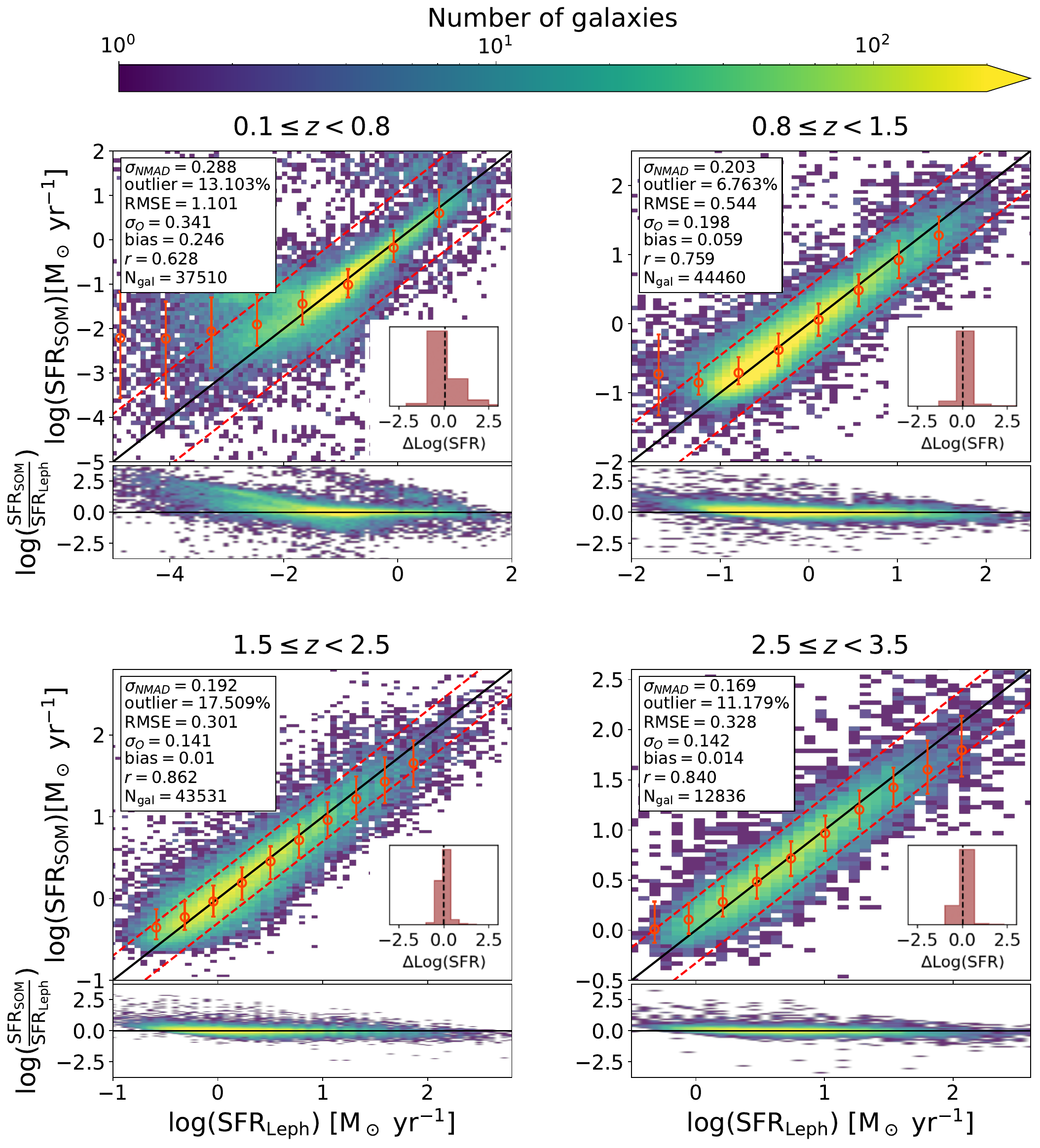}
    \caption{Comparison of the SFR predicted by the SOM trained on CW data with the SFR estimated using SED fitting (LePhare). Most of the data points are well estimated. See the caption of Figure~\ref{fig:mass_sim} for details about the lines.
}
    \label{fig:sfr_obs}
\end{figure}

For the SFR, after renormalizing the values estimated by the SOM, we analyze the results of SOM prediction. Figure~\ref{fig:sfr_obs} shows the predicted SFR compared to the SED fitting output. Most data points are align along the 1:1 relation line, as reflected in the Pearson correlation coefficient, which ranges from $0.63$ to $0.86$. However, dispersion is observed across all redshift ranges, particularly at low SFR values in the lower redshift bins. 

In the redshift bin $0.1 \leq z < 0.8$, the largest $\sigma_{\mathrm{NMAD}}$ $(0.29)$ and RMSE ($1.101$) are observed, along with a relatively large bias ($0.246$). Significant overestimation occurs at low SFR values. In the redshift bin $0.8 \leq z < 1.5$, the dispersion is lower, as reflected in the reduced $\sigma_{\mathrm{NMAD}}$ and RMSE, although overestimation at low SFR remains evident. In the bin $1.5 \leq z < 2.5$, about $18\%$ of the points are outliers beyond $1\sigma$, yet the overall predictions are reasonable and the bias remains low $(0.01)$. In the bin $2.5 \leq z < 3.5$, the lowest bias indicates that most of the data align with the 1:1 relation, although overestimation occurs at low SFR.

For the sSFR, we also divided the predicted SFR by the predicted stellar mass. Figure \ref{fig:ssfr_obs} (see Appendix~\ref{ap:ssfr_estimation}) shows the predicted sSFR in all redshift bins, compared to the output of the SED fitting. More than $75\%$ of the data points have values within $1\sigma$ dex, indicating that the predictions are accurate for most of the data, as also reflected in the relatively low biases. However, the dispersion in the predictions is significant, as indicated by the large $\sigma_{\mathrm{NMAD}}$ and RMSE values, as well as the low Pearson's correlation coefficient.  

In all redshift bins, both overestimation and underestimation occur at low and high sSFR values, respectively. In the redshift bin $0.1$–$0.8$, the large RMSE value ($1.189$) indicates a high level of dispersion and the bias is also high ($0.213$). In the redshift bin $0.8$–$1.5$, the Pearson correlation coefficient ($r = 0.725$) is slightly higher than in the other redshift bins. In the redshift range $1.5$–$2.5$, the lowest bias is observed, with a value of $-0.001$, which is also reflected in the RMSE. In the redshift bin $2.5$–$3.5$, the predictions are more dispersed, as indicated by a slightly higher outlier fraction.

\begin{figure}
	\includegraphics[width=1\columnwidth]{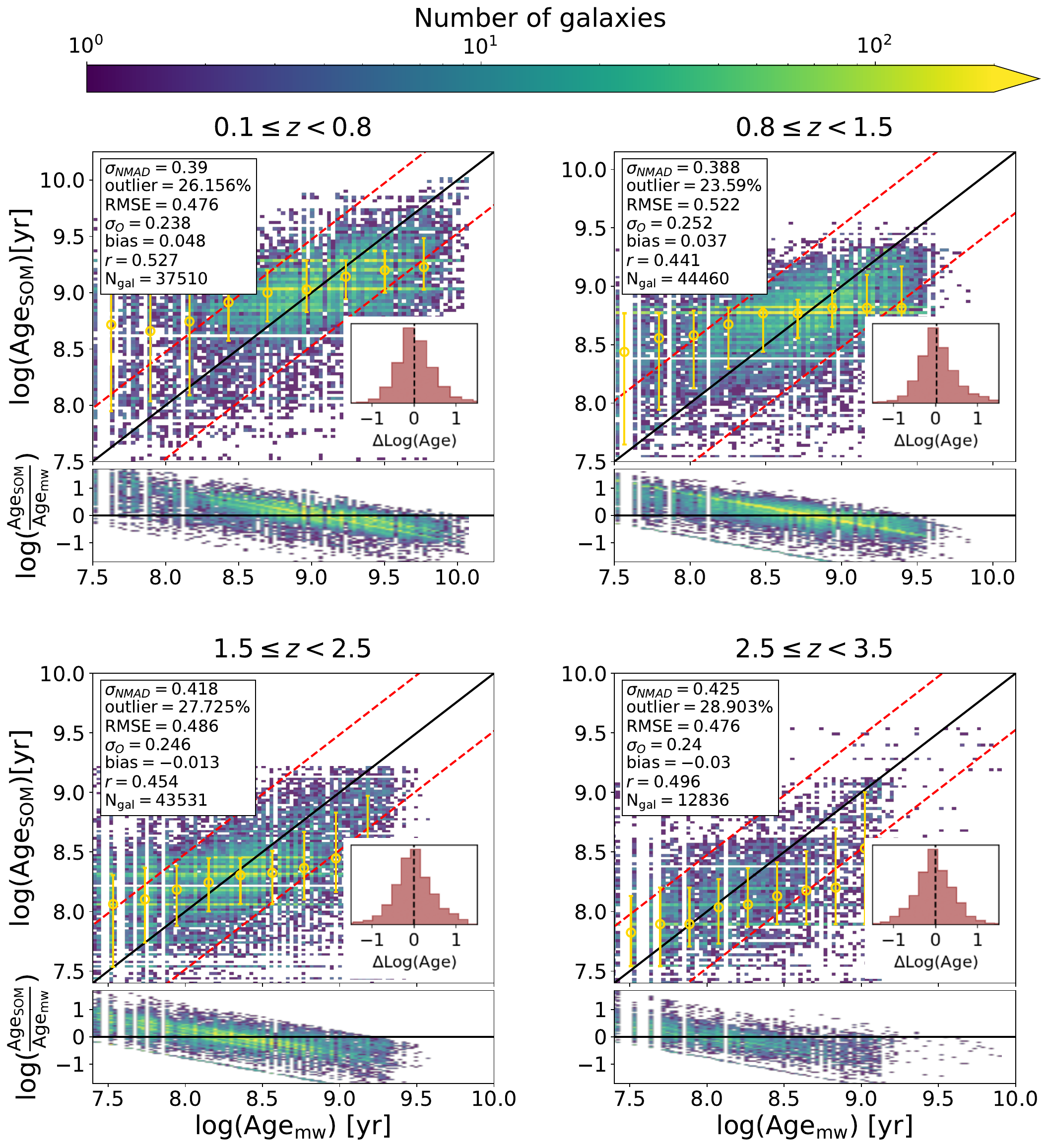}
    \caption{Comparison of the mass-weighted age predicted by the SOM trained on CW data with the age estimated using SED fitting (LePhare). Overestimation occurs at low ages and underestimation at high ages. See the caption of Figure~\ref{fig:mass_sim} for details about the lines.  
}
    \label{fig:age_obs}
\end{figure}

The last parameter we estimated with the SOM is mass-weighted age. Figure~\ref{fig:age_obs} shows the predicted ages compared to the output of the SED fitting. About $70\%$ of the data fall within the $1\sigma$ dex range, although a large dispersion is still visible, which is evident in the low Pearson correlation coefficient ($0.5$). The bias is small, indicating that the average difference between predictions and references is minimal. Overestimation occurs at younger ages, while underestimation occurs at older ages.

\begin{figure}
	\includegraphics[width=1\columnwidth]{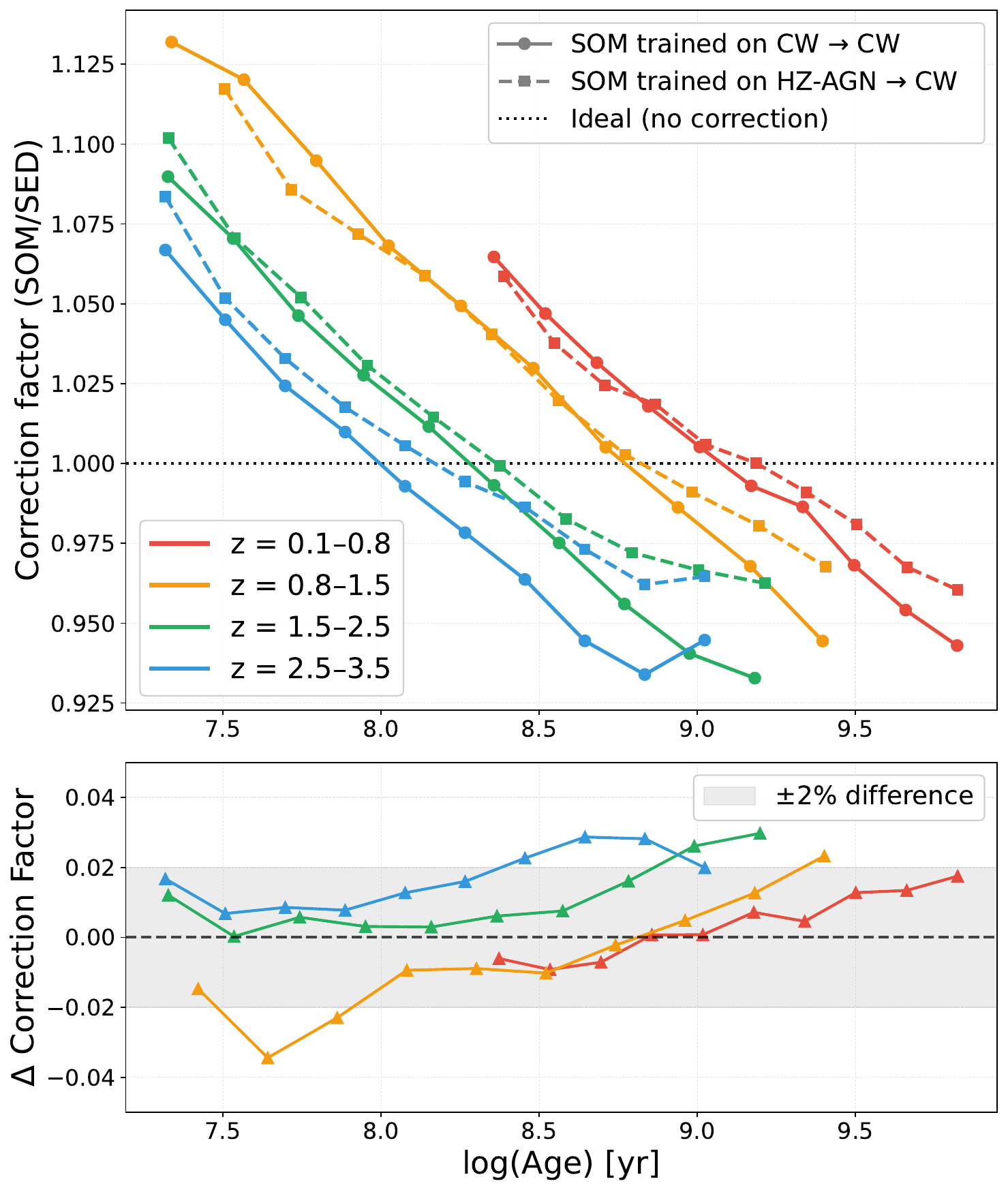}
    \caption{The correction factor represents the median ratio of SOM to mass-weighted ages in bins of SED-derived age. Top panel: Circles with solid lines show the correction factors when applying the CW to the SOM trained on CW, while squares with dashed lines represent results from applying the CW to the SOM trained on HZ-AGN data. Bottom panel: The difference between the correction factors as a function of the mean age of the bins, with the shaded region indicating the $\pm 0.02$ difference.
}
    \label{fig:correction_factor}
\end{figure}

\begin{figure}
	\includegraphics[width=1\columnwidth]{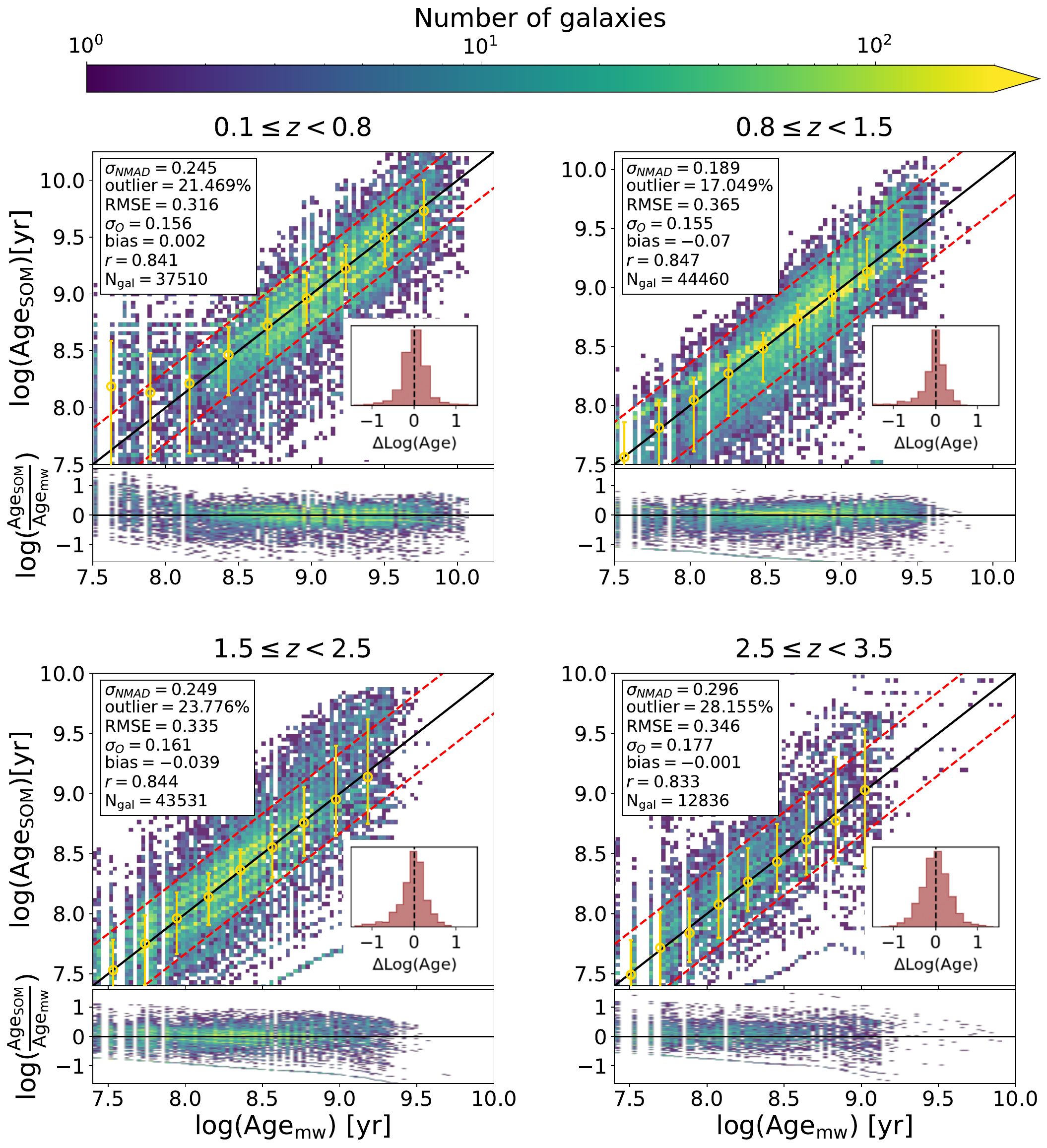}
    \caption{Comparison of the mass-weighted age predicted by the SOM trained on CW data, after applying the correction factor, with the age estimated using SED fitting (LePhare). The $\sigma_{\mathrm{NMAD}}$, RMSE, and the Pearson correlation coefficient values are improved. See the caption of Figure~\ref{fig:mass_sim} for details about the lines. 
}
    \label{fig:age_obs_corr}
\end{figure}

To improve the alignment between SOM-predicted and LePhare-derived ages, we calculated age-dependent correction factors. These were derived by taking the median ratio of SOM to LePhare ages within bins of LePhare age. In Figure~\ref{fig:correction_factor}, each circle represents the correction factor for a specific age range. These factors were then interpolated across the entire age range and used to scale the SOM predicted ages accordingly. The corrected ages are shown in Figure~\ref{fig:age_obs_corr}, where $\sigma_{\mathrm{NMAD}}$, RMSE, and the Pearson correlation coefficient all show improvement, while the outlier fraction remains unchanged. However, a slight increase in bias is observed after correction.

In summary, the SOM trained with CW data exhibits more degeneracy compared to the SOM trained on HZ-AGN data. Redshift prediction presents more challenges than other parameters, as indicated by the larger dispersion in the data.
Stellar mass estimation is generally accurate, although a large overestimation is observed for masses below the mass completeness limit in the redshift bin $0.1-0.8$. For SFR, sSFR, and age, both overestimation and underestimation are seen at low and high values, with some dispersion. Overall, the prediction of physical parameters is acceptable, though not as accurate as in the simulation data.

\subsection{Applying the SOM Trained on HORIZON-AGN to the COSMOS-Web Data}
\label{subsec:som_sim_obs}

\begin{figure}
	\includegraphics[width=1\columnwidth]{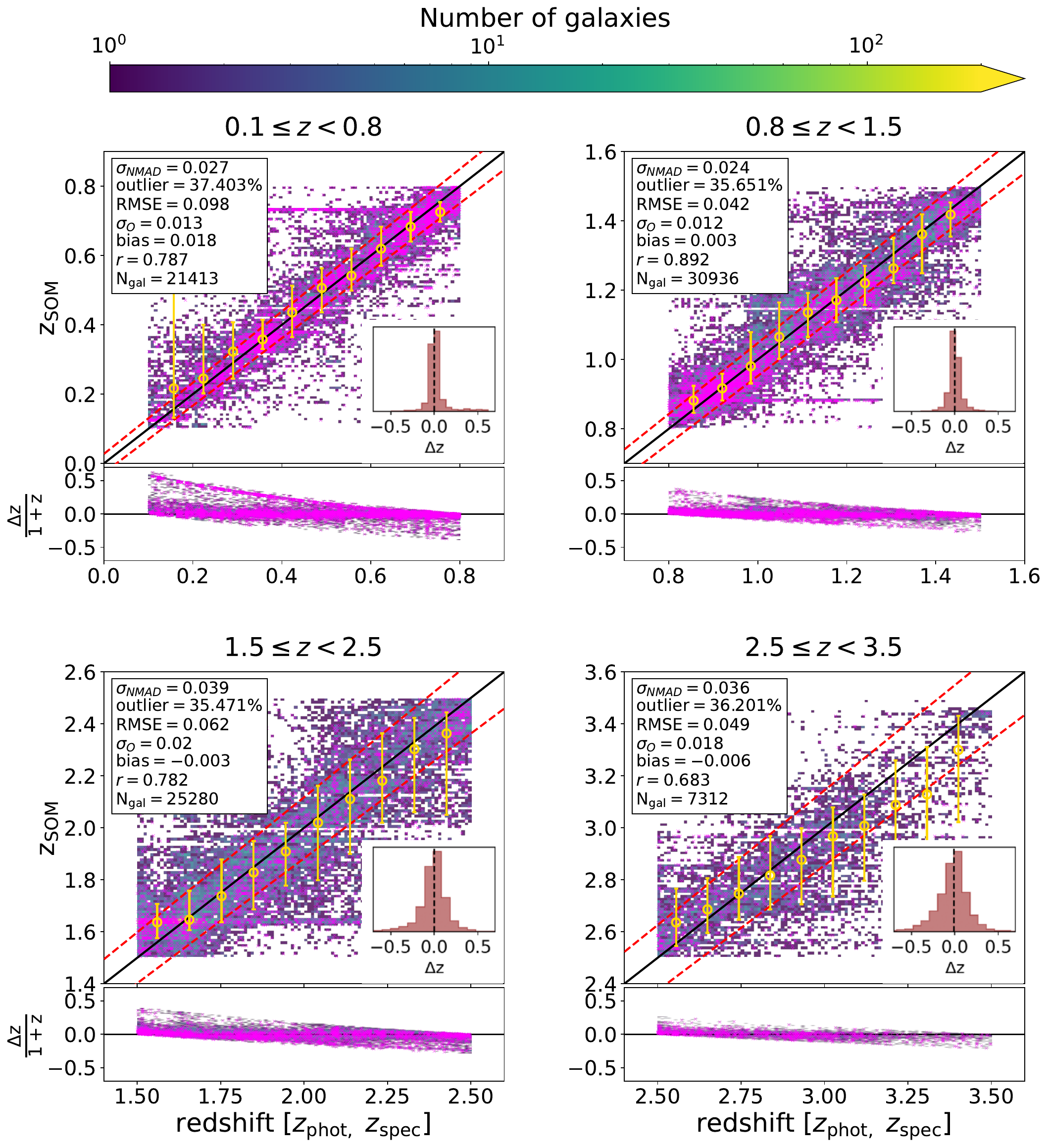}
    \caption{Comparison of the redshift predicted by applying the CW data to the SOM trained on HZ-AGN data with the redshift estimated using SED fitting (LePhare) and the spectroscopic redshifts, where available. Magenta points correspond to galaxies with spectroscopic redshifts, which align with the 1:1 line as expected. The points are dispersed, with overestimation at low redshift and underestimation at high redshift. See the caption of Figure~\ref{fig:redshift_sim} for details about the lines.
}
    \label{fig:redshift_sim_obs}
\end{figure}

In this section, we aim to predict the physical parameters of the CW data by applying the SOM trained on the simulation data. We began with redshift, incorporating both photometric and spectroscopic redshifts for galaxies where available.
Figure~\ref{fig:redshift_sim_obs} shows the redshift predictions, where the magenta points represent galaxies with spectroscopic redshifts. About $35\%$ of the data fall outside the dashed lines, which represent the $1\sigma_{\mathrm{NMAD}}(1+z)$ range, indicating noticeable dispersion. Galaxies with spectroscopic redshifts align more closely with the 1:1 relation.

In the redshift bin $0.1-0.8$, the overestimation is observed at redshifts below $0.4$. In the $0.8-1.5$ bin, the median redshift is well estimated below $z = 1.0$, but the redshifts between $1.0$ and $1.2$ are overestimated and the values above $1.2$ tend to be underestimated. The Pearson correlation coefficient also shows a strong linear correlation  ($0.9$).  In the $1.5-2.5$ bin, an overestimation is observed for redshifts below $1.7$, while an underestimation occurs above $2.1$. For the $2.5-3.5$ bin, the number of galaxies is low and the predictions are less accurate. There is greater dispersion, and a similar pattern of overestimation at lower redshifts and stronger underestimation at higher redshifts is evident.

\begin{figure}
	\includegraphics[width=1\columnwidth]{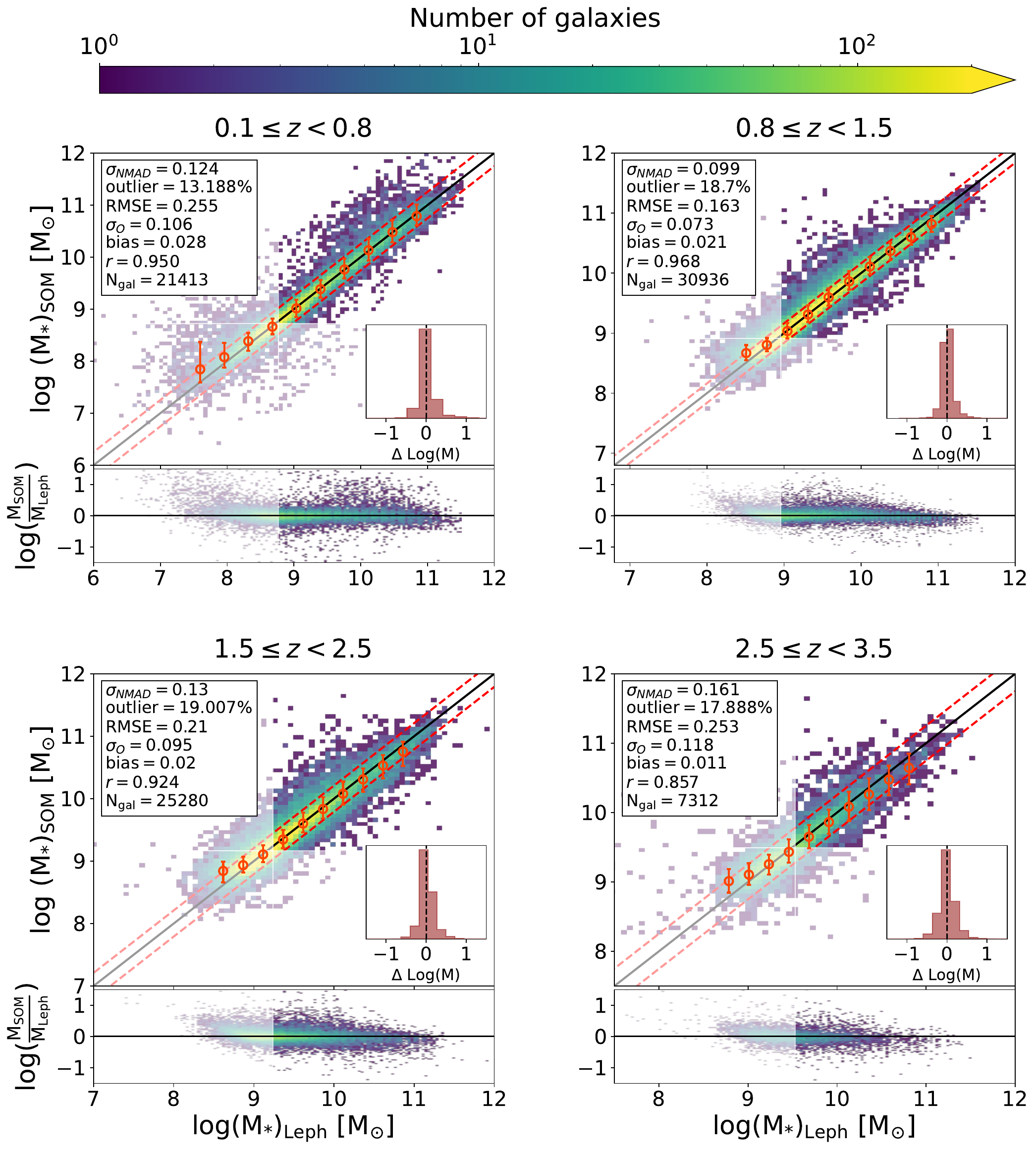}
    \caption{Comparison of the total stellar mass predicted by applying the CW data to the SOM trained on HZ-AGN data with the total stellar mass estimated using SED fitting (LePhare). The shaded regions denote the mass completeness limit corresponding to $m_\mathrm{F444}^\mathrm{lim} = 24.8$ for both the LePhare and SOM estimates. The overall prediction is close to the 1:1 relation, with some scatter. See the caption of Figure~\ref{fig:mass_sim} for details about the lines.
}
    \label{fig:mass_sim_obs}
\end{figure}

For the stellar mass, after renormalizing the estimated values, we compare the estimated stellar mass to the output of the SED-fitting stellar mass. Figure \ref{fig:mass_sim_obs} shows the stellar mass estimated by LePhare compared to the prediction from the SOM. In all redshift bins, the estimation is acceptable and close to the 1:1 relationship, as evidenced by the Pearson correlation coefficient ($r= 0.9$). The most scattered points appear below the mass completeness limit calculated for the $m_\mathrm{F444}^\mathrm{lim}=24.8$.

In the redshift bin $0.1$–$0.8$, some points are scattered from the SED fitting values, which is reflected in the high $\sigma_\mathrm{NMAD}$ and RMSE. However, the overall prediction aligns with the 1:1 relation, with a small bias (bias = $0.028$). In the redshift bin $0.8$–$1.5$, the $\sigma_{\mathrm{NMAD}}$ is lower ($0.099$), and the bias is close to zero  ($0.021$), although about $19\%$ of the points are scattered outside the $1\sigma$ boundaries. This is also evident in the redshift bin $1.5$–$2.5$. In the redshift bin $2.5$–$3.5$, a high $\sigma_{\mathrm{NMAD}}$ is observed ($0.161$), indicating poor estimation for some points, although the bias remains small.

\begin{figure}
	\includegraphics[width=1\columnwidth]{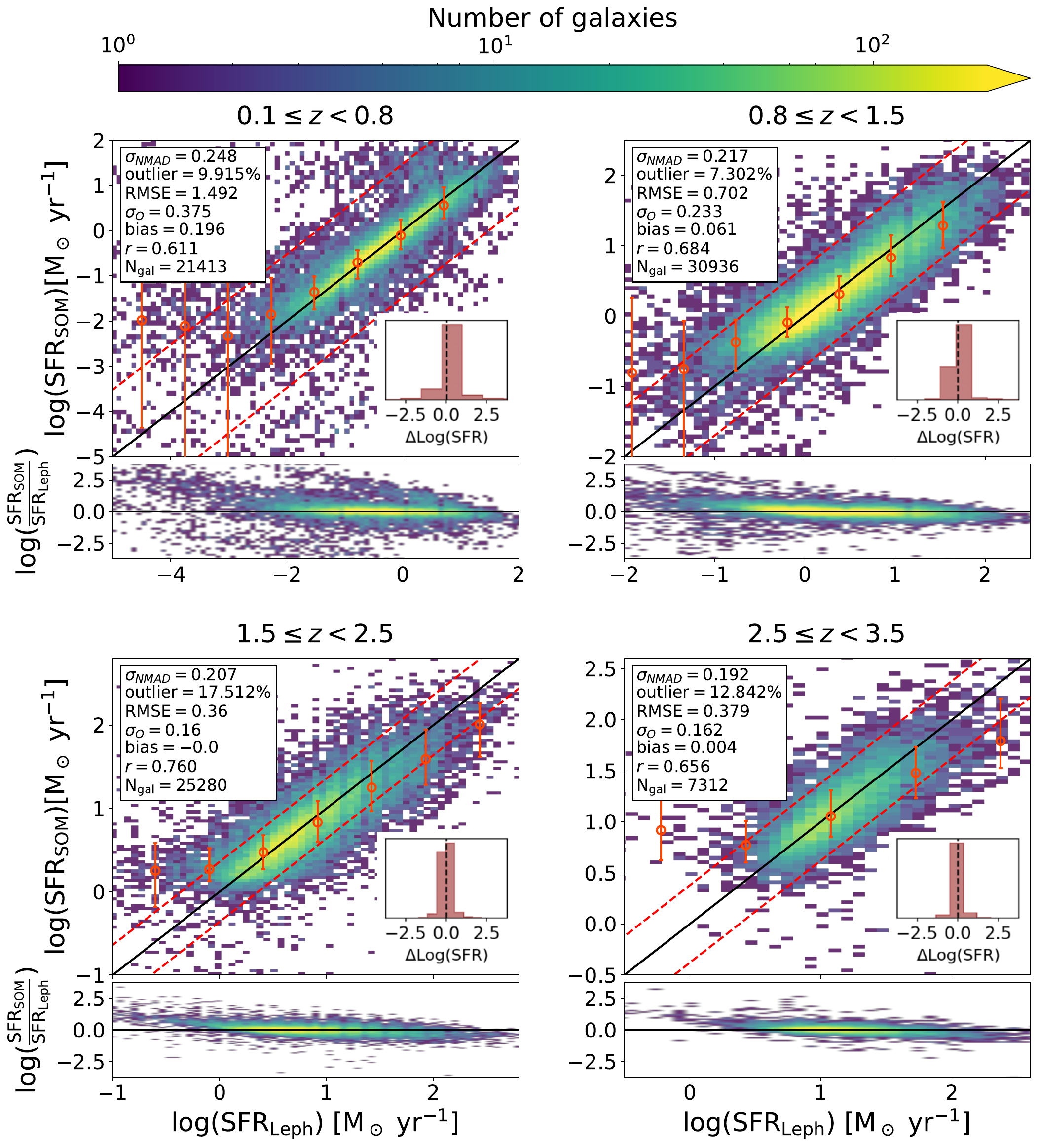}
    \caption{Comparison of the SFR predicted by applying the CW data to the SOM trained on HZ-AGN data with the SFR estimated using SED fitting (LePhare). The overall prediction is acceptable, with overestimation in low-SFR galaxies. See the caption of Figure~\ref{fig:mass_sim} for details about the lines.
}
    \label{fig:sfr_sim_obs}
\end{figure}

After renormalizing the SFR predicted by the SOM, we compare it to the output of the SED fitting. Figure \ref{fig:sfr_sim_obs} shows the SFR estimates from the SOM. It is evident that there is more scatter at low SFRs, especially in the redshift bins $0.1-0.8$ and $0.8-1.5$. The overestimation and underestimation are also seen at low and high SFRs. In general, $85\%$ of the data show a difference between LePhare and SOM SFRs within 1$\sigma$. The Pearson correlation coefficient is also between $0.6 < r < 0.8$, indicating that the linear correlation is not as strong as in the previous SFR predictions (see Figures \ref{fig:sfr_sim} and \ref{fig:sfr_obs}).

In the redshift bin $0.1$–$0.8$, a large $\sigma_{\mathrm{NMAD}}$ is observed in the predictions ($0.248$), which is due to scattering in galaxies with low SFR. In the redshift bin $0.8-1.5$, $\sigma_{\mathrm{NMAD}}$ is lower ($0.217$), and the bias is also lower ($0.061$). In the redshift bin $1.5-2.5$, about $18\%$ of the data are outliers beyond 1$\sigma$, indicating slightly more scatter than in the other redshift bins. In the redshift bin $2.5-3.5$, $\sigma_{\mathrm{NMAD}}$ is slightly lower($0.192$), but some points show large prediction errors (RMSE = $0.379$).

For the sSFR, we divided the estimated SFR by the estimated stellar mass. Figure \ref{fig:ssfr_sim_obs} (see Appendix~\ref{ap:ssfr_estimation}) shows the SOM predictions compared to the results of the SED fitting. About $80\%$ of the points lie within the $1\sigma$ boundaries, with dispersion present across all redshift bins. Overestimation at lower sSFR and underestimation at higher sSFR are observed. In higher redshift bins, the scattering is greater than in lower redshift bins, which is evident in the $\sigma_\mathrm{NMAD}$ and $r$ values.

\begin{figure}
	\includegraphics[width=1\columnwidth]{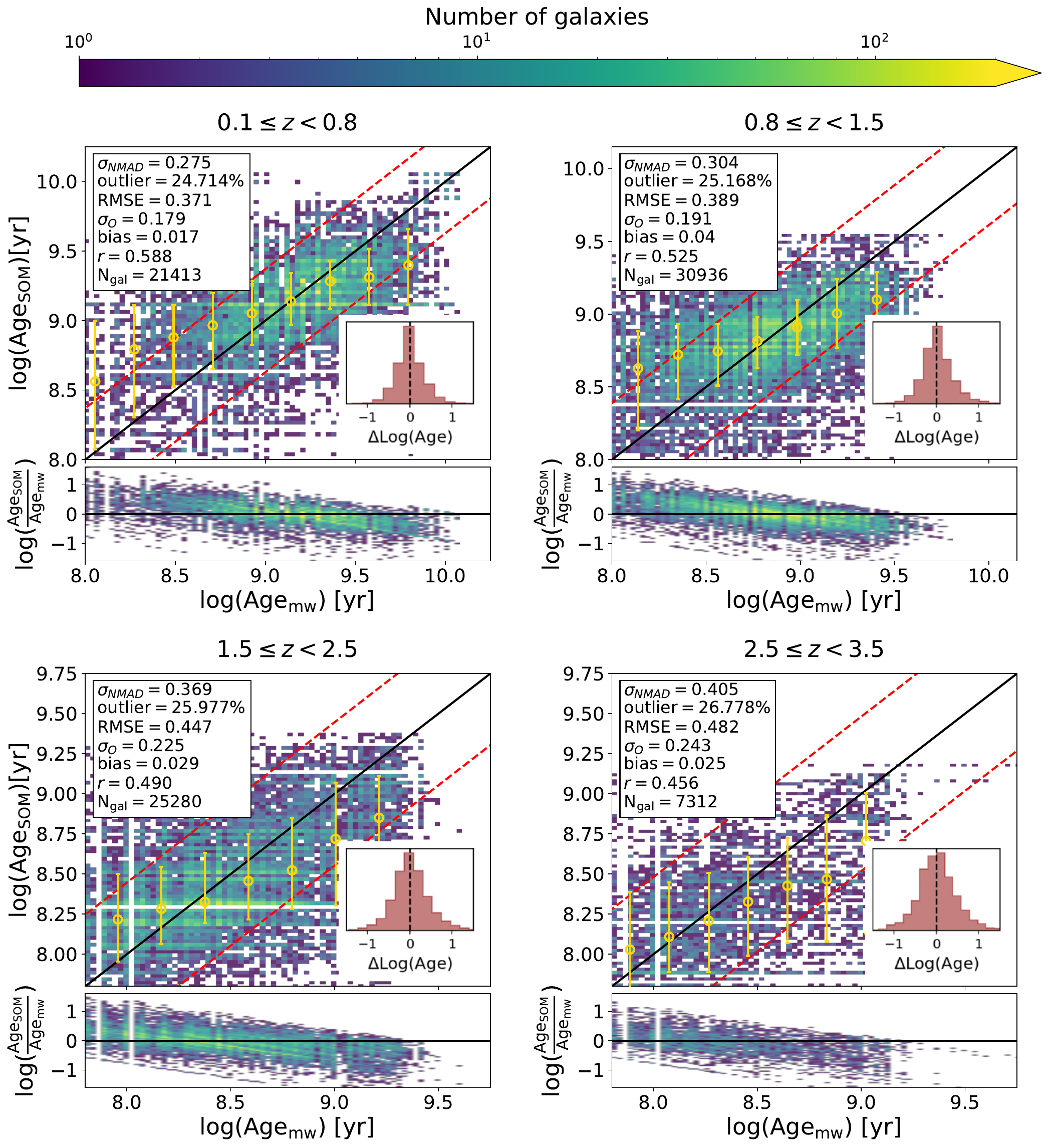}
    \caption{Comparison of the mass-weighted age predicted by applying the CW data to the SOM trained on HZ-AGN data with the age estimated using SED fitting (LePhare). About $75\%$ of the data points fall within the $\pm1\sigma$ boundaries. See the caption of Figure~\ref{fig:mass_sim} for details about the lines.
}
    \label{fig:age_sim_obs}
\end{figure}

Next, we estimated the mass-weighted age from the SOM. Figure \ref{fig:age_sim_obs} shows the age predictions from the SOM compared to the SED fitting output. The large scattering is seen in the prediction of age, although about $75\%$ lie within 1$\sigma$ dex. The median of the estimates shows overestimation at low ages and underestimation at high ages. In redshift bins $0.1-0.8$ and $0.8-1.5$ the $\sigma_{\mathrm{NMAD}}$ is lower while in the redshift bins $1.5-2.5$ and $2.5-3.5$, the $\sigma_{\mathrm{NMAD}}$ is larger.

\begin{figure}
	\includegraphics[width=1\columnwidth]{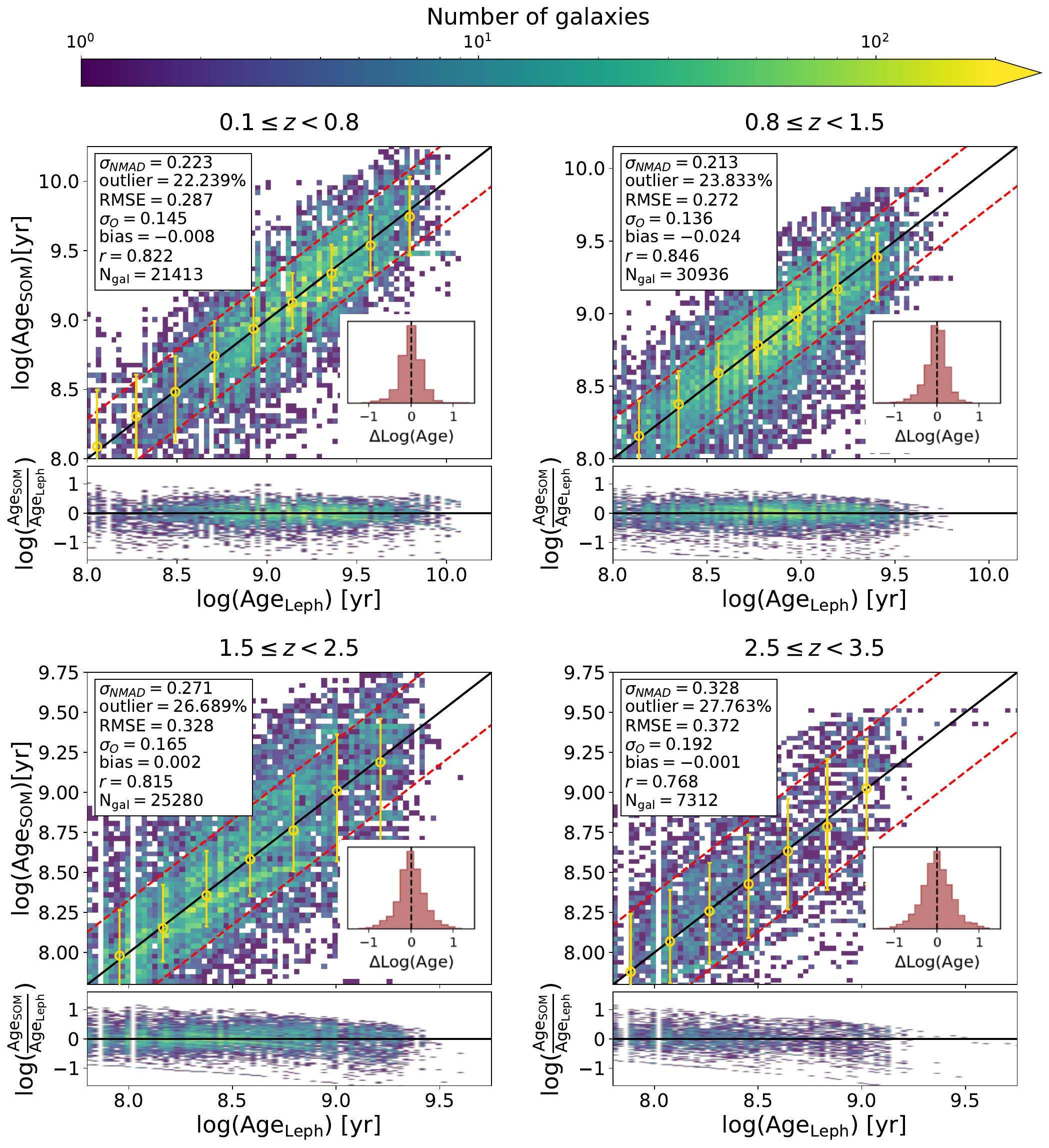}
    \caption{Comparison of the mass-weighted age predicted by applying the CW data to the SOM trained on HZ-AGN data, after applying the correction factor, with the age estimated using SED fitting (LePhare). The $\sigma_{\mathrm{NMAD}}$, RMSE, and Pearson correlation coefficient values show improvement. See the caption of Figure~\ref{fig:mass_sim} for details about the lines.
}
    \label{fig:age_sim_obs_corr}
\end{figure}

Similarly to the previous subsection (\ref{subsec:som_obs}), we derived age-dependent correction factors by calculating the median ratio of predicted SOM to LePhare-derived ages in bins of LePhare age. The squares with dashed lines in Figure~\ref{fig:correction_factor} show the applied correction factors. These were interpolated across the age range and then used to scale the SOM-predicted ages to better match the LePhare estimates. Figure~\ref{fig:age_sim_obs_corr} presents the age estimates after applying these corrections, indicating an overall improvement in performance. The values of the $\sigma_{\mathrm{NMAD}}$, RMSE, and Pearson correlation coefficient improved.

In summary, the prediction of CW galaxies using the SOM trained on HZ-AGN data leads to acceptable estimations with some scatter. For redshift, the SOM does not perform as well as it does on the simulation data, and notable scattering is evident. Galaxies with spectroscopic redshifts yield better predictions. For stellar mass and SFR, more data points lie close to the 1:1 relation, although some scatter is present. In the case of low-SFR galaxies, the SOM predictions tend to be overestimated. For sSFR, overestimation and underestimation are observed at low and high sSFR, respectively. In predicting age, about $75\%$ of the data fall within 1$\sigma$ dex, although a large dispersion is still observed.

\begin{figure}
	\includegraphics[width=1\columnwidth]{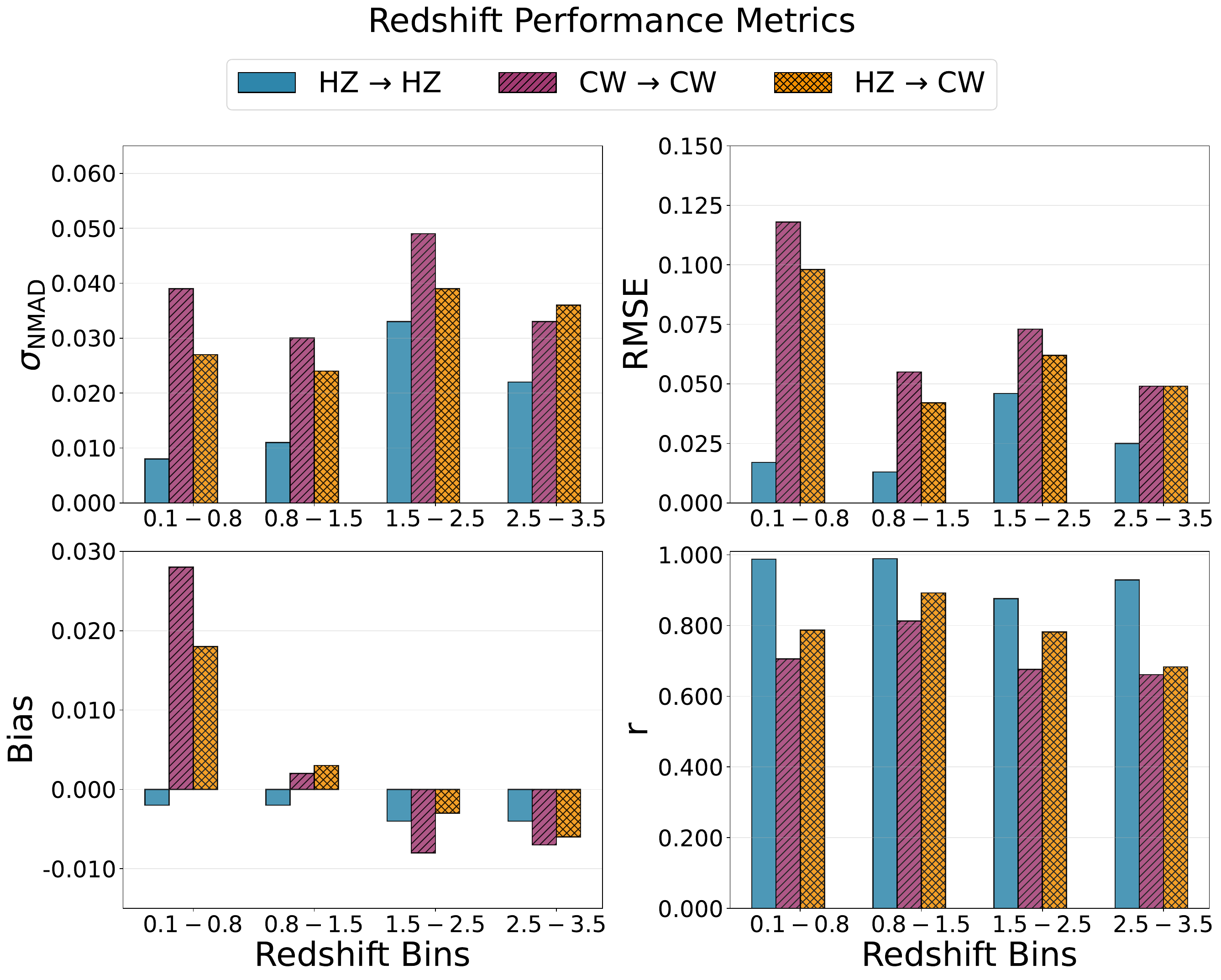}
    \caption{Summary of redshift estimation using the SOM for four metrics: $\sigma_\mathrm{NMAD}$, RMSE, bias, and Pearson correlation ($r$), for HZ applied to the SOM trained on HZ, CW applied to the SOM trained on CW, and CW applied to the SOM trained on HZ.
}
    \label{fig:sum_redshift}
\end{figure}

\begin{figure}
	\includegraphics[width=1\columnwidth]{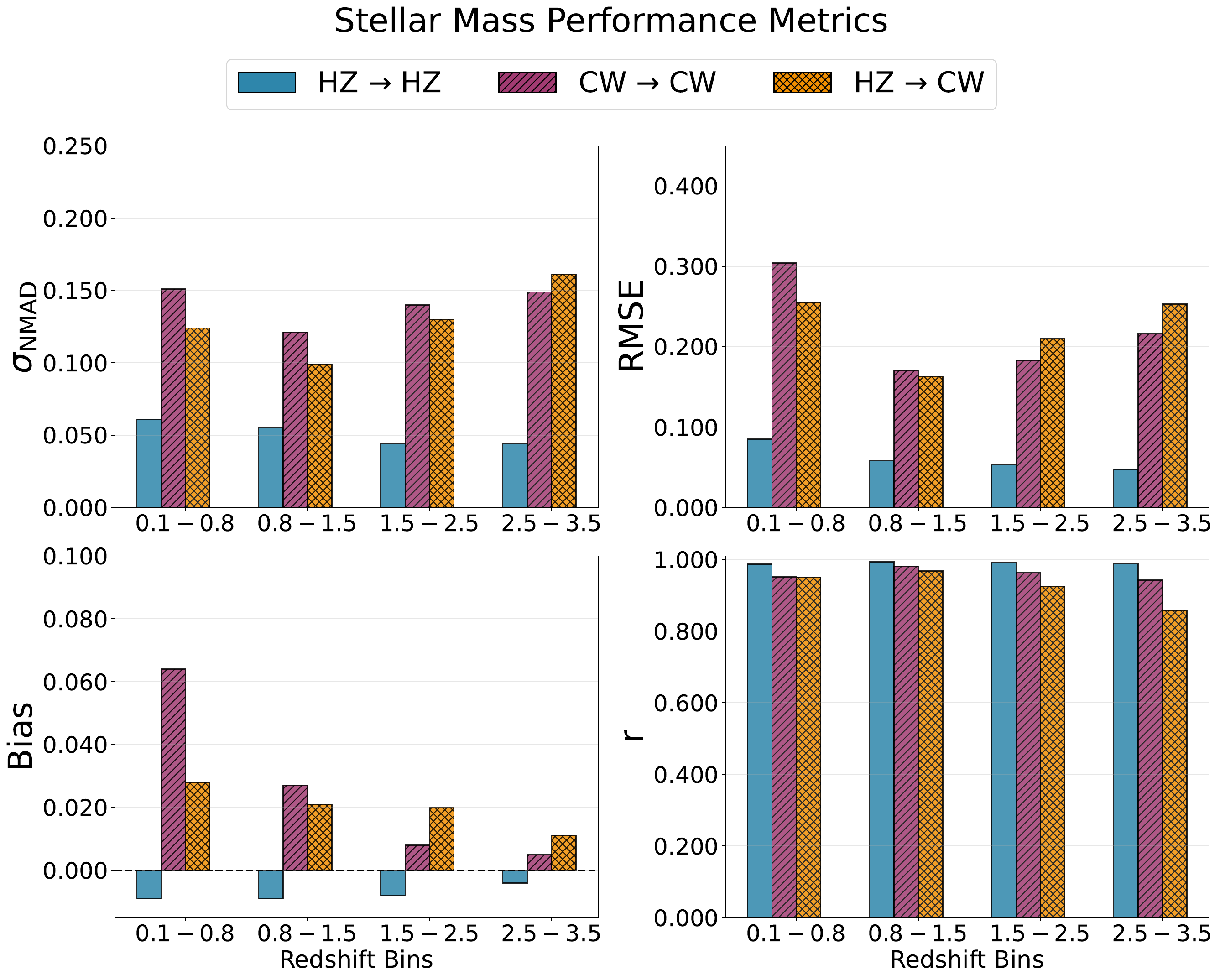}
    \caption{Summary of stellar mass estimation using the SOM for four metrics: $\sigma_\mathrm{NMAD}$, RMSE, bias, and Pearson correlation ($r$), for HZ applied to the SOM trained on HZ, CW applied to the SOM trained on CW, and CW applied to the SOM trained on HZ.
}
    \label{fig:sum_mass}
\end{figure}

\begin{figure}
	\includegraphics[width=1\columnwidth]{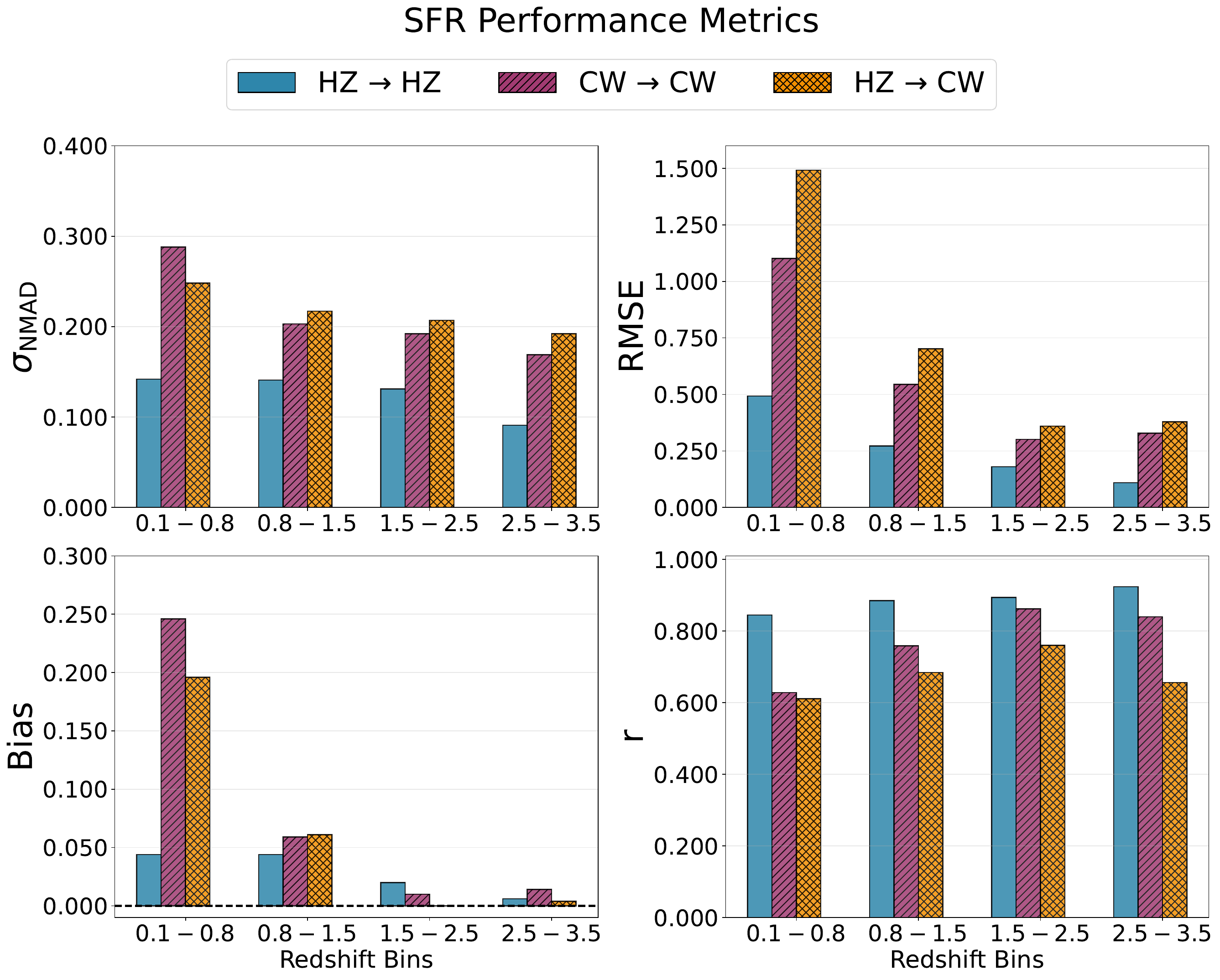}
    \caption{Summary of SFR estimation using the SOM for four metrics: $\sigma_\mathrm{NMAD}$, RMSE, bias, and Pearson correlation ($r$), for HZ applied to the SOM trained on HZ, CW applied to the SOM trained on CW, and CW applied to the SOM trained on HZ.
}
    \label{fig:sum_sfr}
\end{figure}

\begin{figure}
	\includegraphics[width=1\columnwidth]{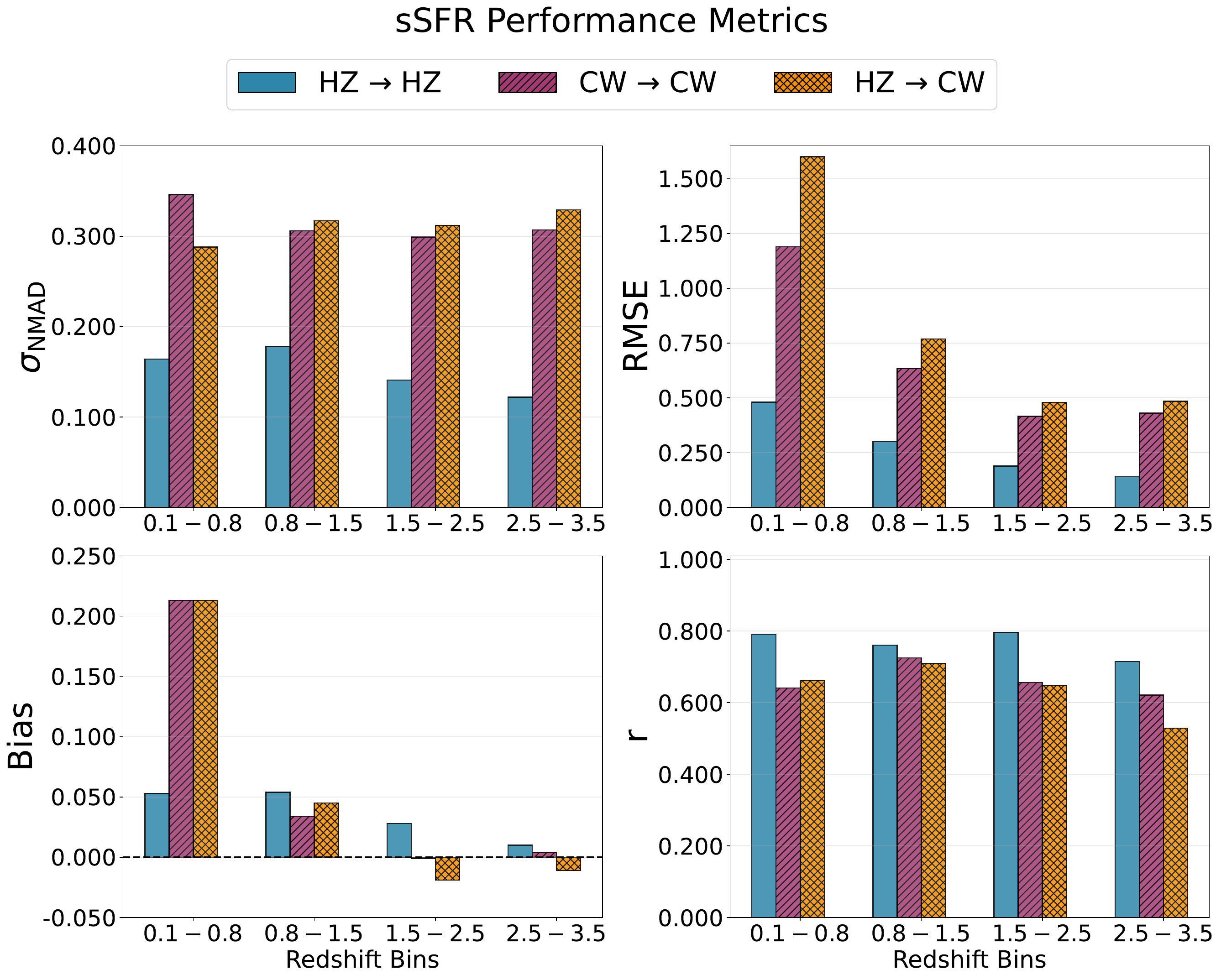}
    \caption{Summary of sSFR estimation using the SOM for four metrics: $\sigma_\mathrm{NMAD}$, RMSE, bias, and Pearson correlation ($r$), for HZ applied to the SOM trained on HZ, CW applied to the SOM trained on CW, and CW applied to the SOM trained on HZ.
}
    \label{fig:sum_ssfr}
\end{figure}

\begin{figure}
	\includegraphics[width=1\columnwidth]{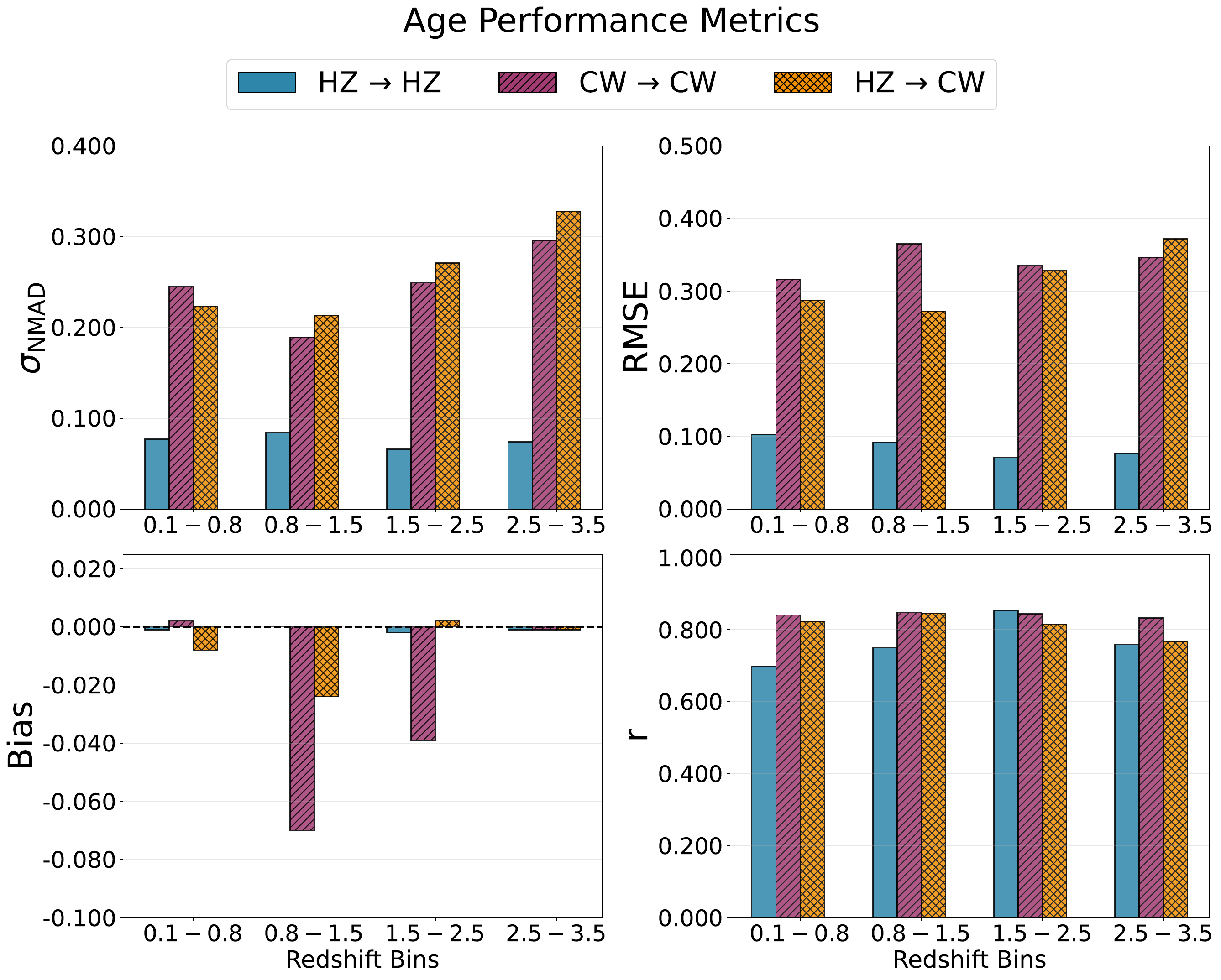}
    \caption{Summary of mass-weighted age estimation using the SOM for four metrics: $\sigma_\mathrm{NMAD}$, RMSE, bias, and Pearson correlation ($r$), for HZ applied to the SOM trained on HZ, CW applied to the SOM trained on CW, and CW applied to the SOM trained on HZ.
}
    \label{fig:sum_age}
\end{figure}

Figures \ref{fig:sum_redshift}, \ref{fig:sum_mass}, \ref{fig:sum_sfr}, \ref{fig:sum_ssfr}, and \ref{fig:sum_age} show the summary of all predictions of the physical parameters from the SOM training for redshift, stellar mass, SFR, sSFR, and age (after applying the correction factor), respectively. For each redshift bin, the metrics are presented for the HZ applied to SOM trained on HZ, CW applied to SOM trained on CW, and CW applied to SOM trained on HZ. HZ applied to HZ-trained SOM provides the best predictions, while the other SOM trainings show similar results.

\section{Discussion}
\label{sec:discussion}
\subsection{Comparison with Previous Studies}

In this study, we aim to predict physical parameters, such as redshift, stellar mass, SFR, sSFR, and age, for both simulated and observational data using a SOM. We use two data sets: HZ-AGN and CW. The filters applied are $u-g$, $g-r$, $r-i$, $z-Y$, $Y-J$, $J-H$, $H-K_\mathrm{s}$, $K_\mathrm{s}-\mathrm{F277}$, and $\mathrm{F277}-\mathrm{F444}$. We train the SOM on the HZ-AGN and CW datasets to estimate the physical parameters of each dataset, respectively. Furthermore, we applied the SOM trained on the HZ-AGN data to the CW data to predict its physical parameters. By propagating the photometric uncertainties into the likelihood for each SOM cell, the peak of the resulting PDF is taken as the predicted physical parameter. In this section, we compare our results with previous studies to identify similarities and differences.

In \citet{davidzon2019}, the authors used HZ-AGN data to train the SOM and predict redshift and SFR. After accounting for photometric errors, they trained the SOM with a grid size of $80 \times 80$ and labeled it with physical parameters. The physical parameters were then estimated by considering the 10 nearest cells and weighting them based on the distance in color space. The results were compared with the true redshift and SFR values from the simulation data, as well as with the predictions from the LePhare SED fitting.

Since we train the SOM in individual redshift bins, whereas that study trains it on the entire dataset, we also train the SOM over the full redshift range from $0.1$ to $3.5$ to enable a direct comparison and assess whether our improvements in data are effective. The difference in filter sets between that study and ours lies in the optical and infrared bands: we use the g, r, i, and z bands from Subaru/HSC, while they used B, V, r, $i^+$, and $z^{++}$. Additionally, their infrared data came from Spitzer/IRAC channels $1$ and $2$, whereas we use the F277W and F444W filters from JWST. The reference magnitude used to normalize the stellar mass and the SFR was $m_{i^+} = 22$, since we use $m_\mathrm{F444}=22$. Another difference between our work and the paper is the total number of galaxies: we use $257,110$ to train the SOM whereas the paper used $371,168$ for estimating the redshift and SFR. The optimal grid size for our data was set to $100 \times 90$.

To compare the predicted redshift with the true redshift of the simulated data, the article reported a scatter of $\sigma_{\text{NMAD}} = 0.044$, while our result is $0.021$. Their outlier fraction was $6.13\%$ for $\Delta z/(1+z) > 0.15$, compared to our $1.05 \%$. The bias in their results was $-0.001$, while ours is $-0.004$.

The authors also compared the predictions of the SOM to the estimates from SED fitting. The $\sigma_\mathrm{NMAD}$ and bias for the redshift estimated using SED fitting, compared to the true redshift values, were $0.024$ and $-0.011$, respectively. In our results, the $\sigma_{\mathrm{NMAD}}$ is $0.021$, which is lower than that of the LePhare SED fitting, and the bias is also smaller. This indicates that the redshift predictions from the SOM are comparable to those from SED fitting, despite the SOM using only $11$ bands compared to the larger number used in LePhare.

For SFR estimation, their results were presented in four redshift bins ($0.2 < z < 0.8$, $0.8 < z < 1.5$, $1.5 < z < 2$, and $2 < z < 3$), and they reported $\sigma_{\text{NMAD}}$ values consistently below $0.2$ dex. When training in all redshift bins, $\sigma_{\text{NMAD}}$ is $0.135$ dex, indicating an improvement in our results. The paper also reported an overestimation of the SFR for low-SFR galaxies at $z > 0.8$. This overestimation is smaller in our results. Additionally, an underestimation was observed for highly star-forming galaxies in both studies, but the offset is smaller in our case.


In \citet{latorre2024}, the SOM was trained with HZ-AGN simulation data to predict redshift, stellar mass, SFR, and sSFR. The paper mainly focused on the choice of filters and the effects of missing bands on the prediction of these physical parameters. Starting with a mass-limited simulated data set from the HZ-AGN lightcone, the authors performed color calibration on the noiseless magnitudes using the UltraVISTA catalog to provide realistic color-parameter correlations. An SOM on the $80 \times 80$ grid was then trained using a random subset of galaxies selected from this mass-limited, color-calibrated data set, based on their 10 observed frame colors. A separate sample limited in Ks ($m_\mathrm{Ks} < 23.5$) was constructed from the mass-limited data to mimic the HSC-Deep survey and was used as the test data set to predict the parameters. The comparison with the true simulation values was first performed by projecting this test sample onto the trained SOM and using the best-matching SOM unit for the noiseless case and then repeated after adding photometric uncertainties to the magnitudes to simulate a more realistic scenario. 

The choice of filters in this paper was based on their sensitivity to physical parameters, using the distance correlation between 55 two-color combinations. The color combinations used to train the SOM were as follows: u -- g, g -- Y, g -- J, r -- i, r -- z, r -- Y, i --z, J -- H, K -- ch2 and ch1 -- ch2. Compared to our filters, the magnitudes we use are mostly the same, with the main differences being the use of JWST filters instead of Spitzer filters, as well as differences in color combinations. Our approach combines colors in order to increase the wavelength to allow for better comparison with other studies. Testing other combinations or filters is the focus of another paper (Kalantari et al., in preparation), which investigates which colors are most effective for estimating physical parameters using mutual information analysis.

In the paper, the number of data points used for training the SOM was $228,524$, and $227,365$ were used for prediction. In our study, we used $257,110$ data points for training and $257,113$ to predict the physical parameters. The redshift range in their work spans up to $3.5$, the same as in our study. The reference magnitude used to normalize the stellar mass and the SFR was $m_J = 23$, since we use $m_\mathrm{F444}=22$. 

To compare the results of the paper that includes photometric errors with our results, we start with redshift predictions. The paper reported the following metrics: $\sigma_{\text{NMAD}} = 0.026$, which is slightly higher than our result  ($0.021$); the outlier fraction was $1.206\%$, which is comparable to our value ($1.05\%$). The RMSE was $0.099$ in the paper, while in our study it is somewhat better ($0.039$). The bias reported in the paper was $0.003$, while ours is slightly better at $-0.004$. These results indicate that we have some data points with poor redshift predictions at $z > 1$, while the number of the outliers was fewer in \citet{latorre2024}. However, the overall scatter, particularly in the low-redshift range from $0.1$ to $1$, is smaller in our case, contributing to a lower RMSE and improved bias.

For stellar mass estimation, the paper reported a $\sigma_{\text{NMAD}}$ of $0.090$, while in our study it is $0.051$. The outlier fraction was $4.376\%$ in the paper and $3.78\%$ in our case. RMSE and bias were $0.110$ and $0.010$, respectively, while in our study they are $0.064$ and $-0.007$. The stellar mass prediction in our study exhibits less scatter, indicating better performance.  For SFR and sSFR, our results are close to those reported in the paper. Our $\sigma_\mathrm{NMAD}$ are $0.135$ and $0.159$, lower than $0.169$ and $0.194$ in the paper, suggesting that our results are more accurate. We note that while the paper excludes quenched galaxies ($\log(\mathrm{sSFR})<-13 {\text{yr}}^{-1}$), we retain them in our analysis.

In summary, the prediction of the physical parameters improves, as evidenced by most of the metrics and the reduced scatter in the data points. The improvements are likely due to the use of JWST filters, which are deeper and enable the SOM to be better trained and to predict physical parameters more accurately with less degeneracy.

In \citet{davidzon2022}, the SOM was used to predict redshift, stellar mass, and SFR. It was trained on the COSMOS2020 dataset using 12 observed frame bands, with a Ks-band limit of $m_\mathrm{Ks}<24.8$. SFRs in this dataset were derived from MIPS-detected galaxies to provide more robust constraints. The SOM, with a grid size of $80\times80$, was trained and then labeled using the median values of the physical parameters. A separate observational dataset, the SXDF2018 catalog, was used as test data to estimate physical parameters such as stellar mass and SFR. These were inferred using the five nearest SOM cells, weighted by the inverse of the Euclidean distance between the colors and the weights of each cell.

The colors used in this paper are the same as those we used, with the same combinations. The only difference lies in the JWST filters: we use $K_S - F277$ and $F277 - F444$ instead of $K_S - \text{ch1}$ and $\text{ch1} - \text{ch2}$. They also use the Subaru/HSC y-band filter, which we do not use.  The reference magnitude used to normalize the stellar mass and the SFR was $m_{i^+} = 22.5$, while we used $m_{\mathrm{F444}} = 23$. The redshift cut in the paper is $z < 1.8$, whereas in our case it is $z < 3.5$. The COSMOS2020 dataset used for training the SOM contained $174,522$ galaxies, and the SXDF2018 catalog used for testing had $208,404$ galaxies. In our study, both training and test data are from the CW, using separate galaxies, with $N^\mathrm{CW}_\mathrm{CW,train}=147,884$ for training and $N^\mathrm{CW}_\mathrm{CW,test}=147,884$ for testing. The optimal grid size for our data was set to $70 \times 80$.

To allow a better comparison with \citet{davidzon2022}, we trained the SOM over the entire redshift range. For stellar mass, \citet{davidzon2022} reported a relative scatter of $0.25$~dex, while the standard deviation in our results for stellar masses greater than $10^8\,M_\odot$ is  $0.2$~dex, suggesting a slight improvement. An underestimation in the predicted stellar mass was observed in their study for $3\%$ of the data, which is not present in our results. However, in our study, an overestimation appears at low stellar masses. In contrast, their results show a higher scatter at low stellar mass compared to ours, even though we include masses down to $10^6\,M_\odot$ and redshift up to $3.5$. This improvement is also evident in the Pearson correlation coefficient, which is close to $1$ ($0.94$).

For SFR, the paper used values derived from UVIR bands and H$\alpha$ indicators. The number of data points for ${\mathrm{SFR}_{\mathrm{UVIR}}}$ was $608$ in the SXDF-CANDELS field, and for ${\mathrm{SFR}_{\mathrm{H\alpha}}}$, it was $3718$ in COSMOS. The standard deviations reported for ${\mathrm{SFR}_{\mathrm{UVIR}}}$ and ${\mathrm{SFR}_{\mathrm{H\alpha}}}$ were $0.28$~dex and $0.4$~dex, respectively. In our case, the standard deviation is $0.4$ dex for $\mathrm{SFR}_\mathrm{LePh}$ values higher than $\log(\text{SFR}[M_\odot,\text{yr}^{-1}]) > -1$. This standard deviation is slightly higher than that reported in the paper. In our study, there is a strong correlation between our predictions and the LePhare output, as indicated by a Pearson correlation coefficient of $0.74$.

Despite these differences, the use of JWST filters and the selection of the best-matching SOM cell appear to yield comparable results in standard deviation in the estimation of stellar mass and SFR, although with higher scatter seen in the SFR. It should be noted that our study used a larger sample of galaxies for SFR estimation ($N_\mathrm{pred}=138,337$ galaxies).

We also included redshift predictions across all redshift bins. The figure shows that $\sim80\%$ of the data fall within the $\pm0.15(1+z)$ range, although the scatter is still evident. The galaxies with spectroscopic redshifts align well with the 1:1 relation line. An underestimation of redshift is observed for $z > 2$.  For sSFR, the RMSE is high, and the outlier fraction beyond a $\pm0.5$~dex threshold is $21\%$, with an RMSE of $0.835$. An overestimation is also observed at low sSFR values.


In \citet{hemmati2019b}, the SOM was trained on a grid of size $80\times60$ using templates generated with BC03 and the LePhare package at a fixed redshift of $1$. The COSMOS2015 dataset, within the redshift range $0.8-1.2$, was then used as test data to predict stellar mass. The difference in filter sets between that paper and our study lies in the optical and infrared bands: we use the $g$, $r$, $i$, and $z$ bands from Subaru/HSC, whereas the paper uses $B$, $V$, $r$, $i^+$, and $z^{++}$. Additionally, while they use Spitzer/IRAC channels 1 and 2, we use the F277W and F444W filters from JWST. Although we do not have templates corresponding exactly to this redshift bin, we compared our results to those obtained from a SOM trained on simulated HZ-AGN data in the redshift bin $0.8-1.5$ to evaluate whether stellar mass predictions for COSMOS data have improved.

The number of templates used for training in the paper was $13,776$, while in our case, within the redshift range $0.8-1.5$, it is $41,115$ galaxies. Figure 1 of that paper shows the labeling of SOM with four parameters: age, sSFR, $E(B - V)$, and $M/L_K$, using extrapolation without considering the median of each parameter. It should be noted that in the $M/L_K$ plot, there appears to be an issue in the code, as zero values are shown, even though they do not exist in the actual $M/L_K$ data. 

The number of observational galaxies used in their study was $13,781$, whereas in our case, the final number of galaxies considered for prediction is $30,936$. In their paper, scaling coefficients and manual shifts were applied to align the observed colors with SOM weights/templates, whereas we use the Wasserstein–KDE alignment method. They estimated the stellar mass from the $M/L_K$ ratio and luminosity of the observational data, while we used the stellar mass from the simulation data. The paper reported an outlier fraction of $0.03$ for stellar mass prediction (with $\Delta \log(M) > 0.5$), while our results yield a similar outlier value using the same criterion ($0.02$). The $\sigma_{\text{NMAD}}$ is $0.18$ in their work, while in our case it is $0.099$. The RMSE was $0.22$ for them, compared to $0.163$ for us. Finally, $\sigma_{\text{O}}$ is reported as $0.19$ in their case, while in ours it is $0.13$. These comparisons suggest that our results represent a slight improvement and are comparable to the results presented in the paper.

\subsection{Limitation of the Predictions}

In this subsection, we examine the conditions that may affect the SOM’s ability to predict the physical parameters, and assess whether the resulting estimates can be considered reliable.

\subsubsection{Model Misspecification}
\label{subsubsec:model_misspecification}
First, we discuss the errors in the physical parameters in the simulation data, and in particular, in the observational data, where the uncertainties mainly arise from the SED fitting process.

When applying the simulation data to the SOM trained on the HZ-AGN data, and focusing on the prediction of redshift, we observe a larger scatter in the high-redshift bins. Although redshift is an initial input parameter in the simulation data, the degeneracy in the colors of high-redshift galaxies can lead to galaxies with similar colors but different redshift values being assigned to the same SOM cell.

For the other physical parameters, the predictions in the lower-redshift bins are slightly more scattered than in the higher-redshift bins. This is likely due to the broader range of galaxy types, ages, metallicities, and star formation histories in the local universe, which can cause galaxies with similar colors to have different stellar masses, SFRs, and ages.

When the CW data are applied to the SOM trained on CW and used to calibrate the SOM trained on HZ-AGN, the spread of the data within each SOM cell becomes broader than in the SOM trained solely on HZ-AGN. In other words, galaxies with similar observed colors correspond to a wider range of physical properties, leading to increased dispersion. In the prediction of the redshift, the prediction of the spectroscopic redshift are more accurate than the photometric redshift. This suggests that the prediction accuracy may be affected by uncertainties in photometric redshifts derived from SED fitting.

For other parameters, such as SFR, we compare our predictions with those reported by \citet{davidzon2022}. It is important to note that our analysis uses $\mathrm{SFR}_{\mathrm{LePh}}$, which is more uncertain than $\mathrm{SFR}_{\mathrm{UVIR}}$ or $\mathrm{SFR}_{\mathrm{H\alpha}}$. This likely explains why their results show better agreement than ours. We observe a similar behavior for age in predictions based on both simulation and observational data. Comparing the SOM trained on HZ and applied to the HZ sample (Figure~\ref{fig:age_sim}) with the age predictions for the CW data (Figures~\ref{fig:age_obs}, \ref{fig:age_sim_obs}, and \ref{fig:sum_age}), we find that the HZ→HZ predictions achieve a mean $\sigma_\mathrm{NMAD} = 0.075 \pm 0.007$ and a mean $\mathrm{RMSE} = 0.086 \pm 0.015$, whereas predictions involving CW data show a mean $\sigma_\mathrm{NMAD} \approx 0.252 \pm 0.045$ and a mean $\mathrm{RMSE} \approx 0.328 \pm 0.035$, i.e., approximately $3.3$ times larger.

We also find a systematic trend in the CW-based predictions, with ages overestimated at the young end and underestimated at the old end. This behavior is likely driven by biases inherent to SED fitting, where mass-weighted age is weakly constrained and strongly affected by the age–metallicity degeneracy and star formation history priors \citep{gallazzi,Gallazzi2006}, often pushing solutions toward extreme values. In contrast, the SOM assigns galaxies with similar SEDs to common nodes and returns a typical age for each region of color space, which naturally compresses the dynamic range and pulls extreme SED-based ages toward intermediate values. In the simulations, ages are model inputs rather than fitted parameters and are not subject to these degeneracies, resulting in substantially better agreement.

The bottom panel of Figure~\ref{fig:correction_factor} shows the difference between the age correction factors obtained by applying the CW data to the SOM trained on CW data (CW$\rightarrow$CW) versus on HZ-AGN data (HZ$\rightarrow$CW). Across redshift bins, the mean differences range from $0.003$ to $0.017$ dex (medians $0.003-0.016$ dex), with standard deviations of $0.008-0.016$ dex and $70-100\%$ of points within $\pm2\%$. To quantify the relative contributions of SED-fitting systematics versus training dataset choice, we computed the following quantities in each redshift bin:
(i) the mean absolute deviation of the correction factors from unity, averaged over the CW$\rightarrow$CW and HZ$\rightarrow$CW cases, representing the total correction required, and (ii) the mean absolute difference between the CW$\rightarrow$CW and HZ$\rightarrow$CW correction factors.

The fraction of the total correction arising from the choice of training dataset was then calculated as the ratio of (ii) to (i), with the remainder attributed to SED-fitting systematics. We find that the training dataset choice contributes $26-28\%$ of the total correction in the three lower redshift bins ($z < 2.5$) and $48\%$ in the highest redshift bin ($2.5 < z < 3.5$), with an overall contribution of $\sim32\%$ across all redshifts. This indicates that the majority ($52-74\%$) of the bias in age estimates arises from inherent systematic effects in the SED-fitting process.

To better identify model degeneracies and evaluate whether they can be accounted for in the SOM predictions, we define the total uncertainty as $\sigma_{\rm tot, param} = \sqrt{\sigma_{\rm stat, param}^2 + \sigma_{\rm sys, param}^2}$ where $\sigma_{\rm stat, param}$ is the standard deviation of the likelihood-based PDF, which reflects both photometric uncertainty and color–template degeneracy, and $\sigma_{\rm sys, param}$ is half of the 84th–16th percentile range of the physical parameters of the SOM BMU, which also traces color–template degeneracy \citep{latorre2024}.

\begin{figure}
	\includegraphics[width=1\columnwidth]{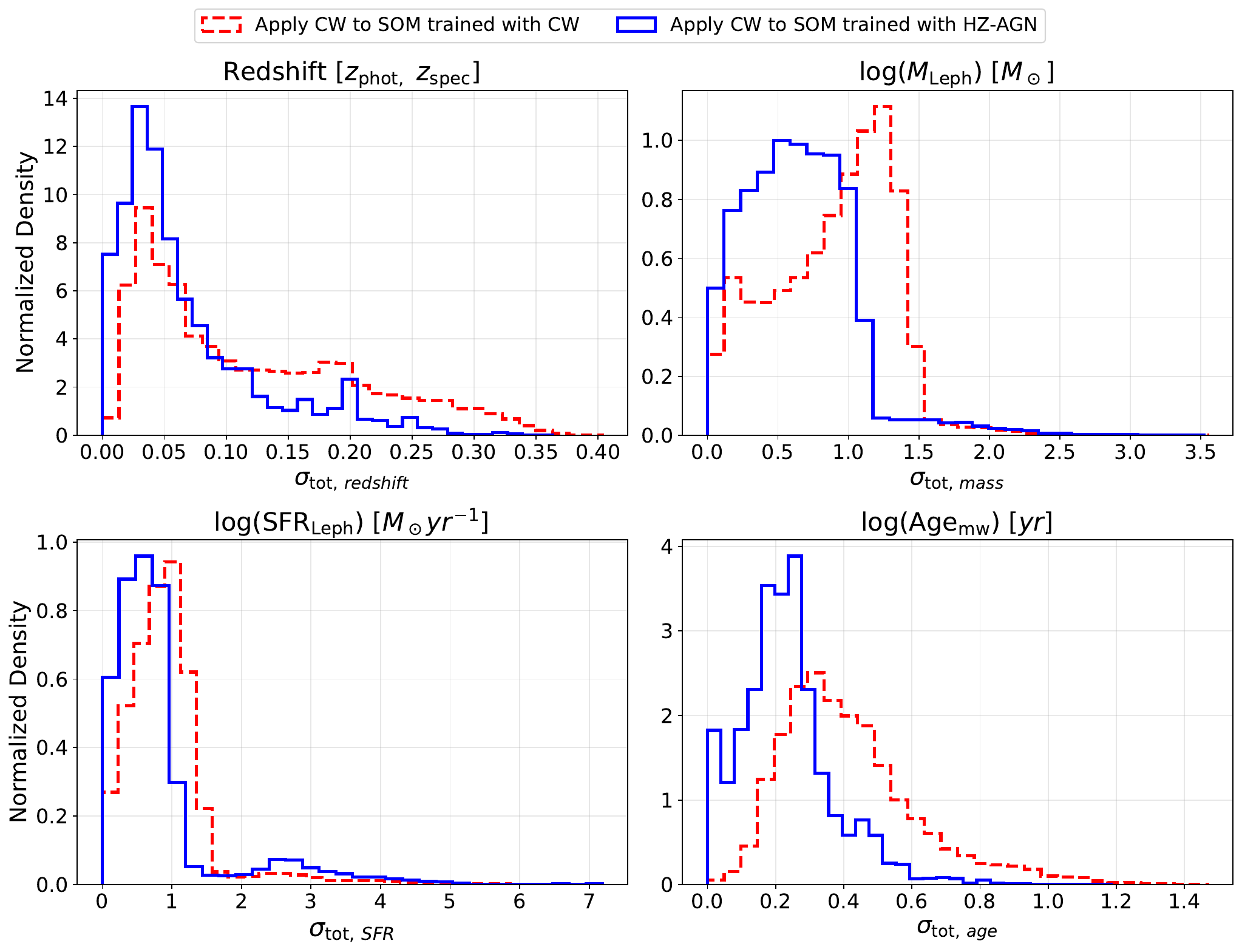}
    \caption{The total uncertainty ($\sigma_{\rm tot, param}$) for redshift, stellar mass, SFR, and mass-weighted age in the $0.1$–$0.8$ redshift bin is shown for the CW data applied to the SOMs trained on both the CW and HZ-AGN datasets. Although a long tail is present for stellar mass and SFR, most of the data have relatively low uncertainties.
}
    \label{fig:model_histogram}
\end{figure}

Figure \ref{fig:model_histogram} shows the distribution of $\sigma_{\rm tot, param}$ for redshift, stellar mass, SFR, and mass-weighted age in the redshift bin $0.1-0.8$. For redshift, the uncertainties are acceptable. Although the CW applied to the SOM trained on the CW data results in lower uncertainties compared to applying it to the SOM trained on the HZ-AGN data, this difference can be seen in the redshift predictions in Figure \ref{fig:sum_redshift}. For stellar mass and SFR, the uncertainties exhibit longer tails, which may result from scattered points in the predictions of stellar mass and SFR (Figures \ref{fig:sum_mass} and \ref{fig:sum_sfr}), although most data have uncertainties below 1. For instance, when applying the CW to the SOM trained on the CW, $64.5\%$ of the outlier points in the mass estimation have $\sigma_{\rm tot, mass}$ greater than 1. For the SFR, $63.5\%$ of the outlier points have $\sigma_{\rm tot, SFR}$ greater than $0.8$, which suggests that the scatter may be caused by color–template degeneracies. The uncertainty in the mass-weighted age prediction is relatively low, prior to applying the correction factor. The long tail in the uncertainties is primarily due to limitations in SED template fitting (e.g., bursty SFR, dust geometry, AGN contamination), which can produce the same colors for different models.

For all parameters, the CW applied to the SOM trained on the HZ-AGN data is lower than the CW applied to the SOM trained on the CW data. This suggests that using simulation data and calibrating it with CW data may help the SOM to be trained more effectively. In future work, using templates could be examined to more precisely explore the effects of the choice of initial functions and templates on the prediction of physical parameters with the SOM.

\subsubsection{Attenuation Bias}
\label{subsec:attenuation_bias}
In this subsection, we examine whether photometric magnitude errors can affect the results. In Appendix \ref{ap:SOM_sim}, we trained the SOM using the intrinsic (error-free) magnitudes of the HZ-AGN sample and then applied the same intrinsic magnitudes to the trained SOM. We found that both the quantization error and the redshift prediction improved, although the predicted physical parameters did not change significantly. This improvement suggests that the diversity in magnitudes introduced by simulated observational errors limits the SOM's ability to cluster galaxies effectively, leading to greater dispersion, particularly in redshift estimation.

When training the SOM with the CW data, we also observe significantly more degeneracy compared to the SOM trained on the simulation. This degeneracy may be due to differences between the simulation and observational data, as the CW sample contains more complex, noisier colors and possibly greater photometric uncertainties. When applying the CW data to the SOM trained and calibrated on the HZ-AGN data, the effects of photometric uncertainties become more important in determining whether the results can be trusted.

\begin{figure}
	\includegraphics[width=1\columnwidth]{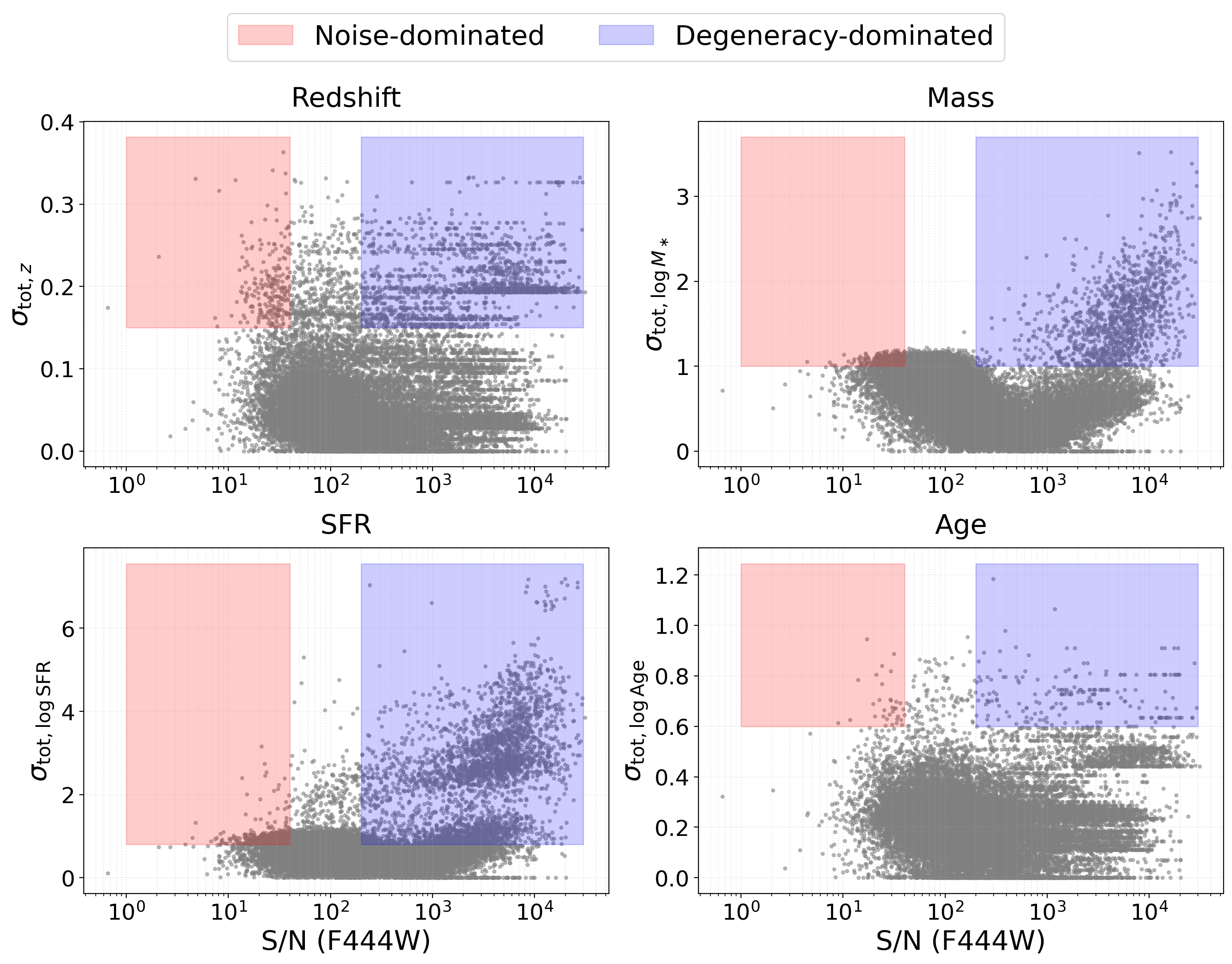}
    \caption{$\sigma_{\rm tot,param}$ versus the signal-to-noise ratio (S/N) in the F444W band for the CW predictions applied to the SOM trained on HZ-AGN data in the redshift bin $0.1-0.8$. The pink region shows low S/N and high $\sigma_{\rm tot,param}$, indicating the effect of photometric uncertainty on the predictions, while the blue region shows high S/N and high $\sigma_{\rm tot,param}$, reflecting the degeneracy of the templates in the SED-fitting parameters.
}
    \label{fig:attenuation_plot}
\end{figure}

Considering the $\sigma_{\rm tot,param}$ in the redshift bin $0.1-0.8$, where more scattered points are present, we examine the signal-to-noise ratio (S/N) in the F444 band (measured from native-resolution, i.e., non-PSF-homogenized, images in 0.2" diameter apertures) to determine whether high $\sigma_{\rm tot,param}$ values occur at low or high S/N. Figure \ref{fig:attenuation_plot} shows $\sigma_{\rm tot,param}$ as a function of S/N for the CW data applied to the SOM trained on the HZ-AGN data. The pink and blue regions correspond to low S/N and high S/N in areas of high $\sigma_{\rm tot,param}$, respectively. The pink regions show points where the effect of photometric uncertainty on the SOM prediction is small, whereas the blue regions indicate points where model degeneracies lead to high $\sigma_{\rm tot,param}$.

\subsubsection{Selection Functions}
\label{subsec:selection_function}
The predictive performance of the SOM also depends on the selection functions applied to the data. Training the SOM separately in each redshift bin, using training and test samples of equal size and applying a magnitude completeness limit, improves the training process. However, this also reduces the total number of available data points, especially for the HZ-AGN dataset, where the original sample size is larger. Although increasing the total size of the training and test samples provides more data, it makes the SOM harder to train efficiently. Training over the full redshift range introduces further challenges in parameter estimation. In contrast, bin-wise training reduces the scatter in parameters such as redshift and stellar mass, even though fewer data points are used in each bin.

When training the SOM with observational data, we include all sources, such as galaxies, stars, and X-ray detected objects. To assess whether including non-galaxy objects introduces noise into the SOM topology, we performed a test training only on galaxies. The results did not show significant changes in the predicted parameters, indicating that including non-galaxy objects does not negatively affect performance and may help reduce potential biases.

One selection function that may affect the results is the choice of the reference magnitude used to normalize stellar mass and SFR. For HZ-AGN data, We adopt $F444W = 22$, and for CW data, $F444W = 23$ (Equation~\ref{eq:normalization}). In previous studies, the $i$ and $J$ bands were used as the reference magnitude \citep{davidzon2019, davidzon2022, latorre2024}. To assess the impact of this choice, we also consider $i = 24$, $J = 23$, and $K_s = 23$, which correspond to limits fainter than $90\%$ of the CW data. We then test these alternative reference magnitudes using a SOM trained on the CW data and applied to the CW dataset.

\begin{figure}
	\includegraphics[width=1\columnwidth]{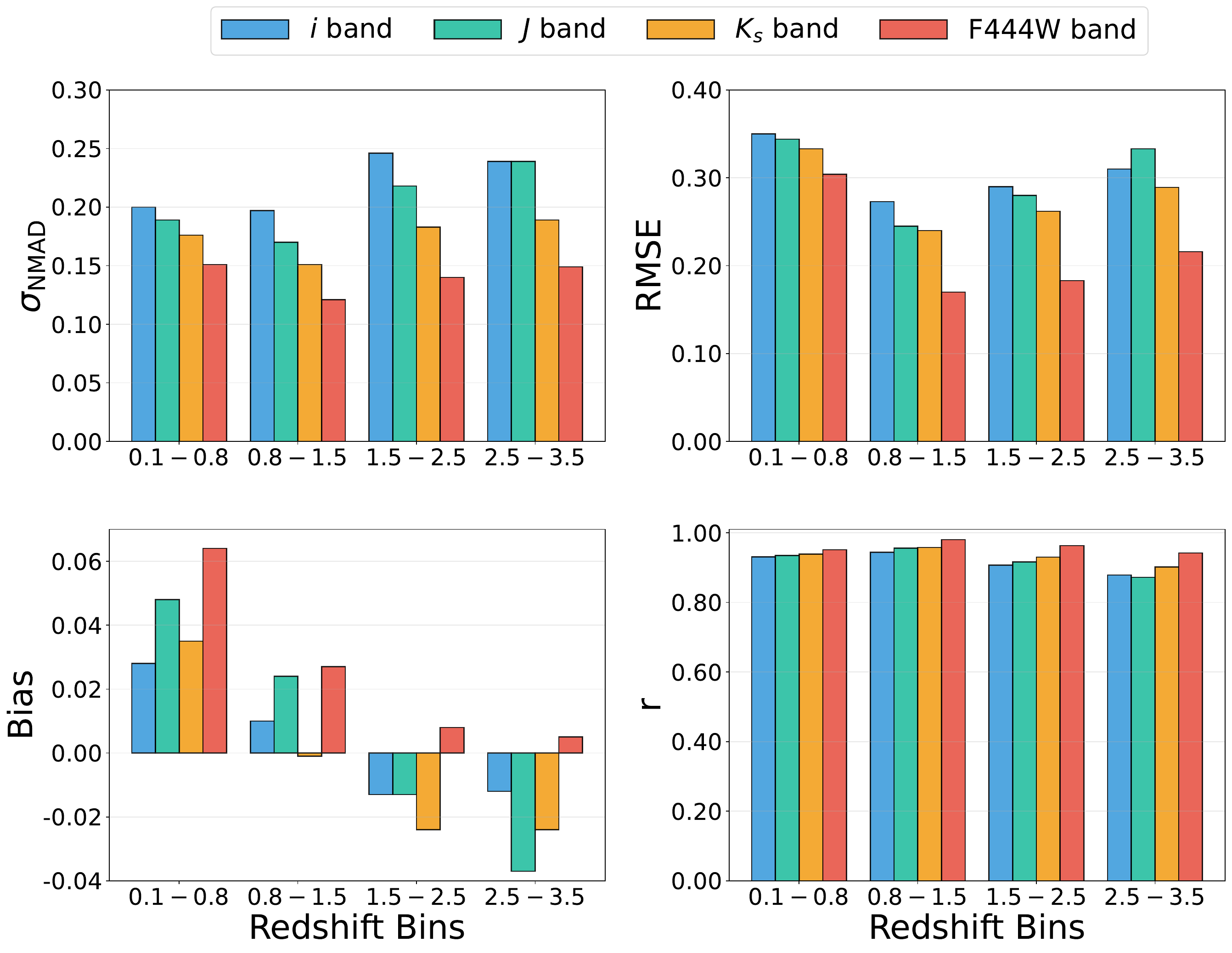}
    \caption{Comparison of stellar mass prediction metrics for different reference magnitudes used for normalization (i, J, $\mathrm{K}_\mathrm{s}$, and F444W), based on a SOM trained on and applied to the CW data. The metrics indicate that F444W provides the best predictive performance.
}
    \label{fig:normalization_com}
\end{figure}

Figure~\ref{fig:normalization_com}  shows the stellar mass prediction metrics for different normalization choices. Normalization using F444W outperforms all other bands, achieving mean $\sigma_\mathrm{NMAD} = 0.140 \pm 0.012$ and $\mathrm{RMSE} = 0.218 \pm 0.052$, compared to i-band ($\sigma_\mathrm{NMAD} = 0.221 \pm 0.022$, $\mathrm{RMSE} = 0.306 \pm 0.029)$, J-band ($\sigma_\mathrm{NMAD} = 0.204 \pm 0.027$, $\mathrm{RMSE} = 0.301 \pm 0.040$), and K-band ($\sigma_\mathrm{NMAD} = 0.175 \pm 0.015$, $\mathrm{RMSE} = 0.281 \pm 0.035$). This represents improvements of $36\%$ in $\sigma_\mathrm{NMAD}$ and $29\%$ in RMSE relative to i-band normalization, with F444W performing better in all four redshift bins tested. When F444W band is unavailable, K-band normalization provides the next-best performance, with $\sigma_\mathrm{NMAD}$ reduced by $21\%$ and $14\%$ relative to i-band and J-band, respectively. For SFR predictions, the choice of reference magnitude has a weaker impact on the results.

The distribution of physical parameters in the training data also affects the SOM prediction. For example, in the case of sSFR, the high-sSFR regime is sparsely populated, and the model performs worse in this range because the sample is dominated by low-sSFR galaxies. When the SOM is trained on the full redshift range using only the CW data, a larger scatter is observed, especially in the predictions of stellar mass and SFR. The sSFR can be predicted directly from the sSFR labels of the SOM, but dividing the predicted SFR by the predicted stellar mass reduces both the systematic offsets and the scatter in the predictions. This indicates that calculating sSFR in this way, similar to the procedure used in SED fitting, yields better results.

Another selection effect is the difference in magnitude completeness between the simulation and the observational data. Since the observational dataset is deeper, applying a magnitude cut such as $m_{\rm F444W} < 22$ removes a larger fraction of objects. This may contribute to differences between the predictions for the HZ-AGN and CW samples. The CW data include fainter galaxies, while the simulation remains more complete at brighter magnitudes.

Considering the median of the physical parameters in each SOM cell may affect the results if the galaxies in a cell have a wide range of physical parameters. We also considered using the mean or percentiles, but these did not provide advantages in improving the parameter estimates. In future studies, considering the full range of physical parameters in each cell and applying some form of probabilistic function may lead to more reliable results.

\subsubsection{Covariate Shifts}
\label{subsubsec:Covariate}
We define three metrics to assess covariate shift and validate our distribution alignment procedure: the Wasserstein distance (measuring distributional similarity), the SOM overlap ratio (quantifying coverage of feature space), and underpopulation flags (identifying samples in underpopulated training regions) (See Appendix \ref{ap:covariate_shfits} for more details about these metrics).

Table~\ref{tab:covariate_shift} presents these metrics before and after applying the covariate shift correction across different redshift bins. The results demonstrate an improvement: the mean Wasserstein distance decreased by $62\%$ in the lowest redshift bin (from $0.113$ to $0.043$) and by $61\%$ in the highest bin (from $0.192$ to $0.074$), indicating excellent distributional alignment. The SOM overlap ratio remained at or near $100\%$ in all redshift bins both before and after correction, confirming that our training data comprehensively spans the observational feature space with no extrapolation required. Underpopulation flagging identified $10$–$12\%$ of samples in sparsely populated regions before correction, which decreased to $4$–$10\%$ after alignment, demonstrating that the correction effectively shifted the observational data into well-sampled regions of the training distribution. Based on this analysis, we remove galaxies in the observational data that lie in underpopulated regions.

\begin{table}
\centering
\small
\caption{Covariate shift metrics (before $\rightarrow$ after alignment) across redshift bins.}
\label{tab:covariate_shift}
\setlength{\tabcolsep}{4pt}
{
\begin{tabular}{lccc}
\hline\hline
Redshift bin & SOM overlap (\%) & Wasserstein distance & Flags (\%) \\
\hline
$0.1\!\leq\!z\!<\!0.8$ & 99.8$\rightarrow$99.9 & 0.113$\rightarrow$0.043 & 10.6$\rightarrow$3.6 \\
$0.8\!\leq\!z\!<\!1.5$ & 100.0$\rightarrow$100.0 & 0.088$\rightarrow$0.050 & 10.2$\rightarrow$5.7 \\
$1.5\!\leq\!z\!<\!2.5$ & 100.0$\rightarrow$99.9 & 0.147$\rightarrow$0.050 & 10.3$\rightarrow$5.9 \\
$2.5\!\leq\!z\!<\!3.5$ & 100.0$\rightarrow$100.0 & 0.192$\rightarrow$0.074 & 11.7$\rightarrow$10.2 \\
\hline
\end{tabular}
}
\end{table}

\subsection{Strengths and Weaknesses of SOM for Predicting Physical Parameters}
In the following, we discuss the strengths and weaknesses of using the SOM to predict the physical parameters of galaxies. One clear advantage of the SOM is its ability to reduce dimensionality and cluster data. In our case, we use $10$ color indices, and the SOM effectively groups galaxies with similar colors into the same regions. This clustering helps identify which datasets exhibit more coherent groupings. This clustering also helps us recognize degeneracies, the cases where galaxies have similar colors but different physical parameters. These degeneracies provide insight into which galaxy properties (e.g., age, metallicity, mass, and SFR) can vary independently of color, especially at lower redshifts.

Another advantage of the SOM is its ability to predict physical parameters using fewer bands than SED fitting. In this study, we used only $11$ bands, and the estimation of physical parameters remains acceptable. Additionally, the SOM required less training time, making it more practical for estimating physical parameters in large datasets.

A key feature of SOM is that it is an unsupervised method; it is trained only on colors, without using physical parameters. This can be both an advantage and a disadvantage. The advantage is that SOM relies purely on color information, allowing us to explore the relationship between colors and physical properties more deeply. However, this approach can introduce challenges, particularly in redshift estimation. Galaxies with similar colors but different physical parameters may be mapped to the same SOM cell, leading to inaccurate redshift predictions.

Stellar mass and SFR predictions are less affected by this limitation because they are renormalized based on galaxy magnitudes. However, redshift estimation suffers more. In many cases, a wide range of true redshift values is compressed into a narrow predicted range due to the SOM labeling process. To mitigate this, we trained the SOM in separate redshift bins to ensure consistency in the redshift distribution. This approach improves the results, though the issue still persists. Predictions of age and sSFR also show a noticeable dispersion.

While training separate SOMs in distinct redshift bins has been employed in previous studies and improves predictions by reducing color-redshift degeneracies within each regime, this approach can introduce a potential circularity when applied to redshift estimation on real datasets. In this study, our primary goal is to test whether SOMs can be used to predict physical parameters (including redshift). This potential circularity can be effectively managed as the volume of available spectroscopic data grows through current and upcoming wide-field surveys. Larger, high-confidence spectroscopic datasets allow for the calibration of the color–redshift relation and the reliable assignment of galaxies to the correct redshift regime without relying solely on photometric priors. \citet{buchs2019} addressed a similar issue by combining wide-field photometry, deep multi-band photometry, and spectroscopic redshifts to calibrate redshift distributions using multiple SOMs in a non-circular way. In future studies, it will be important to further investigate strategies to reduce color degeneracies and to exploit increasingly large spectroscopic datasets, enabling more robust applications of SOMs to purely photometric samples.

In \citet{cadiou2025}, the authors suggested that due to the complexity of the observational data, the latent space for dimensionality reduction techniques should be at least four-dimensional (ideally four to five dimensions) to minimize information loss. In future studies, it will be essential to use higher-dimensional reductions than 2D to investigate whether the prediction of physical parameters, especially redshift, can be improved. In addition, Kalantari et al. (in preparation) are investigating whether deep learning models can improve the results obtained with the SOM.

Another challenge arises when the SOM trained on one dataset is applied to a different one. To achieve reliable results, it is crucial to first align the distribution of the input data with that of the training data or SOM weights, and second, to calibrate the SOM across datasets. Differences between simulation and observational data, as well as the intrinsic structure of the SOM, can contribute to performance gaps. Future work should focus on refining the SOM structure to improve predictive accuracy, particularly by better linking the SOM topology to underlying features such as galaxy colors, or by exploring the use of higher-dimensional SOMs for estimating physical parameters.

For redshift, the metrics became slightly worse and the dispersion increased slightly, though the same patterns were observed. For stellar mass, a similar pattern is seen, with slightly worse metrics. For SFR and sSFR, the metrics are also slightly worse, but overall trends remain consistent. For age, the prediction is slightly worse. After applying the correction factor, the results improve, although they are still not better than those obtained without errors.  Only in the redshift bin $0.1$–$0.8$ is a slight improvement evident. These results indicate that including photometric errors slightly affects the predictions, but does not significantly improve or degrade them.

\subsection{Guidelines for Galaxy Parameter Estimation Using SOMs}

In this section, we provide practical guidelines for observers wishing to apply SOMs to photometric data for estimating galaxy physical parameters, based on key lessons from this work.

\begin{itemize}

\item \textbf{Redshift Binning. }  
Training the SOM in separate redshift bins is more effective than training across the entire redshift range (we used $0.1$–$0.8$, $0.8$–$1.5$, $1.5$–$2.5$, $2.5$–$3.5$). This approach reduces color–redshift degeneracies (Section~\ref{subsec:data_prepration}).

\item \textbf{Choice of Training Dataset. }
Due to inherent biases and model misspecification in SED fitting methods (e.g., LePhare), training on simulated data (here, HZ-AGN) is highly effective (Section~\ref{subsec:som_sim}). Additionally, training the SOM on simulation data without added photometric uncertainties reduces quantization errors and improves redshift predictions (Appendix~\ref{ap:SOM_sim}).

\item \textbf{Use of Spectroscopic Redshifts. }
Substitute photometric redshifts with spectroscopic redshifts for training and labeling whenever available. This reduces color degeneracies in SOM training and leads to more accurate redshift predictions for galaxies with spectroscopic redshifts compared to those relying on photometric estimates (Section~\ref{subsec:data_prepration}).

\item \textbf{Filter Selection. }
We find that using the JWST F277W and F444W filters instead of the Spitzer IRAC channels 1 and 2 substantially improves the quality of SOM training and the accuracy of parameter estimation (Section~\ref{subsec:input_data}).

\item \textbf{Magnitude Completeness. }
We apply the magnitude completeness cut in the F444W filter. We use $F444W<24.8$ for the HZ-AGN data and $F444W<27$ for the CW data (Section~\ref{subsec:data_prepration}).

\item \textbf{Inclusion of Non-Galaxy Objects in Training the SOM on CW Data. }  
When training the SOM on CW data, we include all observational objects—galaxies, stars, and X-ray sources—in the training set. In the test set, only galaxy objects are retained. This inclusion helps reduce biases during the training phase (Sections~\ref{subsec:data_prepration}).

\item \textbf{Normalization of Mass and SFR Using the F444W Filter. }  
For labeling the SOM, the training is performed using colors, while stellar mass and SFR depend on the full SED. Therefore, these quantities must be normalized. We adopt the F444W band as the reference magnitude, as it provides better performance than the i, J, or K bands. In particular, longer-wavelength filters are more effective for normalization (Section~\ref{subsec:visalization_of_som}).

\item \textbf{Grid Size Optimization. }
Determine optimal SOM grid dimensions by balancing quantization error, cell occupancy, and computational cost. We define a quality score (Equation~\ref{eq:quality_score}) weighting these factors and test grids from $10\times10$ to $100\times100$ (Section~\ref{subsec:training_som}).

\item \textbf{Covariate Shift Correction. }  
When applying observational data to a SOM trained on simulations, it is essential to align the test data with the learned SOM weights. We align the color distributions of the observational and simulated datasets using a feature-wise affine transformation, the Wasserstein–KDE alignment method. This alignment promotes statistical consistency between the datasets, improves SOM population, and reduces extrapolation errors (Section~\ref{subsec:covaraite_shift_correction}).

\item \textbf{Considering Underpopulated Regions. }  
For reliable results, after aligning CW data to a SOM trained on HZ-AGN data, it is important to remove galaxies projected onto underpopulated regions, which can be identified using the Wasserstein distance or BMU metrics (Appendix~\ref{ap:covariate_shfits}). Parameter estimation performs better in regions of the parameter space that are well sampled by the training data, and vice versa—this may seem obvious, but it is critical to verify.

\item \textbf{Calibration Subset. }
When applying a simulation-trained SOM to observations, reserve $20\%$ of the observational data for calibration, independent of the test set used for final predictions. Use this calibration subset to adjust SOM cell labels to match observational properties (Section~\ref{subsec:covaraite_shift_correction}).

\item \textbf{Probabilistic Prediction and Considering Photometric Redshift. }  
During the prediction phase, instead of using point estimates, we compute a probabilistic prediction by evaluating the likelihood function while considering photometric uncertainties. The peak of the likelihood-weighted posterior PDF is then taken as the SOM-predicted value (Section~\ref{subsec:Prediction}).

\item \textbf{Deriving sSFR. }  
For sSFR estimation, we divide the predicted SFR by the predicted stellar mass, rather than directly predicting sSFR. This approach provides more accurate results and reduces systematic offsets (Appendix~\ref{ap:ssfr_estimation}).

\item \textbf{Age Corrections. }
For mass-weighted age estimates, the SOM exhibits systematic biases (For mass-weighted age estimates, the SOM tends to overestimate young ages and underestimate old ages, likely due to SED fitting limitations in the training labels. We derive empirical age-dependent correction factors and apply them to the predictions (Section~\ref{subsec:som_obs}).

\item \textbf{Total Uncertainty Estimates. }
For each predicted parameter, we compute the total uncertainty as the root sum square of the statistical and systematic uncertainties. The statistical uncertainty is defined as the standard deviation of the likelihood-based PDF, while the systematic uncertainty is taken as half of the 84th–16th percentile range of parameter values within the BMU cell. Galaxies with large total uncertainties are likely affected by color–parameter degeneracies (Section~\ref{subsubsec:model_misspecification}).

\end{itemize}


\section{Conclusions}
\label{sec:conclusions}

In this study, we used two datasets: HZ-AGN data and CW data, using JWST filters and manifold learning. Our goal was to examine whether the use of JWST filters could improve the prediction of physical parameters of galaxies. When training the SOM on HZ-AGN and CW data, the quantization errors of the two SOM types were similar. However, when training the SOM on simulation data, the variation in physical parameters among galaxies with similar colors is lower than in the observational data. In contrast, the SOM trained on observational data showed a higher degeneracy, where galaxies with similar colors but a wide range of physical parameters are assigned to the same cell. This difference may be due to the fact that the CW sample has more complex and noisier colors.

In the simulation data, the predicted physical parameters are close to the true values. This shows that these filters have a strong correlation with physical parameters. For the observational data, we explored two scenarios. The first was training the SOM with CW data, and the second was applying the SOM trained on HZ-AGN data. For the CW data, a clear increase in degeneracy was observed among physical parameters with similar colors. This affected the prediction accuracy of redshift, sSFR, and age, although the estimation of stellar mass and SFR remained acceptable. For redshift predictions, spectroscopic redshifts aligned more closely with the 1:1 relation than photometric redshifts. The dispersion may result from possible uncertainties in the physical parameters of the observational data.

We also investigated how the predictions vary across different redshift bins. In the simulation data, redshift estimation was slightly less accurate at higher redshifts. For other parameters, the predictions tended to be slightly less accurate at lower redshifts. In the CW data, predictions in the low-redshift bin were less reliable for galaxies with low stellar mass and low SFR. Some dispersion was observed, suggesting that degeneracies in the local universe may affect the performance of the SOM, particularly when there is greater diversity in physical parameters. A key novelty of this work is the use of JWST/NIRCam filters, especially F444W, which enhances SOM training and enables more accurate estimation of physical parameters compared to previous studies that relied on Spitzer/IRAC channels 1 and 2.

The performance of the SOM suggests that it can also be used for future surveys. It requires less training time than the SED fitting, but should be optimized to improve predictions in observational data, particularly for redshift estimation. This could involve adding more dimensions beyond the 2D SOM map or incorporating additional mid-infrared filters as input data. Furthermore, extending the method to unseen survey data will require additional refinement. In particular, improving the training procedure and the prediction strategy, and assessing the influence of the selection function, photometric uncertainties, and a stronger focus on templates, will be important for ensuring reliable performance. With these developments, the SOM has the potential to become a robust predictive tool for upcoming observational datasets.

\section*{Acknowledgements}

We acknowledge the contribution of the COSMOS collaboration, consisting of more than 200 scientists. More information about the
COSMOS survey can be found at https://cosmos.astro.caltech.edu/. This work was made possible by utilizing the CANDIDE cluster at the Institut d’Astrophysique de Paris. The cluster was funded through grants from the PNCG, CNES, DIM-ACAV, the Euclid Consortium, and the Danish National Research Foundation Cosmic Dawn Center (DNRF140). It is maintained by Stephane Rouberol. French COSMOS team members are partly supported by the Centre National d’Etudes Spatiales (CNES). We acknowledge the funding of the French Agence Nationale de la Recherche for the project iMAGE (grant ANR-22-CE31-0007). We thank Clotilde Laigle, Yohan Dubois, and Grégoire Aufort for providing access to the Horizon-AGN catalog used in this study. We thank the anonymous referees for their constructive and helpful comments, which improved the quality of this paper.

\section*{Data Availability}

The data underlying this article will be shared upon reasonable request to the corresponding author. The source code for training the SOM, applying the covariate shifts and predicting the physical parameters is available at the following GitHub link: \\
\url{https://github.com/flsabedini/SOM_ON_COSMOSWEB}




\bibliographystyle{mnras}
\bibliography{biblio} 

\section*{Affiliations}
\begin{itemize}
  \item[$^1$] Institute for Advanced Studies in Basic Sciences (IASBS) 444 Prof. Yousef Sobouti Blvd., Zanjan 45137-66731, Iran
  \item[$^2$] Department of Computer Science, Aalto University, PO Box 15400, Espoo 00 076, Finland
  \item[$^3$] Department of Physics, University of Helsinki, PO Box 64, 00014 Helsinki, Finland
  \item[$^4$] Helmholtz-Institut für Strahlen-und Kernphysik (HISKP), Universität Bonn, Nussallee 14-16, D-53115 Bonn, Germany
  \item[$^5$] Aix Marseille Univ, CNRS, LAM, Laboratoire d'Astrophysique de Marseille, Marseille, France
  \item[$^6$] Department of Astronomy, The University of Texas at Austin, 2515 Speedway Blvd Stop C1400, Austin, TX 78712, USA
  \item[$^7$] Cosmic Dawn Center (DAWN), Denmark
  \item[$^8$] Niels Bohr Institute, University of Copenhagen, Jagtvej 128, DK-2200, Copenhagen, Denmark
  \item[$^9$] Department of Physics, University of California, Santa Barbara, Santa Barbara, CA 93106, USA
  \item[$^{10}$] Department of Physics and Astronomy, University of Hawaii, Hilo, 200 W Kawili St, Hilo, HI 96720, USA
  \item[$^{11}$] Caltech/IPAC, MS 314-6, 1200 E. California Blvd. Pasadena, CA 91125, USA
  \item[$^{12}$] Laboratory for Multiwavelength Astrophysics, School of Physics and Astronomy, Rochester Institute of Technology, 84 Lomb Memorial Drive, Rochester, NY 14623, USA
  \item[$^{13}$] Université Paris-Saclay, Université Paris Cité, CEA, CNRS, AIM, 91191 Gif-sur-Yvette, France
  \item[$^{14}$] Department of Physics and Astronomy, University of California, Riverside, 900 University Avenue, Riverside, CA 92521, USA
  \item[$^{15}$] Department of Physics and Astronomy, University of Kentucky, 505 Rose Street, Lexington, KY 40506, USA
  \item[$^{16}$] Space Telescope Science Institute, 3700 San Martin Dr., Baltimore, MD 21218, USA
  \item[$^{17}$] Institute for Computational Cosmology, Department of Physics, Durham University, South Road, Durham DH1 3LE, United Kingdom
  \item[$^{18}$] Purple Mountain Observatory, Chinese Academy of Sciences, 10 Yuanhua Road, Nanjing 210023, China
  \item[$^{19}$] Institut d'Astrophysique de Paris, UMR 7095, CNRS, and Sorbonne Université, 98 bis boulevard Arago, 75014 Paris, France
  \item[$^{20}$] Institute of Cosmology and Gravitation, University of Portsmouth, Dennis Sciama Building, Burnaby Road, Portsmouth, PO13FX, United Kingdom
  \item[$^{21}$] Jet Propulsion Laboratory, California Institute of Technology, 4800 Oak Grove Drive, Pasadena, CA 91001, USA
  \item[$^{22}$] Department of Astronomy and Astrophysics, University of California, Santa Cruz, 1156 High Street, Santa Cruz, CA 95064, USA
  \item[$^{23}$] University of Geneva, 24 rue du Général-Dufour, 1211 Genève 4, Switzerland
  \item[$^{24}$] DTU Space, Technical University of Denmark, Elektrovej 327, 2800 Kgs. Lyngby, Denmark
  \item[$^{25}$] Thüringer Landessternwarte, Sternwarte 5, 07778 Tautenburg, Germany
  \item[$^{26}$] Department of Astronomy, University of Massachusetts, Amherst, MA 01003, USA
\end{itemize}







\appendix

\section{Data Distribution and Distance Metrics in the SOM}
\label{ap:SOM}
\begin{figure}
	\includegraphics[width=1\columnwidth]{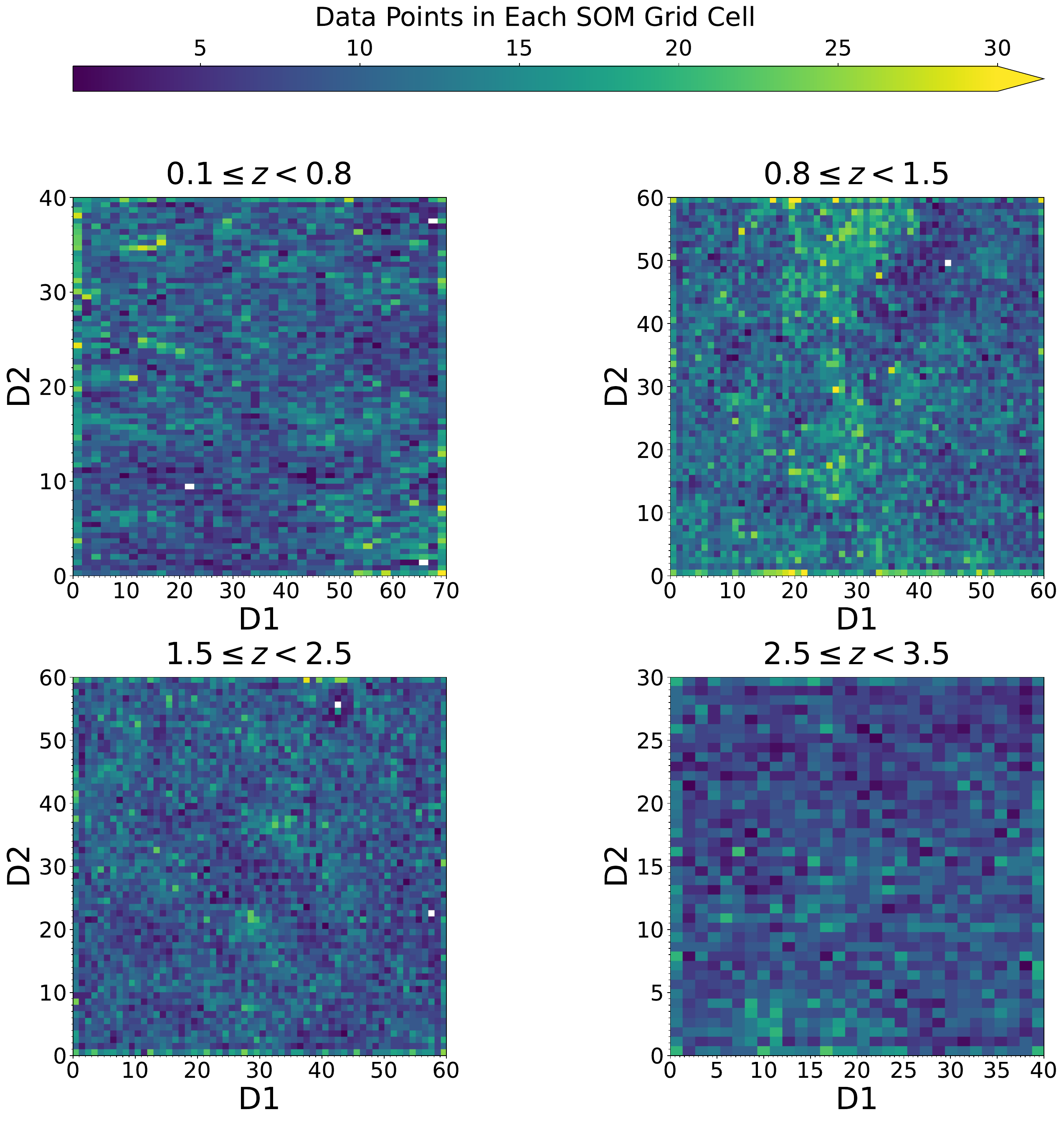}
    \caption{Number of data points in each cell of the SOM trained on simulation data (HZ-AGN). Most cells contain more than five galaxies.}
    \label{fig:datapoint_sim}
\end{figure}

\begin{figure}
	\includegraphics[width=1\columnwidth]{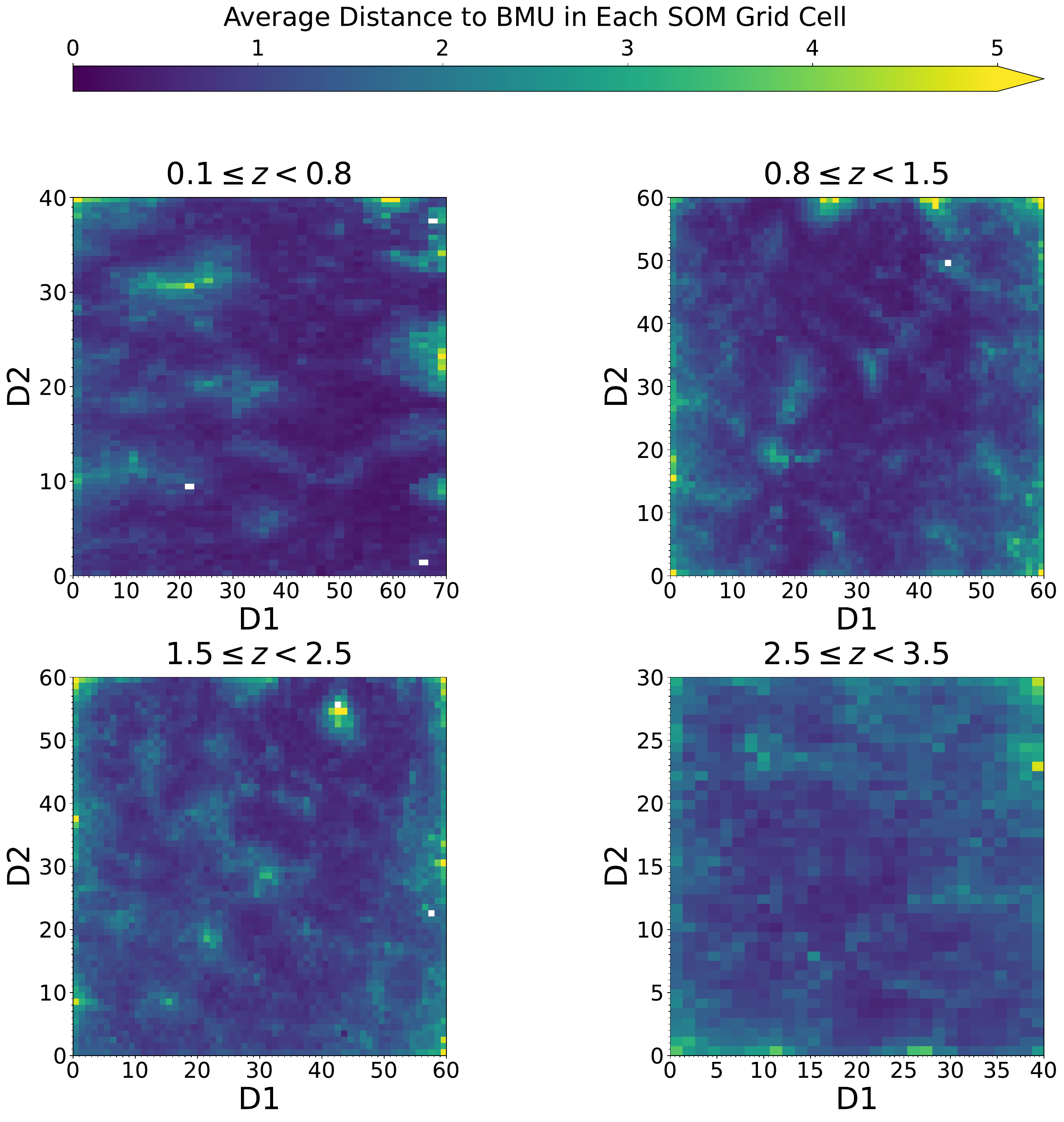}
    \caption{Average Euclidean distance between the weights of the BMU and the normalized color inputs in each cell of the SOM trained on simulation data (HZ-AGN). Most cells have an average distance below $2$.}
    \label{fig:distance_sim}
\end{figure}

\begin{figure}
	\includegraphics[width=1\columnwidth]{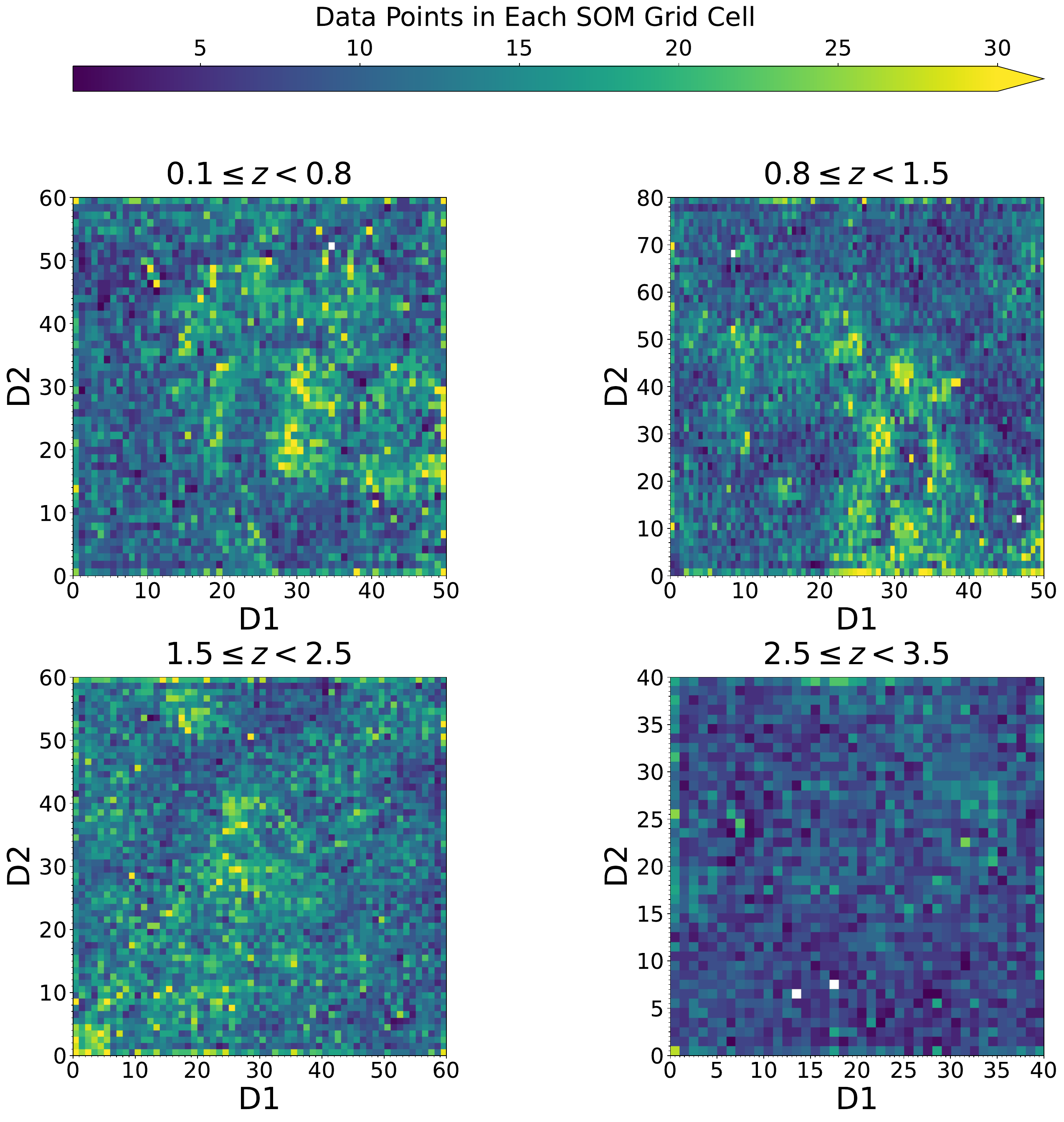}
    \caption{Number of data points in each cell of the SOM trained on observational data (CW).}
    \label{fig:datapoint_obs}
\end{figure}

\begin{figure}
	\includegraphics[width=1\columnwidth]{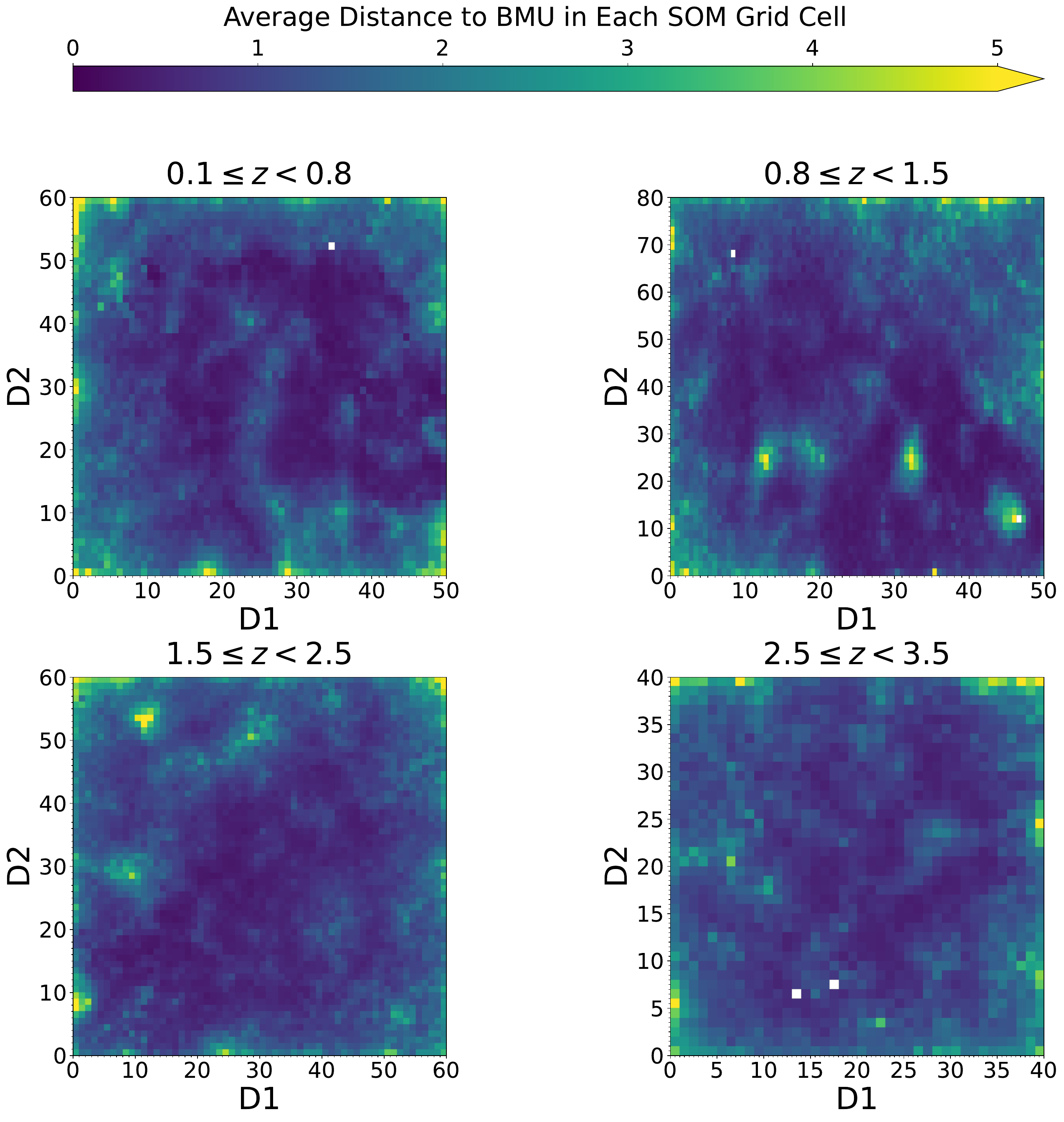}
    \caption{Average Euclidean distance between the weights of the BMU and the normalized color inputs in each cell of the SOM trained on observational data (HZ-AGN). Cells with high average distance are located at the borders of the SOM and tend to have a higher number of data points.}
    \label{fig:distance_obs}
\end{figure}

To use the trained SOM for estimating physical parameters, it is important to ensure that the number of data points in each cell is not too low and that the average Euclidean distance to the input data is not too high. Figures \ref{fig:datapoint_sim} and \ref{fig:distance_sim} show the number of data points and the average Euclidean distance between the weights of the BMU and the normalized input points for each SOM trained on simulation data within each redshift bin. These figures indicate that most cells contain more than $5$ data points, and the average distance for most cells is less than $5$. It is important to note that high average distances typically occur at the borders of the map, where the number of data points is higher. This effect has also been observed in previous studies, such as \citet{davidzon2019, davidzon2022, latorre2024}, and is referred to as a boundary effect, which can also cause a large spread in the differences between the physical parameters of galaxies assigned to that cell, as well as systematic overestimation. Figures \ref{fig:datapoint_obs} and \ref{fig:distance_obs} present the corresponding results for the SOM trained on observational data, showing similar patterns, although a larger number of cells exhibit high quantization error.

\section{SOM Training With HZ-AGN Intrinsic Magnitudes}
\label{ap:SOM_sim}

We tested how the results change when using the intrinsic magnitudes of the HZ-AGN data to train the SOM. For the same grid size as that used in the SOM trained on the HZ-AGN data, the quantization error improves. Both the quantization error and cell occupation are also acceptable when using intrinsic magnitudes. The average quantization error for each redshift bin is as follows: $0.44$ for $0.1-0.8$, $0.43$ for $0.8-1.5$, \textbf{$0.44$} for $1.5-2.5$, and $0.55$ for $2.5-3.5$. These quantization errors clearly indicate that galaxies are better clustered compared to the SOM trained with magnitudes that include observational errors.

\begin{figure}
	\includegraphics[width=1\columnwidth]{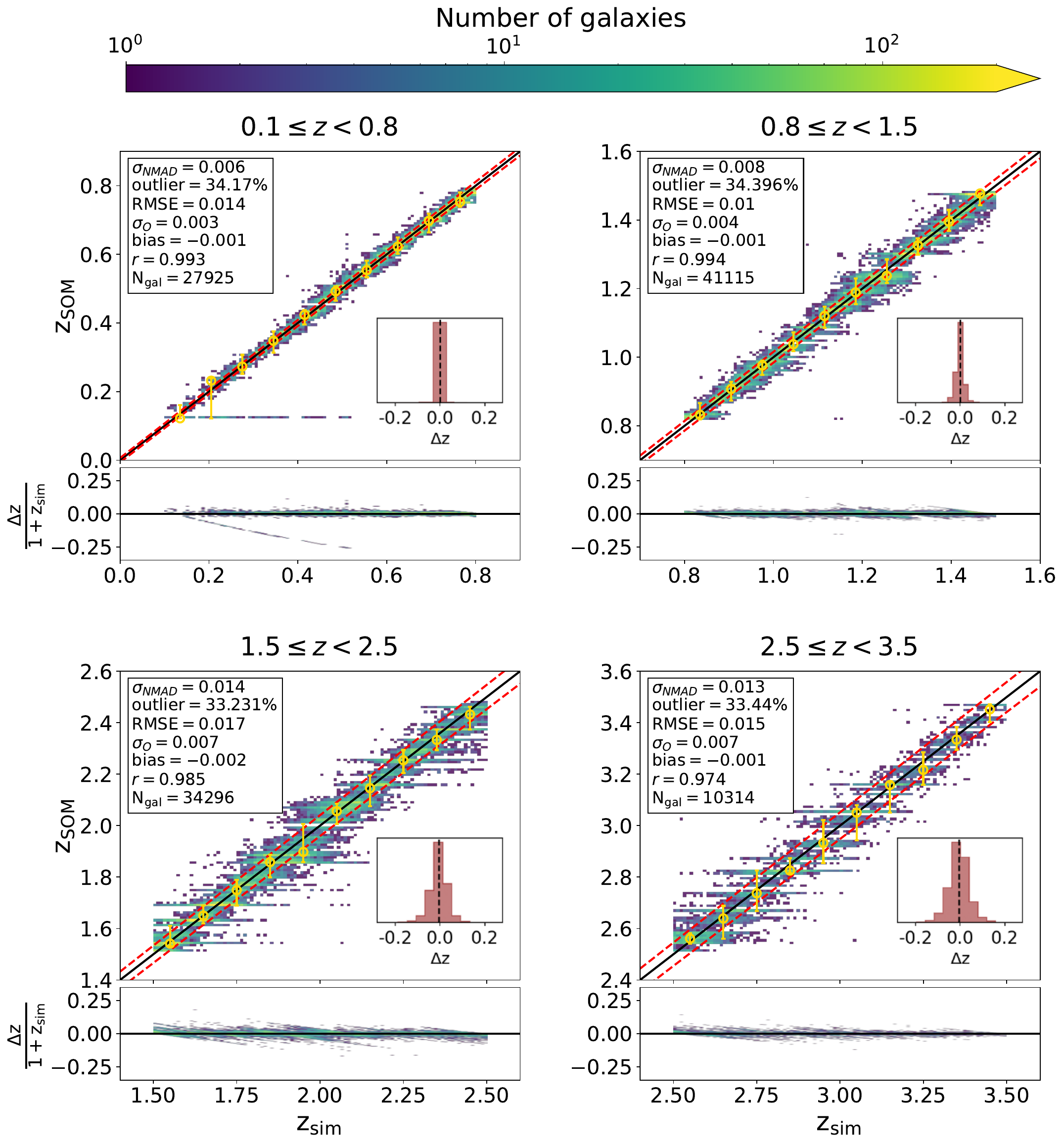}
    \caption{Comparison between predicted and true redshifts using the SOM trained on intrinsic magnitudes from HZ-AGN data. See Figure~\ref{fig:redshift_sim} for details about the lines.}
    \label{fig:redshift_sim_noerr}
\end{figure}

For physical parameter estimation, predictions for stellar mass, SFR, sSFR, and age slightly improved. However, the prediction for redshift improved significantly, especially in the higher redshift bins, as seen in Figure \ref{fig:redshift_sim_noerr}, compared to the SOM trained on magnitudes with errors shown in Figure \ref{fig:redshift_sim}.

\section{Covariate Shift Metrics}
\label{ap:covariate_shfits}
This appendix describes the metrics used to quantify the covariate shift between the training distribution $P$ and the observational distribution $Q$.

The one-dimensional Wasserstein distance between two distributions $P$ and $Q$ is defined as:
\begin{equation}
W_1(P, Q) = \int_{-\infty}^{\infty} |F_P(x) - F_Q(x)| \, dx,
\end{equation}
where $F_P$ and $F_Q$ are the cumulative distribution functions. For data with $d$ features, we report the mean Wasserstein distance: 
$
\bar{W}_1 = \frac{1}{d}\sum_{i=1}^{d} W_1(P_i, Q_i).
$

For the SOM overlap ratio, $\mathcal{C}_{\rm train}$ is defined as the set of SOM cells containing at least one training sample, and $\mathcal{C}_{\rm obs}$ as the set containing observational samples. The overlap ratio is defined as:
\begin{equation}
    R_{\rm overlap} = \frac{|\mathcal{C}_{\rm train} \cap \mathcal{C}_{\rm obs}|}{|\mathcal{C}_{\rm obs}|}.
\end{equation}
A value of 100\% indicates that all observational samples lie within regions covered by the training data.

 A sample is flagged as potentially unreliable due to underpopulation if it satisfies at least three of the following four criteria:
\begin{enumerate}
    \item The distance from the sample to its best-matching SOM unit exceeds the 95th percentile of training-sample distances, reflecting an unusual combination of features not well represented by the SOM prototypes.

    \item The Euclidean distance to the nearest training sample exceeds the 90th percentile of all observational-to-training distances, indicating a region of feature space sparsely populated by training data.

    \item The SOM cell to which the sample is assigned contains fewer training samples than the 5th percentile of training-cell populations, implying insufficient training coverage in that region.

    \item At least one feature value falls outside the 2.5--97.5 percentile range of the training data, indicating an extreme or atypical value.
\end{enumerate}
Samples satisfying at least three of these four criteria are considered to lie in underpopulated or poorly sampled regions of the training distribution.

\section{sSFR Estimation}
\begin{figure}
	\includegraphics[width=1\columnwidth]{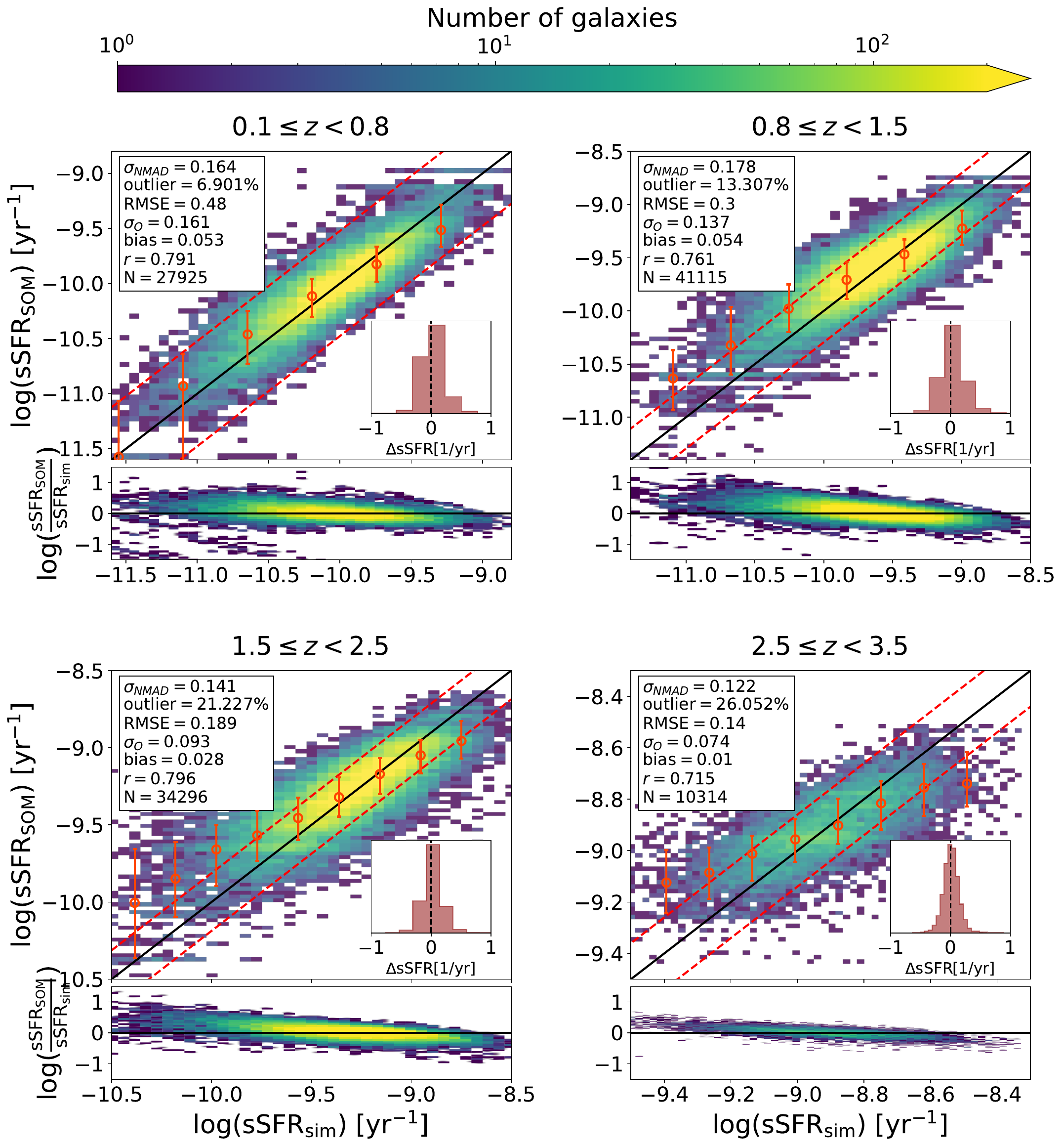}
    \caption{Comparison of the sSFR obtained by dividing the predicted SFR by the predicted stellar mass from the SOM trained on HZ-AGN data with the true values. Overestimation and underestimation occur at low and high sSFR. See the caption of Figure~\ref{fig:mass_sim} for details about the lines. }
    \label{fig:ssfr_sim}
\end{figure}

\begin{figure}
	\includegraphics[width=1\columnwidth]{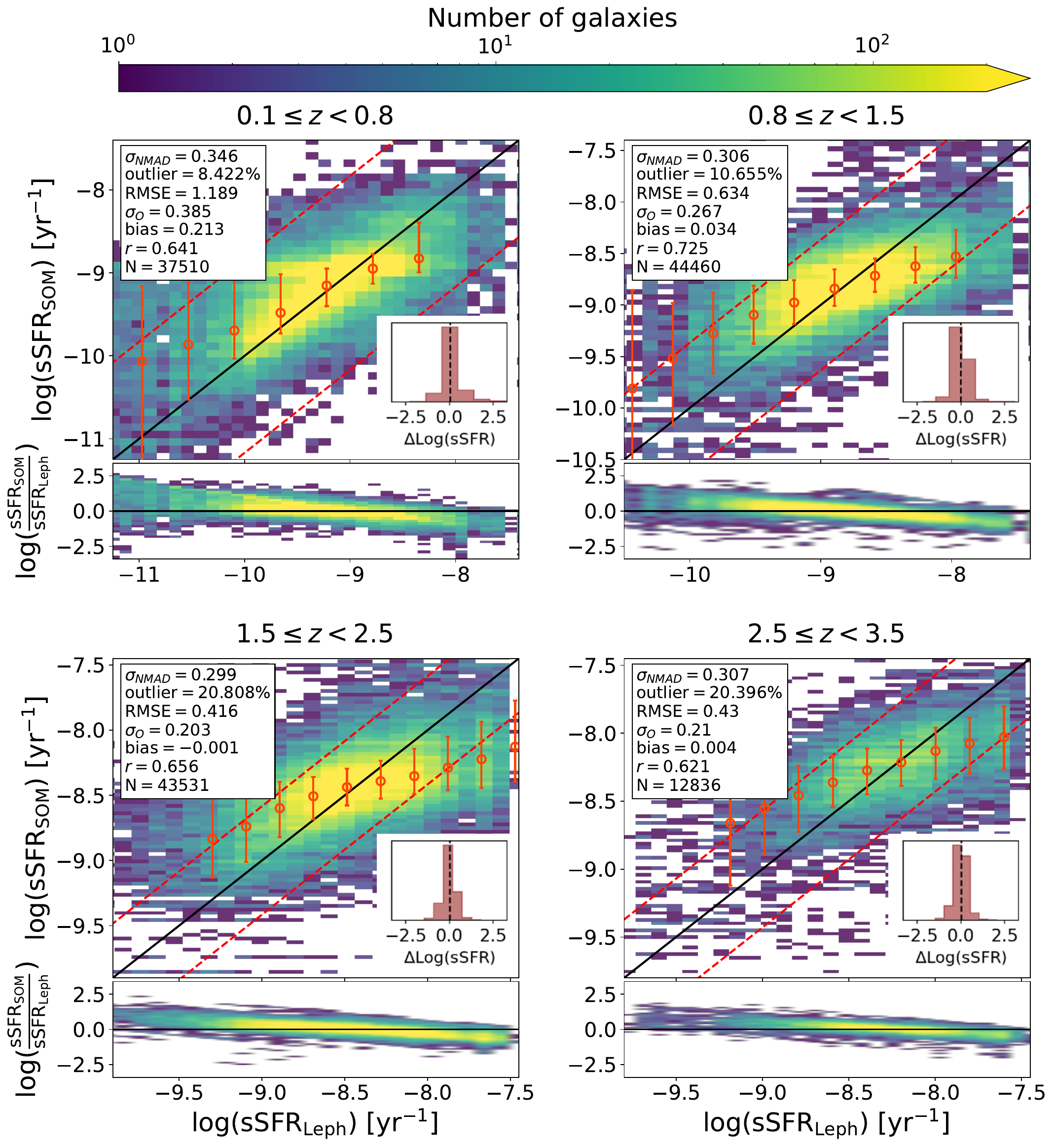}
    \caption{Comparison between the sSFR obtained by dividing the predicted SFR by the predicted stellar mass from the SOM trained on CW data and the sSFR estimated using SED fitting (LePhare). See the caption of Figure~\ref{fig:mass_sim} for details about the lines.
}
    \label{fig:ssfr_obs}
\end{figure}

\begin{figure}
	\includegraphics[width=1\columnwidth]{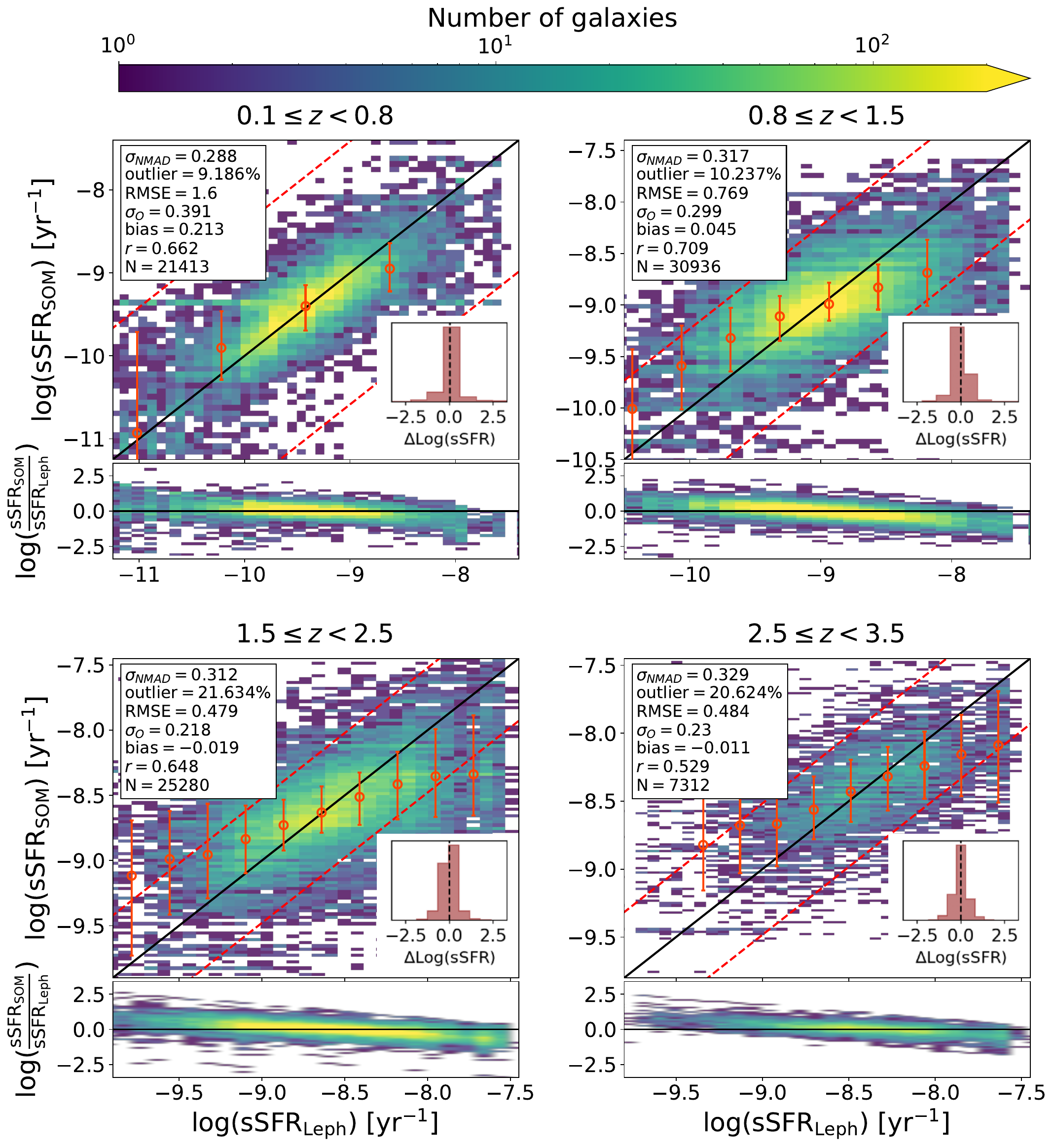}
    \caption{Comparison of the sSFR obtained by dividing the predicted SFR by the predicted stellar mass, using the CW data applied to the SOM trained on HZ-AGN data, with the sSFR estimated using SED fitting (LePhare). The overestimation at lower sSFR and underestimation at higher sSFR are observed. See the caption of Figure~\ref{fig:mass_sim} for details about the lines.
}
    \label{fig:ssfr_sim_obs}
\end{figure}

In this appendix, we present the estimated sSFR, obtained by dividing the estimated SFR by the estimated stellar mass, for the HZ-AGN and CW datasets in three cases: (i) applying the SOM trained on HZ-AGN data to the HZ-AGN sample (Figure~\ref{fig:ssfr_sim}); (ii) applying the SOM trained on CW data to the CW sample (Figure~\ref{fig:ssfr_obs}); and (iii) applying the SOM trained on HZ-AGN data to the CW sample (Figure~\ref{fig:ssfr_sim_obs}). These cases correspond to the three SOM configurations discussed in Sections~\ref{subsec:som_sim}, \ref{subsec:som_obs}, and \ref{subsec:som_sim_obs}, respectively.
\label{ap:ssfr_estimation}

\bsp	
\label{lastpage}
\end{document}